\documentclass[final,authoryear,3p,times]{elsarticle}
\usepackage{afterpage}
\usepackage{algorithm}
\usepackage{algpseudocode}
\usepackage{amsmath}
\usepackage{amssymb}
\usepackage{amsthm}
\usepackage{bm}
\usepackage{booktabs}
\usepackage{calc}
\usepackage{caption}
\usepackage{subcaption}
\usepackage[table]{xcolor} % Required for \rowcolor
\usepackage{amsmath,amssymb}
\usepackage{booktabs}
\usepackage{tabularx}

\usepackage{enumerate}
\usepackage{graphicx}
\usepackage{ifthen}
\usepackage[parfill]{parskip} % no indent after line break
\usepackage{multicol}
\usepackage{siunitx}
\usepackage{ulem}

\usepackage[colorlinks]{hyperref}
\usepackage[nameinlink,capitalise]{cleveref}
\crefname{appendix}{}{}
\usepackage{tikz}
\usetikzlibrary{arrows}
\usetikzlibrary{automata}
\usetikzlibrary{calc}
\usetikzlibrary{chains}
\usetikzlibrary{positioning}
\usetikzlibrary{shapes}

\usepackage{varwidth}
\usepackage{makecell}

\usepackage{tcolorbox}
\newtcolorbox[auto counter]{problem}[2][]{colframe=blue!30, colback=blue!5, coltitle=black, title=Problem~\thetcbcounter~ #2,#1}

\usepackage{spverbatim} % like verbatim with linebreaks
\usepackage{listings} % like verbatim with linebreaks

\definecolor{Demiray}{rgb}{0.1725, 0.6275, 0.1725}
\definecolor{Gent}{rgb}{1.0, 0.4902, 0.0431}
\definecolor{Holzapfel}{rgb}{0.1216, 0.4667, 0.7059}
\definecolor{MooneyRivlin}{rgb}{0.8392, 0.1529, 0.1569}
\definecolor{Ogden}{rgb}{0.5804, 0.4039, 0.7412}
\definecolor{BlatzKo}{rgb}{0.5490, 0.3373, 0.2941}
\definecolor{NeoHooke}{rgb}{0.7373, 0.7412, 0.1333}

\usepackage{cleveref}

\edef\svtheparindent{\the\parindent}
\usepackage{parskip}
\usepackage{ulem}
\parindent=\svtheparindent\relax

\definecolor{ForestGreen}{RGB}{34,139,34}
\definecolor{InternationalOrange}{rgb}{1.0, 0.31, 0.0}
\definecolor{WineRed}{RGB}{139,0,0}

\newcommand{\boldface}[1]{\boldsymbol{#1}}  % italic (slanted)
\newcommand{\bfa}{\boldface{a}}
\newcommand{\bfb}{\boldface{b}}
\newcommand{\bfc}{\boldface{c}}
\newcommand{\bfd}{\boldface{d}}
\newcommand{\bfe}{\boldface{e}}
\newcommand{\bff}{\boldface{f}}

\newcommand{\bfu}{\boldface{u}}

\newcommand{\bfx}{\boldface{x}}

\newcommand{\bfA}{\boldface{A}}
\newcommand{\bfB}{\boldface{B}}
\newcommand{\bfC}{\boldface{C}}

\newcommand{\bfF}{\boldface{F}}

\newcommand{\bfI}{\boldface{I}}

\newcommand{\bfK}{\boldface{K}}

\newcommand{\bfM}{\boldface{M}}
\newcommand{\bfN}{\boldface{N}}

\newcommand{\bfP}{\boldface{P}}
\newcommand{\bfQ}{\boldface{Q}}
\newcommand{\bfR}{\boldface{R}}
\newcommand{\bfS}{\boldface{S}}

\newcommand{\bfU}{\boldface{U}}
\newcommand{\bfV}{\boldface{V}}
\newcommand{\bfW}{\boldface{W}}
\newcommand{\bfX}{\boldface{X}}

\newcommand{\bftheta}{\boldsymbol{\theta}}

\newcommand{\bfsigma}{\boldsymbol{\sigma}}

\newcommand{\bfphi}{\boldsymbol{\phi}}

\newcommand{\bfpsi}{\boldsymbol{\psi}}

\newcommand{\bfSigma}{\boldsymbol{\Sigma}}

\newcommand{\Rset}{\mathbb{R}}

\newcommand{\transposed}{^{\top}}
\newcommand{\mtransposed}{^{-\top}}

\newcommand{\argmin}{\operatornamewithlimits{arg\ min}}

\newcommand{\be}{\begin{equation}}
\newcommand{\ee}{\end{equation}}
\newcommand{\bea}{\begin{equation}\begin{aligned}}
\newcommand{\eea}{\end{aligned}\end{equation}}
\newcommand{\beq}{\begin{eqnarray}}
\newcommand{\eeq}{\end{eqnarray}}
\newcommand{\bem}{\begin{multline}}
\newcommand{\eem}{\end{multline}}
\newcommand{\ba}{\begin{align}}
\newcommand{\ea}{\end{align}}
\newcommand{\bcase}{\left\{ \begin{array}{ll}}
\newcommand{\ecase}{\end{array} \right.}
\begin{document}

\begin{frontmatter}

\title{Neural operators solve inverse problems for constitutive model discovery}

% \title{Neural operators for rapid hyperelastic material model discovery from full-field data}

% \title{Physics-augmented neural operators for accelerated material characterization}
% \title{Physics-augmented neural operators for material model discovery}
% \title{Physics-augmented neural operators for the fast inference of material models}
% \title{Physics-augmented neural operator learning for the fast solution of inverse problems in computational material modeling}

% \author{Anonymous}

\cortext[cor1]{Correspondence: moritz.flaschel@fau.de}
\author[fau]{Moritz Flaschel\corref{cor1}}
\author[cam]{Burigede Liu}
\author[fau,stan]{Ellen Kuhl}
\address[fau]{Institute of Applied Mechanics, Egerlandstraße 5, Friedrich-Alexander-Universität Erlangen-Nürnberg, 91058 Erlangen, Germany}
\address[cam]{Department of Engineering, University of Cambridge, Trumpington Street, Cambridge, CB2 1PZ, UK.}
\address[stan]{Department of Mechanical Engineering, Stanford University, 440 Escondido Mall, California 94305, United States.}

\begin{abstract}

Characterizing the mechanical response of materials traditionally requires solving optimization problems in which model parameters are calibrated or trained to minimize the discrepancy between model predictions and experimental data. This process can be computationally expensive and time-consuming. To overcome this limitation, we propose two neural operator architectures that directly map experimentally measured data to the constitutive functions governing the mechanical response of the material: \textbf{Physics-Augmented Neural Operators (PANO)} and \textbf{Constitutive Artificial Neural Operators (CANO)}. The proposed neural operators approximate the mapping between the infinite-dimensional input space of full-field displacement measurements and net reaction forces, and the infinite-dimensional output space of hyperelastic strain-energy density functions. 
The displacement fields are encoded through Laplacian eigenfunctions to obtain discretization-independent and noise-robust predictions.
Our framework constrains the output space to physically admissible material models that satisfy fundamental physical requirements by design.
% Here we apply this neural operator frameworks in the context of hyperelasticity.
The neural operators are trained on simulated data tuples of displacement fields and reaction forces for a range of material models. %, considering a standardized experiment
Once trained, the neural operators enable near-instantaneous material characterization and require only a single forward pass to infer the strain-energy density function from a given experimental dataset.
We test the predictive power of the neural operators for unseen data, noisy data, data with missing information, data from different spatial discretizations, and data from geometries of different sizes.
We finally discuss the ill-posedness of the inverse material characterization problem and show that constraining the output function space of our neural operator framework sufficiently regularizes the problem.
The trained neural operators enable rapid and robust discovery of polyconvex strain-energy density functions, while avoiding the need to solve computationally expensive inverse problems.

\end{abstract}


\end{frontmatter}

%%%%%%%%%%%%%%%%%%%%%%%%%%%%%%%%%%%%%%%%%%%%%%%%%%%%%%%%%%%%%%%%%%%%%%%%%%%%%%%%%%%%%%%%%%%%%%%%%%%%%%%%%%%%%%%%%%%%%%%

% < to be removed later
% \newpage
% To-Do:
% \begin{itemize}
%     \item 
% \end{itemize}

% \NOTE{Note: The table of contents will be removed later.} % \\[4.pt]
% \tableofcontents
% \newpage
% to be removed later >

% \clearpage

\begin{figure}[!h]
    \centering
    \includegraphics[width=0.6\textwidth]{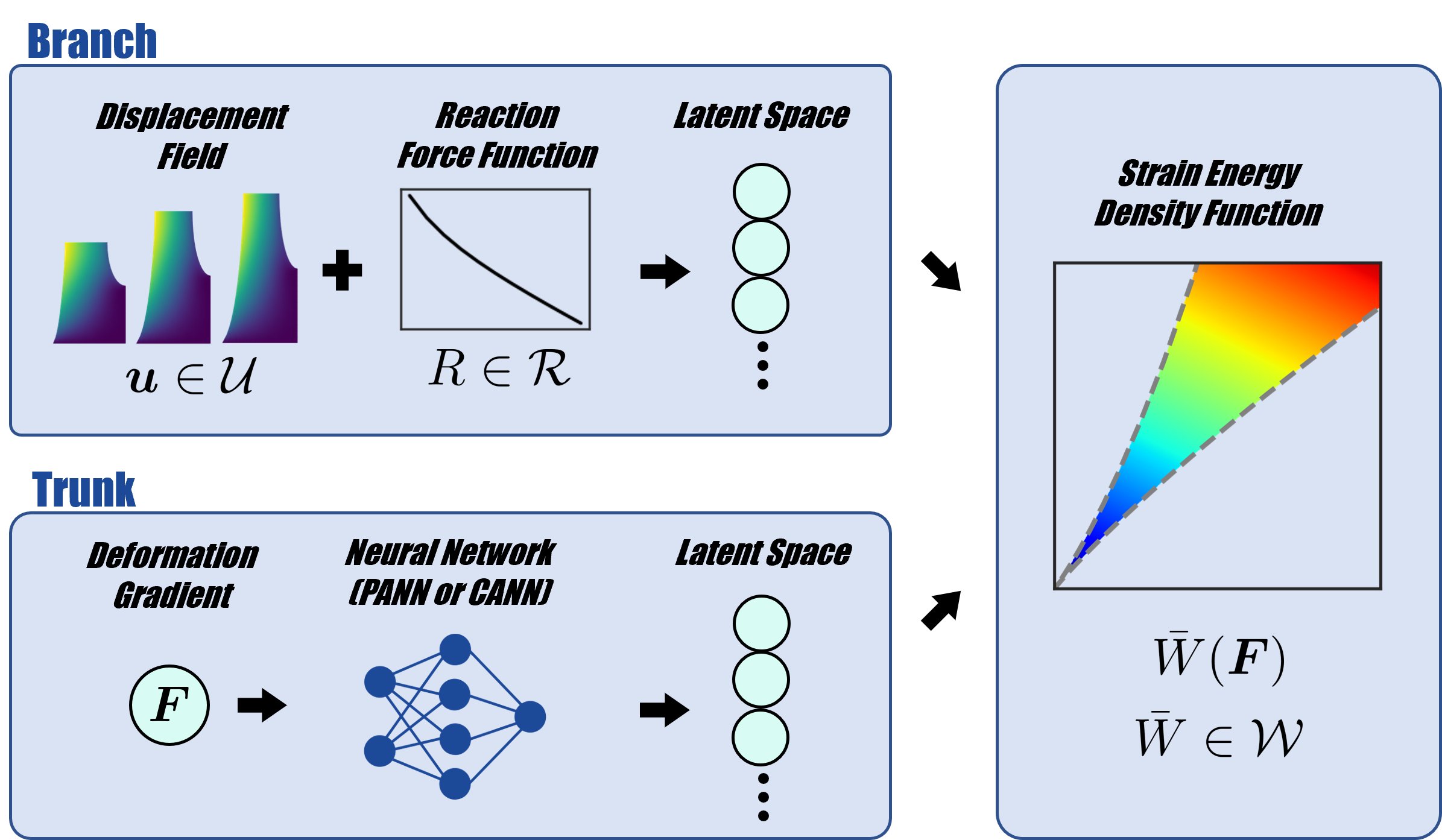}
    \caption{
    Graphical abstract: The \textbf{Physics-Augmented Neural Operator (PANO)} and \textbf{Constitutive Artificial Neural Operator (CANO)} are composed of a trunk and branch net. The trunk net maps the deformation gradient to a latent space. The \textbf{Physics-Augmented Neural Network (PANN)} or \textbf{Constitutive Artificial Neural Network (CANN)} architecture chosen for the trunk net guarantee physical admissibility of the material model. The discretization-independent and noise-robust branch net maps the given displacement and reaction force data to another latent space. The inner product of the latent spaces yields the strain-energy density. After a trained neural operator is queried with a given set of displacement and reaction force data, the branch net can be discarded, and the strain-energy density can be evaluated for any arbitrary deformation.
    }
    \label{fig:abstract}
\end{figure}

% \section*{Notes}

\section{Introduction}
\label{sec:Introduction}

% THE GENERAL OBJECTIVE

Finding suitable mathematical descriptions of mechanical material behavior is central to simulation-based engineering. In many material modeling frameworks, especially those grounded in thermodynamics \citep{onsager_reciprocal_1931,biot_thermoelasticity_1956,ziegler_thermodynamik_1957,halphen_sur_1975}, the material behavior is fully characterized by a set of constitutive functions, such as the strain-energy density potential or the dissipation density potential. The objective of mechanical material characterization is to infer these constitutive functions from experimental data, which typically include reaction forces and displacement fields of deformed specimens over time. Both the experimental observations and the underlying material behavior can naturally be interpreted as elements of infinite-dimensional function spaces. Consequently, material characterization can be formulated as the problem of identifying a mapping between such infinite-dimensional spaces. This perspective motivates the use of neural operator learning \citep{chen_universal_1995}, which aims to learn mappings between infinite-dimensional function spaces. In this work, we take a first step in this direction by developing a neural operator architecture that maps full-field displacement data and reaction forces directly to the strain-energy density function of hyperelastic materials. Once trained, the neural operator characterizes a new material via a single forward pass, without solving any optimization problem which makes it significantly faster than optimization-based approaches. Our proposed framework is graphically illustrated in \cref{fig:abstract}. Before presenting the framework in more detail, we briefly review recent advances in neural operator learning and machine learning-based material modeling.

% NEURAL OPERATOR LEARNING
Conventional neural networks approximate mappings between finite-dimensional spaces. For instance, a standard feed-forward neural network can be viewed as a mapping $\text{NN} : \Rset^n \rightarrow \Rset^m$, where $\Rset^n$ and $ \Rset^m$ are Euclidean spaces with dimensions $n$ and $m$, respectively. However, many problems in physics and engineering involve mappings between infinite-dimensional function spaces. These mappings are commonly called operators. A prominent example is the solution operator of a partial differential equation (PDE), which maps input functions, such as spatially varying parameters, geometric descriptions, boundary conditions, and initial conditions, to solution functions defined over space and time. Recognizing that conventional neural networks are inherently limited in approximating mappings between infinite-dimensional spaces, \cite{chen_universal_1995} proposed in their seminal work a neural network architecture specifically designed for this purpose. In contrast to standard networks, their approach aims to learn operators of the form $\mathrm{NO} : \mathcal{U} \rightarrow \mathcal{W}$, where $\mathcal{U}$ and $\mathcal{W}$ denote infinite-dimensional function spaces. The proposed architecture consists of two coupled neural networks: one that encodes the input function in the space $\mathcal{U}$, and another that represents the dependence on the evaluation point in the output space $\mathcal{W}$. By combining these two components, the network effectively parameterizes a nonlinear operator between function spaces. \cite{chen_universal_1995} further showed that this architecture possesses a universal approximation property. It can approximate a broad class of continuous nonlinear operators between infinite-dimensional spaces to arbitrary accuracy, provided that the network complexity is sufficiently increased, see also \cite{sandberg_approximations_1992} for earlier results on the approximation of functionals. The framework introduced by \cite{chen_universal_1995} was later adopted and further developed by \cite{lu_learning_2021}, who popularized it under the name DeepONet. \cite{lu_learning_2021} established the terminology \textit{branch net} and \textit{trunk net} for the two aforementioned subnetworks in the architecture. DeepONets have since found widespread use across scientific and engineering domains, including phase-field modeling \citep{li_phase-field_2023}, porous media and geology \citep{diab_u-deeponet_2024}, fluid dynamics \citep{bai_hybrid_2026}, and robotics \citep{veil_infinite-dimensional_2026}.  Other widely recognized architectures for approximating mappings between infinite-dimensional function spaces have been proposed by \cite{li_neural_2020}, who proposed neural operator architectures based on the composition of integral operators with nonlinear activation functions. In subsequent work, the same authors developed a variant based on Fourier operators \citep{li_fourier_2021}, leading to the formulation of the Fourier Neural Operator (FNO) whose universal approximation abilities are discussed by \cite{kovachki_universal_2021}. An overview of different choices for the integral operators in the network is provided by \cite{kovachki_neural_2023}. FNOs and related architectures have since been applied across a wide range of problems in physics and engineering \citep{pathak_fourcastnet_2022,li_fourier_2023,lehmann_fourier_2024,li_boundary-based_2025}. A fundamental challenge in learning the solution operator of a PDE is achieving independence from both the underlying geometry and its discretization \citep{bahmani_resolution_2025}. A popular approach to address this are graph-based neural operators \citep{li_neural_2020,li_geometry-informed_2023,mousavi_rigno_2025}, which are adaptations of Graph Neural Networks (GNN) \citep{scarselli_graph_2009,sanchez-gonzalez_learning_2020,pfaff_learning_2021} that are discretization-convergent. The neural operators discussed above can be trained either in a supervised fashion using labeled input-output pairs, or in a physics-informed manner by incorporating the PDE residual directly into the training loss \citep{wang_learning_2021,goswami_physics-informed_2022,faroughi_physics-guided_2024,moon_physics-informed_2025-1}.

% NEURAL OPERATOR LEARNING IN COMPUTATIONAL SOLID MECHANICS
Owing to their universal approximation capabilities, neural operators have become increasingly popular in computational solid mechanics. For instance, \cite{goswami_physics-informed_2022-1} proposed a DeepONet for crack path prediction in brittle materials. \cite{he_novel_2023} use a DeepONet to predict stresses in elastoplastic structures. \cite{kiyani_predicting_2024} proposed and compared three DeepONet architectures for learning the solution operator of phase field fracture problems. \cite{koric_deep_2024} use DeepONets for predicting the stress field in small strain plasticity problems. \cite{zhang_operator_2024} use FNOs to predict the deformation of hyperelastic microstructures on the unit domain, which is then used to compute the homogenized material response. \cite{hildebrand_comparison_2024} investigate the use of neural operators in solid mechanics and draw a comparison to the Finite Element Method (FEM). \cite{kaewnuratchadasorn_physicsinformed_2024} propose physics-informed FNOs for linear isotropic elasticity on the unit domain. \cite{rezaei_finite_2025} use DeepONets to approximate the operator that maps heterogeneous microstructures to the deformation fields in small strain elasticity on the unit domain. \cite{nguyen_universal_2026} use FNOs for predicting the solution of micromechanics problems in linear isotropic or anisotropic elasticity on the unit domain and investigate their relation to methods based on the Fast Fourier Transform (FFT). \cite{eivazi_physics-informed_2026} propose physics-informed reduced-order operator learning for homogenizing hyperelastic microstructures. \cite{mousavi_imposing_2026} propose different neural operator architectures for learning the solution operator of geometrically linear and nonlinear elasticity, with a particular focus on arbitrary boundary conditions.

% INVERSE PROBLEMS
Most existing work on neural operators addresses forward problems and focuses on learning solution operators of PDEs, which map input quantities such as parameters, geometries, boundary conditions, and initial conditions to solution fields. In contrast, the use of neural operators for inverse problems has received comparatively little attention in the literature. Exceptions are the works by \cite{molinaro_neural_2023}, who discuss the use of neural operators for PDE-constrained inverse problems, by \cite{liu_neumann_2025}, who leverage neural operators for solving inverse medium problems such as Magnetic Resonance Imaging (MRI), or by \cite{thorpe_learning_2025}, who propose a special neural operator architecture for tackling the ill-posedness of inverse problems. In the field of computational solid mechanics, only a few works apply neural operator learning to inverse problems. For example, \cite{liu_learning_2023} use recurrent neural operators for learning plasticity models. \cite{moon_physics-informed_2025-1} use physics-informed neural operators for the inference of thermoelectric properties from data. \cite{jin_characterization_2025} leverage DeepONets for the inverse design of metamaterials. \cite{guo_history-aware_2025} use FNO-inspired architectures to learn the material response of elastoplastic materials and materials undergoing damage. \cite{hollenweger_temperature-aware_2026} propose a time-resolution-independent architecture for learning the material response of anisotropic and temperature-dependent plasticity.

% OUR NEURAL OPERATOR IDEA
Prior work has explored the use of neural operators for learning constitutive models \citep{bhattacharya_learning_2022,liu_learning_2023,hollenweger_temperature-aware_2026,sohail_neural_2026}. However, in these works, the neural operators' weights are trained to minimize the mismatch between the model predictions and the data for a specific material. For a new experimental dataset or a different material, the neural operators must be retrained. To the best of our knowledge, no existing study employs neural operators to directly map experimental fields to constitutive functions, enabling the trained neural operator to be repeatedly used for material model inference without retraining. Hence, we propose a neural operator framework that directly maps experimentally measured functions, such as displacement fields and reaction forces over time, to the material's characteristic thermodynamic potentials. In this work, we focus on hyperelastic materials whose mechanical response is fully characterized by the strain-energy density function. Hence, our neural operators directly map from the spaces of the experimental data to the space of admissible strain-energy density functions. The latter space is constrained to include only mathematically and physically meaningful functions that fulfill fundamental physical laws, such as invariance to the observer, a stress-free undeformed configuration, the balance of angular momentum, and continuous differentiability, along with other desired properties such as material symmetry and polyconvexity. Imposing physical constraints on the material model reduces the design space, makes models less data-hungry, simplifies training, improves extrapolation, and prevents unphysical material responses \citep{masi_thermodynamics-based_2021,linka_constitutive_2021,klein_polyconvex_2022,asad_mechanics-informed_2022,thakolkaran_nn-euclid_2022,linka_new_2023,flaschel_automated_2023,rosenkranz_comparative_2023,linden_neural_2023,jadoon_automated_2024,kalina_fetextrmann_2023,bleyer_learning_2025,klein_polyconvex_2025,geuken_input_2025,holthusen_complement_2025}. Here, we specifically draw inspiration from two modeling approaches, Physics-Augmented Neural Networks (PANNs) \citep{klein_polyconvex_2022,thakolkaran_nn-euclid_2022,linden_neural_2023} and Constitutive Artificial Neural Networks (CANNs) \citep{linka_constitutive_2021,linka_new_2023}. PANNs are flexible neural network architectures for constitutive modeling that satisfy the fundamental physical constraints. We emphasize that \textit{physics-augmented} should not be confused with \textit{physics-informed}, which typically refers to loss functions informed by governing physical equations rather than labeled data pairs \citep{raissi_physics-informed_2019,wang_learning_2021}. CANNs, on the other hand, incorporate well-established constitutive modeling features into the neural network architecture, which reduces the number of weights and makes the networks more interpretable, generally at the cost of reduced flexibility.

% FAST MATERIAL CHARACTERIZATION
Beyond physical admissibility, our neural operators offer a significant computational advantage over traditional methods for material characterization. Traditional approaches rely on optimization problems in which material model parameters are calibrated to minimize the discrepancy between model predictions and experimental data. This process can be computationally expensive, particularly when each model evaluation requires solving nonlinear equations with the Newton-Raphson method. In contrast, our proposed neural operators require only a single forward pass to infer the material model from experimental data, which is significantly faster than traditional optimization-based characterization methods. In this context, we note that our proposed neural operator framework shares methodological similarities with approaches that directly map experimental data to the parameters of a priori chosen material models \citep{chen_learning_2021}. However, in contrast to such works, our framework significantly reduces a priori assumptions about material behavior. Rather than fixing the functional form of the strain-energy density function, it can predict a variety of functional forms depending on the available experimental data. In addition, to the best of our knowledge, no learning framework has been presented so far that directly predicts the material model for given displacement fields in a discretization-independent manner.

We further note that the issue of computationally expensive inverse problems has been addressed in our recently proposed Material Fingerprinting framework \citep{flaschel_material_2026,flaschel_unsupervised_2026,flaschel_adaptive_2026,martonova_material_2026}, in which material models are selected from a precomputed database of material responses for a given experimental dataset. By replacing the continuous optimization problem with a fast pattern recognition algorithm, Material Fingerprinting enables significantly faster material characterization. While Material Fingerprinting precomputes and stores a database for use with a pattern recognition algorithm, our proposed neural operators can be interpreted as compressing the information contained in such a database into a neural operator that directly maps experimental data to constitutive functions.

% SUPERVISED VS UNSUPERVISED
Methods for mechanical material characterization can be broadly classified into those informed by labeled stress-strain data pairs from simple experiments with homogeneous deformation fields, such as uniaxial tension tests, and those informed by full-field displacement data and net reaction forces from complexly shaped geometries \citep{grediac_principle_1989,pierron_virtual_2012,hild_comparison_2012,perotti_method_2017,pierron_towards_2020,flaschel_unsupervised_2021,romer_reduced_2025,abbasi_discovery_2026,alheit_cann-euclid_2026,knipper_finite_2026,akerson_learning_2026}. The former approach can yield ill-posed inverse problems, as different material models can produce identical stress-strain responses in the high-dimensional stress-strain space \citep{hartmann_numerical_2001,hartmann_identifiability_2018,moreno-mateos_learning_2026}. The latter approach offers the advantage that the space of deformations imposed on the material is enriched, which can reduce the ill-posedness of the inverse problem. Hence, our proposed neural operators follow this approach. However, the framework can be readily adapted to leverage stress-strain data pairs from experiments with homogeneous deformation fields instead.
 
% OVERVIEW
We describe our proposed neural operator framework and its training in \cref{sec:Neural Operators}. In \cref{sec:Results}, we evaluate the predictive performance of the neural operators on unseen test data, noisy data, data with missing information, data from different spatial discretizations, and data from geometries of varying sizes. This is followed by a discussion of the ill-posedness of the infinite-dimensional learning problem in \cref{sec:ill-posedness}, where we show numerically that the restricted output space of the proposed neural operators provides sufficient regularization of the inverse problem. We conclude in \cref{sec:Conclusions} by summarizing the main findings and outlining directions for future work.

\textbf{Notation}:
Parentheses $(\square)$ denote tuples, open intervals, or function dependencies.
Square brackets $[\square]$ denote closed intervals or the order of computation in equations.
Curly brackets $\{\square\}$ denote sets.
% Derivatives in time are denoted by $\dot\square = {\partial \square}/{\partial t}$.
% Besides subscripts, superscripts in parenthesis $^{(\square)}$ are used for indexing, and thus do not denote exponentiation.
Subscripts and superscripts written in upright font do not denote indexing.
Tensors and matrices may appear in compact or index notation.
In compact notation, vectors (first-order tensors) and second-order tensors are denoted by bold letters, e.g., $\bfC$.
When appearing in index notation, the order of the tensor equals the number of indices, e.g., $C_{ij}$.
Unless stated otherwise, the Einstein convention for summation over repeated indices is used in equations appearing in index notation, e.g., $a_i b_i = \sum_i a_i b_i$.
Inner products are denoted by~$\cdot$, e.g., $\bfa \cdot \bfb = a_i b_i$, $\bfA \cdot \bfB = A_{ij} B_{ij}$, and outer products by~$\otimes$, e.g., $[\bfa \otimes \bfb]_{ij} = a_i b_j$.
If no operation is indicated between two tensors, the juxtaposition implies tensor contraction, e.g., $[\bfA\bfB]_{ij}=A_{ik}B_{kj}$.
The second-order identity is denoted by $\bfI$.
The transpose of a tensor is denoted by $\square\transposed$.
The trace of a tensor is denoted by $\text{tr}(\square)$, i.e., $\text{tr}(\bfC) = C_{ii}$.
As we present a purely numerical study, we do not specify any physical units throughout this work.

\section{Neural Operators}
\label{sec:Neural Operators}

% \subsection{Objective}

% Before describing our proposed neural operators in detail, we briefly outline our general objective.

\subsection{Function spaces}
\label{sec:Function spaces}

The general objective of material characterization based on full field data is to infer the material model from available displacement and reaction force data by solving an inverse problem.
Mathematically, the objective is to infer an element in the material model space $\mathcal{W}$ for a given element in the data space $\mathcal{D} = \mathcal{U} \times \mathcal{R}$, where $\mathcal{U}$ is the space of measurable displacements and $\mathcal{R}$ is the space of measurable reaction forces.
Hence, in this work, we propose neural operators that take elements in $\mathcal{D}$ as input and output an element in $\mathcal{W}$.
In the following, we introduce the function spaces $\mathcal{U}$, $\mathcal{R}$, and $\mathcal{W}$ and discuss how they are constrained by the physical principles and modeling assumptions adopted in this work.

% The general objective of material characterization based on full field data is to infer the material model from available displacement and reaction force data by solving an inverse problem.
% This inverse problem implicitly defines a mapping from the data space to the material model space $S: \mathcal{D} \rightarrow \mathcal{W}$.
% Here, $\mathcal{D} = \mathcal{U} \times \mathcal{R}$ denotes the data space, where $\mathcal{U}$ is the space of measurable displacements, $\mathcal{R}$ is the space of measurable reaction forces, and $\mathcal{W}$ is the space of material models.

\subsubsection{Material model space}

Since we assume incompressible, isotropic hyperelasticity in this work, the mechanical behavior is solely characterized by the strain-energy density function $W$, and the space of material models reduces to the space of admissible strain-energy density functions $\mathcal{W}$. In this work, we restrict $\mathcal{W}$ to functions that satisfy fundamental physical constraints while also exhibiting additional desirable properties. Accordingly, we refer to our neural operators as Physics-Augmented Neural Operators (PANO) or Constitutive Artificial Neural Operators (CANO), in alignment with the notion of Physics-Augmented Neural Networks (PANNs) \citep{klein_polyconvex_2022,thakolkaran_nn-euclid_2022,linden_neural_2023,kalina_fetextrmann_2023} and Constitutive Artificial Neural Networks (CANNs) \citep{linka_constitutive_2021,linka_new_2023} for constitutive modeling. Constraining the output space of the neural operators to physically meaningful models ensures that the predicted material models do not violate physical constraints and can hence be readily deployed in forward simulations without convergence, stability, or robustness issues. Furthermore, the predicted material models are expected to generalize better to unseen deformations than their unconstrained counterparts. Finally, restricting the output space reduces the ill-posedness of the problem, as physically inadmissible models that could otherwise explain the data with similar accuracy are excluded a priori.

The deformation of a body is mathematically described by the mapping $\bfx = \phi(\bfX,t)$, where $t$ denotes time, $\bfX$ represents a material point in the undeformed reference configuration, and $\bfx$ denotes its corresponding position in the deformed configuration. The deformation gradient is defined as the Jacobian of the deformation mapping $\bfF = \nabla_{\bfX} \phi$. Its determinant $J = \det \bfF$ characterizes local volume changes and must therefore remain strictly positive to ensure physically admissible deformations. For incompressible material behavior, it is $J=1$. The deformation gradient describes both volume changes and shape changes. We separate these contributions with the Flory split $\bfF=\bfF^{\text{vol}} \bar \bfF$, where $\bfF^{\text{vol}} = J^\frac{1}{3} \bfI$ is the volumetric, shape-preserving part of the deformation gradient, and $\bar \bfF = J^{-\frac{1}{3}} \bfF$ is the isochoric, volume-preserving part. In the incompressible case, it is $\bfF^{\text{vol}}=\bfI$, and the isochoric deformation gradient therefore plays a central role in the constitutive model. Since the isochoric deformation gradient is not invariant to rigid body motions of the deformed configuration and hence does not constitute an objective measure of deformation, we introduce the isochoric right Cauchy-Green strain $\bar \bfC=\bar \bfF\transposed\bar \bfF$, and its principal invariants $\bar I_1=\text{tr}(\bar \bfC)$, $\bar I_2=\frac{1}{2}[\text{tr}(\bar \bfC)^2 - \text{tr}(\bar \bfC^2)]$, and $\bar I_3 = 1$. We further introduce the lifted polyconvex invariants $\bar I_1^* = \bar I_1 - 3$ and $\bar I_2^* = \bar I_2^{3/2} - 3^{3/2}$, where the motivation for the exponent $\frac{3}{2}$ is detailed later in this section. We denote a tuple of these invariants by $\bar\bfI^*=(\bar I_1^*,\bar I_2^*)$. For the region of possible values taken by these invariants, $\bar\bfI^*\in\mathcal{I}_{\text{adm}}\subset \Rset_{\geq 0}^2$, we refer to \cref{app:admissible_invariants}.

To model incompressible hyperelastic material behavior, we consider strain-energy density functions of the form $W(\bfF) = \bar W(\bfF) - p[J-1]$, where $\bar W$ is the isochoric strain-energy and $p$ is a Lagrange multiplier enforcing the incompressibility constraint.
The stress response derives directly from the strain-energy density
\begin{equation}
    \bfP
    = \frac{\partial W}{\partial \bfF}
    = \frac{\partial \bar{W}}{\partial \bfF} - p \bfF\mtransposed.
\end{equation}
% Hence, the material response depends solely on the isochoric strain-energy $\bar W$. To ensure that the material model and its resulting stress-strain response satisfy fundamental physical constraints while also exhibiting additional desirable properties, we assume that the isochoric strain-energy can be expressed as a function of the invariants $\bar I_1$ and $\bar I_2$, and we constrain the space of isochoric strain-energy functions to
% \begin{equation}
% \label{eq:isochoric strain-energy density space}
%     \mathcal{W} = \left\{ \bar W \in C^2\left([3,\infty\right) \times [3,\infty); \Rset_{\geq 0})
%     \ \middle| \
%     \begin{aligned}
%     &\bar{W}(0, 0) = 0,  \\
%     &\bar{W}(\bar I_1, \bar I_2) = \bar{W}^{\text{NN}}(\bar I_1, \bar I_2^{3/2}) \\
%     &\bar{W}^{\text{NN}} \ \text{monotone, convex} \\
%     \end{aligned}
%     \right\},
% \end{equation}
To ensure that the material model and its resulting stress-strain response satisfy fundamental physical constraints while also exhibiting additional desirable properties, we assume that the isochoric strain-energy can be expressed as a function of the two previously defined invariants $\bar W(\bar\bfI^*)$, and we constrain the space of isochoric strain-energy functions to
\begin{equation}
\label{eq:isochoric strain-energy density space}
    \mathcal{W} = \left\{ \bar W \in C^2\left(\mathcal{I}_{\text{adm}}; \Rset_{\geq 0}\right)
    % \mathcal{W} = \left\{ \bar W \in C^2\left(\Rset_{\geq 0} \times \Rset_{\geq 0}; \Rset_{\geq 0}\right)
    % \mathcal{W} = \left\{ \bar W \in C^2\left([3,\infty\right) \times [3^{3/2},\infty); \Rset_{\geq 0})
    \ \middle| \
    \bar{W} \ \text{separable, monotone, convex}, \bar{W}(0, 0) = 0
    % \bar{W} \ \text{monotone, convex}, \bar{W}(3, 3^{3/2}) = 0
    \right\},
\end{equation}
where the domain of the functions $\bar W$ is restricted by the admissible region $\mathcal{I}_{\text{adm}}$ of the invariants, see \cref{app:admissible_invariants}.

By restricting the output space of the neural operators to the constrained space defined above, we ensure that the material model and its resulting stress-strain response fulfill the following properties by construction: % \citep{linden_neural_2023}
\begin{enumerate}
    \item[] \textbf{Thermodynamic consistency}: By defining the stress response as $\bfP={\partial W}/{\partial \bfF}$, the material response is thermodynamically consistent by construction \citep{coleman_foundations_1961}.
    \item[] \textbf{Objectivity}: Since we are confined to strain-energy density functions that can be expressed solely as functions of the right Cauchy-Green tensor, which is invariant under rigid body motions of the deformed configuration, the material behavior is objective by construction, and the principle of material frame indifference is fulfilled. Specifically, it is $W(\bfF) = W(\bfQ\bfF)$ for all rotation tensors in the special orthogonal group $\bfQ \in \text{SO}(3)$.
    \item[] \textbf{Balance of angular momentum}: The balance of angular momentum states that the Cauchy stress tensor must be symmetric $\bfsigma = \bfsigma\transposed$. The Cauchy stress is given by
    \begin{equation}
    \bfsigma = \frac{1}{J} \bfP\bfF\transposed
    = \frac{1}{J} \left[
        \frac{\partial \bar{W}}{\partial \bar I_1}
        \left[2 J^{-2/3} \bfB - \frac{2}{3} \bar I_1 \bfI \right] + 
        \frac{\partial \bar{W}}{\partial \bar I_2^{3/2}}
        \frac{\partial \bar I_2^{3/2}}{\partial \bar I_2}
        \left[2 J^{-4/3} [I_1 \bfB - \bfB\bfB] - \frac{4}{3} \bar I_2 \bfI \right]
        \right] - p \bfI,
    \end{equation}
    % \begin{equation}
    % \begin{aligned}
    % \bfsigma &= \frac{1}{J} \bfP\bfF\transposed = \frac{1}{J} \frac{\partial W}{\partial \bfF} \bfF\transposed = \frac{1}{J} \left[ \frac{\partial \bar{W}}{\partial \bfF} - p J \bfF\mtransposed \right] \bfF\transposed
    % = \frac{1}{J} \frac{\partial \bar{W}}{\partial \bfF} \bfF\transposed - p \bfI, \\
    % &= \frac{1}{J} \left[
    %     \frac{\partial \bar{W}}{\partial \bar I_1}
    %     \frac{\partial \bar I_1}{\partial \bfF} + 
    %     \frac{\partial \bar{W}}{\partial \bar I_2^{3/2}}
    %     \frac{\partial \bar I_2^{3/2}}{\partial \bar I_2}
    %     \frac{\partial \bar I_2}{\partial \bfF}
    %     \right] \bfF\transposed - p \bfI, \\
    % &= \frac{1}{J} \left[
    %     \frac{\partial \bar{W}}{\partial \bar I_1}
    %     \left[2 J^{-2/3} \bfF - \frac{2}{3} \bfF\mtransposed \bar I_1 \right] + 
    %     \frac{\partial \bar{W}}{\partial \bar I_2^{3/2}}
    %     \frac{\partial \bar I_2^{3/2}}{\partial \bar I_2}
    %     \left[2 J^{-4/3} [I_1 \bfF - \bfF \bfC] - \frac{4}{3} \bfF\mtransposed \bar I_2 \right]
    %     \right] \bfF\transposed - p \bfI, \\
    % &= \frac{1}{J} \left[
    %     \frac{\partial \bar{W}}{\partial \bar I_1}
    %     \left[2 J^{-2/3} \bfB - \frac{2}{3} \bar I_1 \bfI \right] + 
    %     \frac{\partial \bar{W}}{\partial \bar I_2^{3/2}}
    %     \frac{\partial \bar I_2^{3/2}}{\partial \bar I_2}
    %     \left[2 J^{-4/3} [I_1 \bfB - \bfB\bfB] - \frac{4}{3} \bar I_2 \bfI \right]
    %     \right] - p \bfI, \\
    % \end{aligned}
    % \end{equation}
    with $\bfB=\bfF \bfF\transposed$, and hence is symmetric by construction.
    % with $\bar\bfB=J^{-2/3}\bfF \bfF\transposed$, and hence is symmetric by construction.
    % $\bfsigma = \frac{1}{J} \bfP\bfF\transposed = 2\frac{\partial \bar{W}}{\partial I_1}\bfB^{-1} - 2 \frac{\partial \bar{W}}{\partial I_2}\bfB - p \bfI$ with $\bfB=\bfF \bfF\transposed$,
    \item[] \textbf{Material symmetry}: Since the strain-energy density is expressed solely as a function of invariants of the right Cauchy-Green tensor, the resulting material response is isotropic by construction. It is $W(\bfF) = W(\bfF\bfQ)$ for all rotation tensors in the special orthogonal group $\bfQ \in \text{SO}(3)$.
    \item[] \textbf{Vanishing strain-energy density in the undeformed configuration}: No deformation $\bfF=\bfI$ implies $\bar I_1 = 3$, $\bar I_2 = 3^{3/2}$, and $J=1$. Therefore, the constraint $\bar{W}(0, 0) = 0$ ensures that the strain-energy density $W$ vanishes in the undeformed configuration. This constraint is not a mathematical necessity. However, it represents a natural and widely accepted normalization of the strain-energy density.
    \item[] \textbf{Stress-free undeformed configuration}: The isochoric contribution to the stress is
    \begin{equation}
        \bar \bfP
        = \frac{\partial \bar{W}}{\partial \bfF}
        = \frac{\partial \bar{W}}{\partial \bar I_1}
        \frac{\partial \bar I_1}{\partial \bfF} + 
        \frac{\partial \bar{W}}{\partial \bar I_2^{3/2}}
        \frac{\partial \bar I_2^{3/2}}{\partial \bar I_2}
        \frac{\partial \bar I_2}{\partial \bfF},
    \end{equation}
    with
    \begin{equation}
        \frac{\partial \bar I_1}{\partial \bfF} = 2 J^{-2/3} \bfF - \frac{2}{3} \bfF\mtransposed \bar I_1 \quad \mbox{and} \quad
        \frac{\partial \bar I_2}{\partial \bfF} = 2 J^{-4/3} [I_1 \bfF - \bfF \bfC] - \frac{4}{3} \bfF\mtransposed \bar I_2.
    \end{equation}
    In the undeformed configuration, $\bfF=\bfI$ implies ${\partial \bar I_1}/{\partial \bfF} = \bf0$ and ${\partial \bar I_2}/{\partial \bfF} = \bf0$. This, in turn, implies vanishing stress $\bar \bfP=\bf0$.
    \item[] \textbf{Polyconvexity}: 
    A simple way to construct polyconvex strain-energy functions is by taking positive sums of monotone and convex functions of polyconvex invariants. The first isochoric principal invariant $\bar I_1$ is polyconvex, however, the second isochoric principal invariant $\bar I_2$ is not polyconvex \citep{hartmann_polyconvexity_2003}. Hence, we follow \cite{hartmann_polyconvexity_2003} and more recently \cite{dammas_when_2025} and express the isochoric strain-energy density as monotone and convex functions of the modified set of polyconvex invariants $\bar\bfI^*$.
    While polyconvexity is not a strict physical requirement, when combined with appropriate coercivity conditions, it provides a sufficient framework to guarantee the existence of minimizers for the associated forward problem \citep{ball_convexity_1976}.
    \item[] \textbf{Differentiability}: By ensuring that the strain-energy density is at least twice differentiable, $\bar W \in C^2\left(\mathcal{I}_{\text{adm}}; \Rset_{\geq 0}\right)$, we guarantee well-defined stress and stiffness tensor computations.
\end{enumerate}

For a more detailed description of how to enforce physical admissibility in hyperelasticity, we refer the reader to the manual provided by \cite{linden_neural_2023}.

% In summary, the space of material models is constrained to satisfy fundamental physical laws as well as additional desired properties.

% our neural operators aim to learn the mapping $S: \ \mathcal{U} \ \times \ \mathcal{R} \rightarrow \mathcal{W}$, which maps measurable data directly to the space of material models. This space is, in turn, constrained to satisfy fundamental physical laws as well as additional desired properties. For a more detailed description of how to enforce physical admissibility in hyperelasticity, we refer the reader to the manual provided by \cite{linden_neural_2023}.

\subsubsection{Data space}

We consider a displacement-controlled experiment on an incompressible hyperelastic material over the time interval $[0,T]$. The undeformed three-dimensional specimen occupies the domain $\Omega^{\text{3D}}$. We denote the Dirichlet boundaries at which the displacements are prescribed in $X_i$-direction by $\Gamma^{\text{3D}}_{i}$. While conducting the experiment, observable quantities such as the displacement fields on a two-dimensional surface $\Omega^{\text{2D}}$ of the specimen and the reaction force function at the boundary can be measured. The observable displacement fields are elements of
\begin{equation}
    \mathcal{U} = \left\{ \bfu(\bfX,t) : \Omega^{\text{2D}} \times [0,T] \rightarrow \Rset^2 \ \middle| \
    \begin{aligned}
    &u_1(\bfX,t) = \bar u_1(\bfX,t) && \forall \bfX \in \Gamma^{\text{2D}}_{1},\ \forall t \in [0,T], \\
    &u_2(\bfX,t) = \bar u_2(\bfX,t) && \forall \bfX \in \Gamma^{\text{2D}}_{2},\ \forall t \in [0,T], \\
    &\bfu(\bfX,0) = \bf0 && \forall \bfX \in \Omega^{\text{2D}}
    \end{aligned}
    \right\},
\end{equation}
% where $\Omega^{\text{2D}}$ denotes the surface on which the displacement fields are measured, and
where $\bar u_1(\bfX,t)$, $\bar u_2(\bfX,t)$ are the displacements prescribed on the Dirichlet boundaries $\Gamma^{\text{2D}}_{i} = \Omega^{\text{2D}} \cap \Gamma^{\text{3D}}_{i}$.
The observable reaction force functions, which are assumed to be scalar-valued in this work, are elements of
\begin{equation}
    \mathcal{R} = \left\{ R(t) : [0,T] \rightarrow \Rset \ \middle| \ R(0) = 0  \right\}.
\end{equation}
We denote a tuple consisting of an observable displacement field and a reaction force function by $\bfd=(\bfu,R)$ and note that not every pair $\bfd \in \mathcal{D} = \mathcal{U} \times \mathcal{R}$ is physically meaningful. The observable displacement field and reaction force function must correspond to the same solution of the forward boundary value problem for an admissible material model. Hence, in the following, we introduce the constrained space $\mathcal{D}_{\text{BVP}}\subset\mathcal{D}$ that only contains physically meaningful measurements.

For a given material $\bar W \in \mathcal{W}$, the displacement $\bfu^{\text{3D}}_{\bar W}$ and pressure field $p^{\text{3D}}_{\bar W}$ over the three-dimensional domain $\Omega^{\text{3D}}$ are determined by the forward boundary value problem of incompressible hyperelasticity (see \cref{app:forward_problem}). The corresponding deformation gradient and stress fields are
\begin{equation}
   \bfF^{\text{3D}}_{\bar W} = \nabla_{\bfX} \bfu^{\text{3D}}_{\bar W} + \bfI, \quad \bfP^{\text{3D}}_{\bar W} = \left. \frac{\partial \bar W}{\partial \bfF}\right\rvert_{\bfF=\bfF^{\text{3D}}_{\bar W}} - p^{\text{3D}}_{\bar W} [\bfF^{\text{3D}}_{\bar W}]\mtransposed.
\end{equation}
During the experiment, the three-dimensional fields are not directly observable. Instead, only selected quantities can be observed. We denote the observable two-dimensional displacement field over the specimen surface by $\bfu^{\text{2D}}_{\bar W}$, and the observable reaction force function by
\begin{equation}
    R_{\bar W} =\int_{\Gamma^{\text{3D}}_{R}} \bfP^{\text{3D}}_{\bar W} \bfN  \cdot \bfe_R \mathrm{d}A,
\end{equation}
where $\Gamma^{\text{3D}}_{R}$ denotes the boundary at which the reaction force is measured, $\bfN$ is the corresponding outward unit normal vector, and $\bfe_R$ is the direction of the measured reaction force. The observed displacement fields and reaction force functions are fully determined by the material $\bar W$. We hence define the observation mapping
\begin{equation}
    O : \mathcal{W} \rightarrow \mathcal{D}_{\text{BVP}}, \quad O : \bar W \mapsto (\bfu^{\text{2D}}_{\bar W},R_{\bar W}).
\end{equation}
Consequently, the constrained space of physically meaningful observations is $\mathcal{D}_{\text{BVP}}=O(\mathcal{W})\subset\mathcal{D}$.
If we assume a material $\bar W_{\text{sim}} \in \mathcal{W}$ and synthetically generate data $\bfd_{\text{sim}}\in\mathcal{D}_{\text{BVP}}$, it is $O(\bar W_{\text{sim}}) = \bfd_{\text{sim}}$. 
In contrast, due to measurement noise and erroneous modeling assumptions, an experimental measurement $\bfd_{\text{obs}}\in\mathcal{D}$ may not lie in the space $\mathcal{D}_{\text{BVP}}$.
Instead, the experimental measurement suggests that the material model lies in
\begin{equation}
    \mathcal{W}_{\bfd_\text{obs},\varepsilon}
    =
    \left\{
    \bar W\in\mathcal{W}
    :
    \left\|
    O(\bar W)-\bfd_{\text{obs}}
    \right\|_{\mathcal{D}}
    \le \varepsilon
    \right\},
\end{equation}
where $\varepsilon=0$ for synthetic data and $\varepsilon>0$ for experimental data.

\subsection{Statistical conditional inverse operator}

% Given a model in $\mathcal{W}$, the corresponding observation in $\mathcal{D}_{\text{BVP}}$ during a simulated experiment is determined through the forward observation mapping $O: \mathcal{W} \rightarrow \mathcal{D}_{\text{BVP}}$.
% In the inverse setting, our objective is to infer a model in $\mathcal{W}$ given an observation lying on or near the physically attainable data manifold $\mathcal{D}_{\text{BVP}}$.
% However, in general, there exists no single-valued inverse mapping $S = O^{-1}$ with $S: \mathcal{D}_{\text{BVP}} \rightarrow \mathcal{W}$.
% Hence, in this section, we introduce a statistical machine learning framework for learning a conditional inverse operator that takes an observation in $\mathcal{D}_{\text{BVP}}$ as input and outputs a model in~$\mathcal{W}$.

In general, $\mathcal{W}_{\bfd_\text{obs},\varepsilon}$ does not contain a unique material model. Thus, the inverse problem we aim to tackle with our proposed neural operators should not be understood as the application of a classical single-valued inverse operator $S=O^{-1}$. Therefore, in this section, rather than postulating a canonical selector from $\mathcal{W}_{\bfd_\text{obs},\varepsilon}$, we interpret material model discovery as a supervised statistical learning problem. We seek to learn a conditional inverse operator $S_{\mu}: \mathcal{D}_{\text{BVP}} \rightarrow \mathcal{W}$ that maps an observation to the Bayes-optimal predictor induced by the training distribution $\mu$. Although the operator is learned using observations in $\mathcal{D}_{\text{BVP}}$, it is expected to generalize to observations lying near $\mathcal{D}_{\text{BVP}}$.

Let $\mu$ be the probability distribution from which training material models are sampled. We denote a random draw from this distribution by $\bar W_{\mu} \sim \mu$ and the corresponding observation by $\bfd_{\mu} = O(\bar W_{\mu})$. Further, let 
\begin{equation}
    \mathcal{H} = L^2(\mathcal{I}_{\mathrm{adm}},\nu_{\mathcal{I}_{\mathrm{adm}}}) \quad \text{with} \quad \|V\|_{\mathcal H}^2 =  \int_{\mathcal{I}_{\mathrm{adm}}} |V(\bar\bfI^*)|^2 \, \mathrm{d}\nu_{\mathcal{I}_{\mathrm{adm}}}(\bar\bfI^*),
\end{equation}
be the space of square-integrable functions over the admissible invariants domain, where $\nu_{\mathcal{I}_{\mathrm{adm}}}$ denotes the probability distribution used to sample invariant query points for evaluating constitutive models. The Bayes-optimal predictor $S_{\mu}: \mathcal{D}_{\text{BVP}} \rightarrow \mathcal{H}$ associated with the squared-error loss is given by
\begin{equation}
    S_{\mu} \in \argmin_{S: \mathcal{D}_{\text{BVP}} \rightarrow \mathcal{W}    }
    \,
    \mathbb{E}
    \left(
    \left\|
    {S}(\bfd_{\mu})-\bar W_{\mu}
    \right\|_{\mathcal{H}}^{2}
    \right).
\end{equation}
Equivalently, for a fixed observation $\bfd_{\text{obs}}\in\mathcal D_{\mathrm{BVP}}$, the Bayes-optimal prediction $\bar W_{\mu}( \, \cdot \, ; \bfd_{\text{obs}}) = S_{\mu}(\bfd_{\text{obs}})$ with $\bar W_{\mu}( \, \cdot \, ; \bfd_{\text{obs}}): \mathcal{I}_{\text{adm}} \to \Rset_{\geq 0}$ satisfies
\begin{equation}
\bar W_{\mu}( \, \cdot \, ; \bfd_{\text{obs}})
\in
\argmin_{V\in\mathcal{W}}
\,
\mathbb{E}
\left(
\left.
\left\|
V-\bar W_\mu
\right\|_{\mathcal{H}}^{2}
\,\right|\,
\bfd_{\mu} = \bfd_{\text{obs}}
\right).
\end{equation}

If no two distinct material models represented by the training distribution produce the same
observation, that is
\begin{equation}
\bar W_{1},\bar W_{2}
\in
\operatorname{supp}\mu,
\quad
O(\bar W_{1})
=
O(\bar W_{2})
\Rightarrow
\bar W_{1}=\bar W_{2},
\end{equation}
then the material model is identifiable within the training class. In this case, $S_{\mu}(O(\bar W_{\mu}))=\bar W_{\mu}$ for $\mu$-almost every $\bar W$.
If the inverse is not identifiable, several material models may correspond to the same observation. In that case, $S_{\mu}(\bfd_{\text{obs}})$ is the prediction jointly preferred by the training distribution $\mu$, the admissible hypothesis class, and the squared-error loss. In an unrestricted linear function space, the squared-error optimal prediction is the conditional mean
\begin{equation}
\bar W_{\mu}( \, \cdot \, ; \bfd_{\text{obs}})
% S_{\mu}(\bfd_{\text{obs}})
=
\mathbb{E}
\left[
\left.
\bar W_\mu
\right|
\bfd_{\mu} = \bfd_{\text{obs}}
\right].
\end{equation}
Such a prediction should not, in general, be interpreted as a unique physical inverse or as a canonical member of $\mathcal{W}_{\bfd_{\text{obs}},\varepsilon}$.

The neural operators considered below provide a parameterized approximation of the statistical predictor $S_{\mu}$
\begin{equation}
    S_{\bftheta}: \mathcal D_{\mathrm{BVP}} \to \mathcal{H},
    \quad S_{\bftheta}(\bfd) = \bar W_{\bftheta} (\, \cdot \, ; \bfd),
\end{equation}
where the parameters $\bftheta$ are learned from labeled training data $(\bfd_i,\bar W_i)$ as discussed subsequently.

\subsection{Neural operator architectures}
\label{sec:Neural operator architectures}

\begin{figure}[!h]
    \centering
    \includegraphics[width=0.9\textwidth]{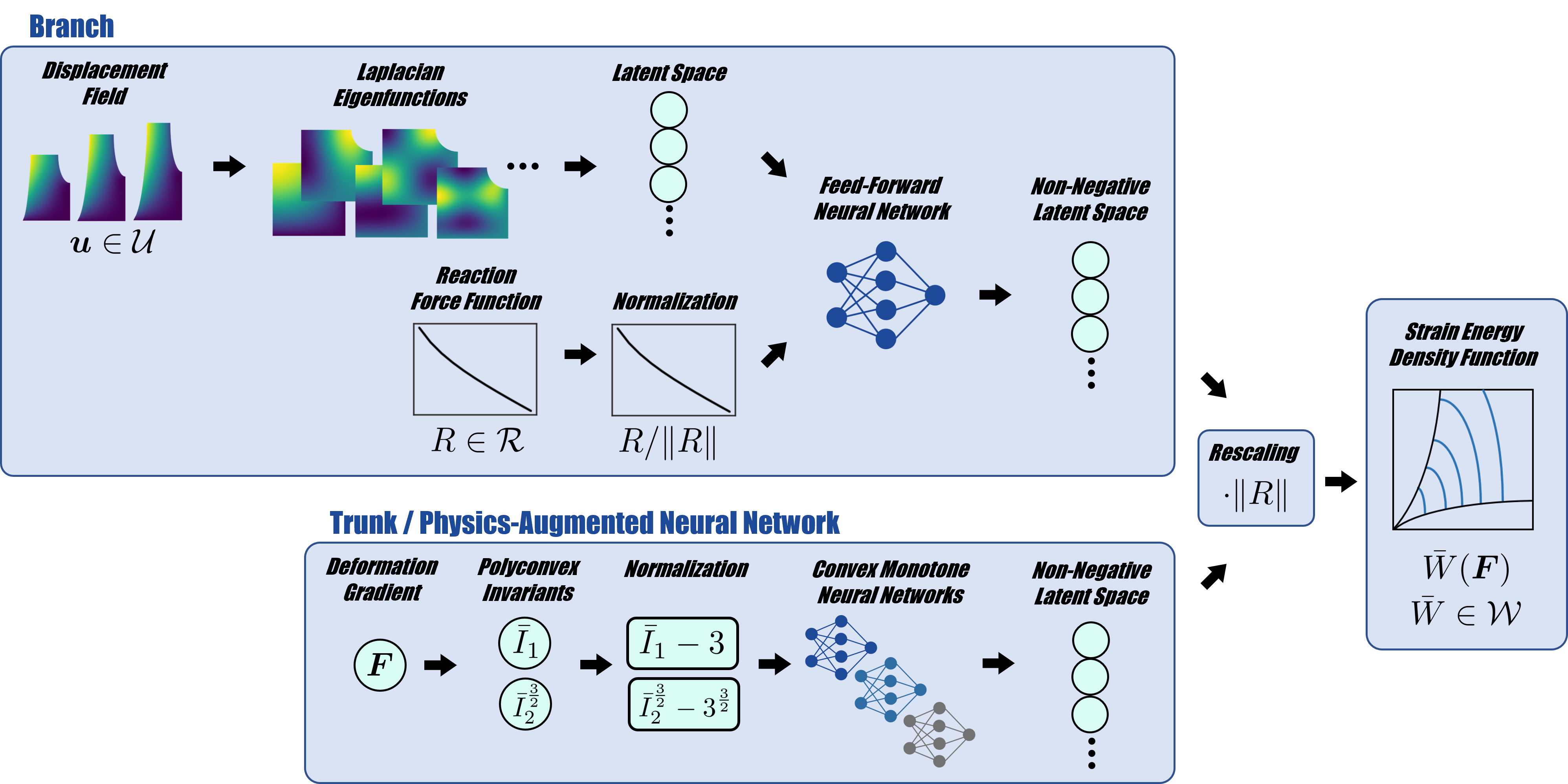}
    \caption{Physics-Augmented Neural Operator (PANO) architecture.}
    \label{fig:pano}
\end{figure}

In the following, we build the neural operators for learning a mapping from elements in $\mathcal{D}$ to elements in $\mathcal{W}$. The architectures chosen in this work are inspired by the DeepONet architectures proposed by \cite{chen_universal_1995,lu_learning_2021}. They consist of a branch net and a trunk net, as illustrated in \cref{fig:abstract}. The trunk net maps the deformation gradient $\bfF$ to a latent space. The discretization-independent and noise-robust branch net maps the given displacement and reaction force data to another latent space. The inner product of the latent spaces yields the strain-energy density $W$. After a trained neural operator is queried with a given set of displacement and reaction force data, the branch net can be discarded, and the $W$ can be evaluated for any arbitrary $\bfF$. In this work, we consider two different architectures, the PANO (\cref{fig:pano}) and the CANO (\cref{fig:cano}), which share the same branch net, but differ in the design of the trunk net. In the following, we describe the shared branch net architecture and the different trunk net architectures.

% \subsubsection{branch net}

The main purpose of the branch net is to encode the available displacement and reaction force data in a latent space (\cref{fig:pano,fig:cano}). We assume the displacement fields over the specimen domain are available through full-field measurement techniques such as Digital Image Correlation. More specifically, we assume that the two-dimensional displacement vector $\bfu$ is known at a total of $N_{\bfX}$ points over the domain $\Omega^{\text{2D}}$ and at a total of $N_t$ time steps in the interval $[0,T]$, resulting in a displacement data array $\bfu_{\text{data}} \in \Rset^{2 \times N_{\bfX} \times N_t}$. However, as we discuss at several points throughout this work, our neural operators are designed such that they can also take displacement fields sampled at arbitrary spatial locations in the domain $\Omega^{\text{2D}}$ as input.

In the first layer of the branch net, the displacement field data are encoded in a lower-dimensional space. Specifically, we represent both components of the displacement field in the reduced basis
\begin{equation}
\label{eq:Laplacian_interpolation}
    u_1(\bfX,t) \approx u_1^{\text{lift}}(\bfX,t) + \sum_{j=1}^{N_{\phi}} a_{1jt} \, \phi_{1j}(\bfX) \quad \mbox{and} \quad
    u_2(\bfX,t) \approx u_2^{\text{lift}}(\bfX,t) + \sum_{j=1}^{N_{\phi}} a_{2jt} \, \phi_{2j}(\bfX), \\
\end{equation}
where $\phi_{1j}(\bfX) : \Omega^{\text{2D}} \rightarrow \Rset$ and $\phi_{2j}(\bfX) : \Omega^{\text{2D}} \rightarrow \Rset$ are time-independent basis functions with non-local support on $\Omega^{\text{2D}}$, and $a_{1jt} \in \Rset$ and $a_{2jt} \in \Rset$ are real-valued coefficients. The number of considered basis functions $N_\phi$ for representing the displacement fields is chosen to be smaller than the number of available displacement measurements $N_{\bfX}$. Hence, the first layer compresses the given displacement data array from $\bfu_{\text{data}} \in \Rset^{2 \times N_{\bfX} \times N_t}$ to the encoded array $\bfa_{\text{data}} \in \Rset^{2 \times N_\phi \times N_t}$. This lower-dimensional representation of the displacement field at each time step serves several purposes. First, the compressed displacement fields are easier for the branch net to process and reduce the number of its trainable weights. Second, because the basis functions have non-local support, the interpolation in \cref{eq:Laplacian_interpolation} reduces the sensitivity of the branch net to local noise in the displacement field. Third, the branch net becomes independent of the spatial resolution because the interpolation in \cref{eq:Laplacian_interpolation} can also be performed when displacements are provided at spatial locations other than the $N_{\bfX}$ points discussed above. We finally note that the interpolation in \cref{eq:Laplacian_interpolation} is fully deterministic and does not depend on any trainable weights.

In this work, we choose the basis functions $\phi_{1j}(\bfX)$ and $\phi_{2j}(\bfX)$ as the first $N_\phi$ Laplacian eigenfunctions of the geometry $\Omega^{\text{2D}}$. These functions are obtained by solving a generalized eigenvalue problem that is derived by discretizing the Laplace partial differential equation with a reference mesh on the geometry. A detailed description of the Laplacian eigenfunction computation is provided in \cref{app:Laplacian_encoding}. The Laplacian eigenfunctions vanish at the Dirichlet boundaries $\Gamma^{\text{2D}}_{1}$, $\Gamma^{\text{2D}}_{2}$, respectively. The fixed lifting fields $u_1^{\text{lift}}(\bfX,t)$ and $u_2^{\text{lift}}(\bfX,t)$, which fulfill the Dirichlet boundary conditions, ensure that the interpolated displacement fields fulfill the Dirichlet boundary conditions. For an illustration of the first Laplacian eigenfunctions $\phi_{1j}(\bfX)$, we refer to \cref{fig:pano,fig:cano}. 

In addition to the displacement field, the scalar-valued vertical reaction force is assumed to be measured by a load cell during the experiment at the same $N_t$ time steps, resulting in a force data array $\bfR_{\text{data}} \in \Rset^{N_t}$. The force data array is normalized as $\bfR_{\text{data}} / \|\bfR_{\text{data}}\|$. Both the displacement data and the normalized force data are mapped to a latent space of non-negative real valued numbers $\Rset_{\geq 0}^{N_{\text{latent}}}$ using a neural network. We denote this neural network by
\begin{equation}
\label{eq:latent_network}
    \text{FFNN}: \Rset^{2 \times N_{\phi} \times N_t} \times \Rset^{N_t} \rightarrow \Rset_{\geq 0}^{N_{b}}.
\end{equation}
Here, we chose a conventional, fully-connected feed-forward neural network that takes the flattened arrays $\bfa_{\text{data}}$ and $\bfR_{\text{data}} / \|\bfR_{\text{data}}\|$ as input. Enforcing the non-negativity of the output $\bfb \in \Rset_{\geq 0}^{N_{b}}$ is straightforward by applying an element-wise, non-negative activation function $\sigma:\Rset \rightarrow \Rset_{\geq 0}$ in the last layer, such as the ReLU function or smooth alternatives. In general, the dimension of the latent space $N_b$ is much smaller than the dimension of the data $N_b \ll 2 \, N_{\phi} \, N_t + N_t$. The dimension of the latent space can be considered a hyperparameter. Its meaning becomes clearer in the following, where we introduce the trunk net for describing the material model.

% This neural network can be, for example, a conventional fully-connected feed-forward neural network, a convolutional neural network, or a graph neural network. We expect this network to be deep and to contain a large number of trainable parameters.

While the branch net encodes the information from available displacement and reaction force data, the trunk net (\cref{fig:pano,fig:cano}) describes the mechanical material behavior. Specifically, the material behavior is governed by the isochoric strain-energy density function in the constrained space $\mathcal{W}$. In this work, we approximate elements in this space using either Physics-Augmented Neural Networks (PANNs) \citep{klein_polyconvex_2022,thakolkaran_nn-euclid_2022,linden_neural_2023} or Constitutive Artificial Neural Networks (CANNs) \citep{linka_constitutive_2021,linka_new_2023}
\begin{equation}
    \text{PANN}: \Rset_{\geq 0}^{2} \rightarrow \Rset_{\geq 0}^{N_{b}} \quad \mbox{or} \quad \text{CANN}: \Rset_{\geq 0}^{2} \rightarrow \Rset_{\geq 0}^{N_{b}},
\end{equation}
which are designed to fulfill the desired physical constraints discussed above. A detailed description of these networks is provided in the subsequent \cref{sec:trunk_pano,sec:trunk_cano}. Note that, as opposed to the original PANN and CANN formulations, our networks do not map directly to the isochoric strain-energy density function but instead to a latent feature vector $\bfpsi \in \Rset_{\geq 0}^{N_{b}}$, which shares the same dimension as the latent output space of the branch net $\bfb \in \Rset_{\geq 0}^{N_{b}}$. Each entry in this vector can be interpreted as a physics-constraint modeling feature for the isochoric strain-energy density function. The isochoric strain-energy density function is then finally obtained through the inner product of the outputs of the branch and trunk nets
\begin{equation}
\label{eq:inner_bt}
    \bar W = \bfb \cdot \bfpsi \cdot \|\bfR_{\text{data}}\|,
\end{equation}
where we use the norm of the reaction force data to appropriately scale the isochoric strain-energy density.

The composition of the branch and trunk nets forms our physics-aware neural operators for constitutive model discovery. The neural operators can be trained in a supervised manner. To this end, labeled data of the form $(\bfu_{\text{data}},\bfR_{\text{data}},\bar\bfI^*_{\text{data}},\bar \bfW_{\text{data}})_i$ with $i=1,\dots,N_{\text{data}}$ are generated in a simulation-based data-mining process. To generate a single data point, an isochoric strain-energy density function $\bar W$ is assumed and evaluated at a total of $N_{\bar I}$ instances of invariant pairs $(\bar I_1, \bar I_2)$. This yields the data arrays $\bar\bfI^*_{\text{data}} \in \Rset^{2 \times N_{\bar I}}$ and $\bar \bfW_{\text{data}} \in \Rset^{N_{\bar I}}$. Next, the displacements and reaction forces are simulated for the standardized experiments, assuming the same isochoric strain-energy density function $\bar W$. This provides the data arrays $\bfu_{\text{data}}$ and $\bfR_{\text{data}}$, where the displacement data $\bfu_{\text{data}}$ is compressed into $\bfa_{\text{data}}$ through the Laplacian encoding (\cref{app:Laplacian_encoding}).

More information regarding training will be provided in \cref{sec:Training}. After training is completed, the neural operator is capable of predicting the isochoric strain-energy density function for a given set of experimentally measured displacements and reaction forces. This function can subsequently be evaluated for arbitrary input pairs of invariants. In this way, the neural operator learns a mapping between function spaces rather than between finite-dimensional Euclidean spaces. When experimental displacements and reaction forces are provided as input, the predicted isochoric strain-energy density function can be directly employed in forward simulations. Notably, for this purpose, it is not necessary to implement the entire neural operator. The branch net can be omitted, as only its outputs are required for the implementation of the material model.

\subsubsection{PANO}
\label{sec:trunk_pano}

In the following, we describe the trunk net architectures in detail. \cref{fig:pano} shows an illustration of the PANN-inspired trunk architecture of the PANO. We assume that the dimension $N_b$ of the latent space is even, and define the trunk architecture as
\begin{equation}
\label{eq:features_pano}
    \begin{aligned}
        \psi_i(\bfF) = \begin{cases}
            \text{CMNN}_i(\bar I_1^*) - \text{CMNN}_i(0) & \text{if } i \leq \frac{N_b}{2} \\
            \text{CMNN}_i(\bar I_2^*) - \text{CMNN}_i(0) & \text{if } i > \frac{N_b}{2}
        \end{cases},
    \end{aligned}
\end{equation}
where $\text{CMNN}_i : \Rset_{\geq 0} \rightarrow \Rset_{\geq 0}$ for $i=1,\dots,N_b$ are neural networks that are convex and monotone with respect to their inputs.
Such networks can be constructed through a slight modification of the input-convex neural networks proposed by \cite{amos_input_2017}. For detailed information on how to impose the desired convexity and monotonicity constraints, we refer to \cite{thakolkaran_nn-euclid_2022,asad_mechanics-informed_2022,klein_polyconvex_2022,flaschel_convex_2025}.
% More information regarding their architecture, which is a modification of the input-convex neural networks proposed by \cite{amos_input_2017}, is provided in \cref{app:CMNN}.
Due to their specifically designed architecture, the features $\psi_i(\bfF)$ are convex and monotone with respect to the normalized polyconvex invariants. Hence, the isochoric strain-energy density in \cref{eq:inner_bt} is polyconvex. The constants $\text{CMNN}_i(0)$ ensure that the features vanish in the undeformed configuration $\psi_i(\bfI)=0$. We note that the convex, monotone neural networks $\text{CMNN}_i$ have independent, trainable weights. The flexibility of the trunk net in fitting the constitutive material response is determined by the number of layers and neurons in $\text{CMNN}_i$, as well as by the number of features $N_b$. 

% which can be designed to fulfill the constraint $\text{ICNN}(3, 3^{3/2}) = 0$ by simply subtracting a constant value after the last layer. Networks of the form above are well-known in the field of physics-augmented learning of material models \citep{asad_mechanics-informed_2022,klein_polyconvex_2022,thakolkaran_nn-euclid_2022}. It has been shown that relatively small neural network architectures, for example, with fewer than $100$ parameters, are sufficient to properly learn material responses in practical applications. We will denote the number of weights in this network by $N_{\text{weight}}$. In classical approaches for physics-augmented learning of material models, these weights are tuned through a training process directly informed by the data. Here, however, we let these weights depend on the outputs generated by the latent space encoder presented in \cref{eq:latent_network}. Specifically, we assume that $N_{\text{latent}} \leq N_{\text{weight}}$, such that a subset of the weights of $\text{ICNN}$ is not trainable, but is instead set equal to the outputs of $\text{NN}$. Since the outputs of $\text{NN}$ are non-negative by construction, the monotonicity and convexity of $\text{ICNN}$ are not violated.

% \subsubsection{trunk net - CANO}
\subsubsection{CANO}
\label{sec:trunk_cano}

\begin{figure}[!h]
    \centering
    \includegraphics[width=0.9\textwidth]{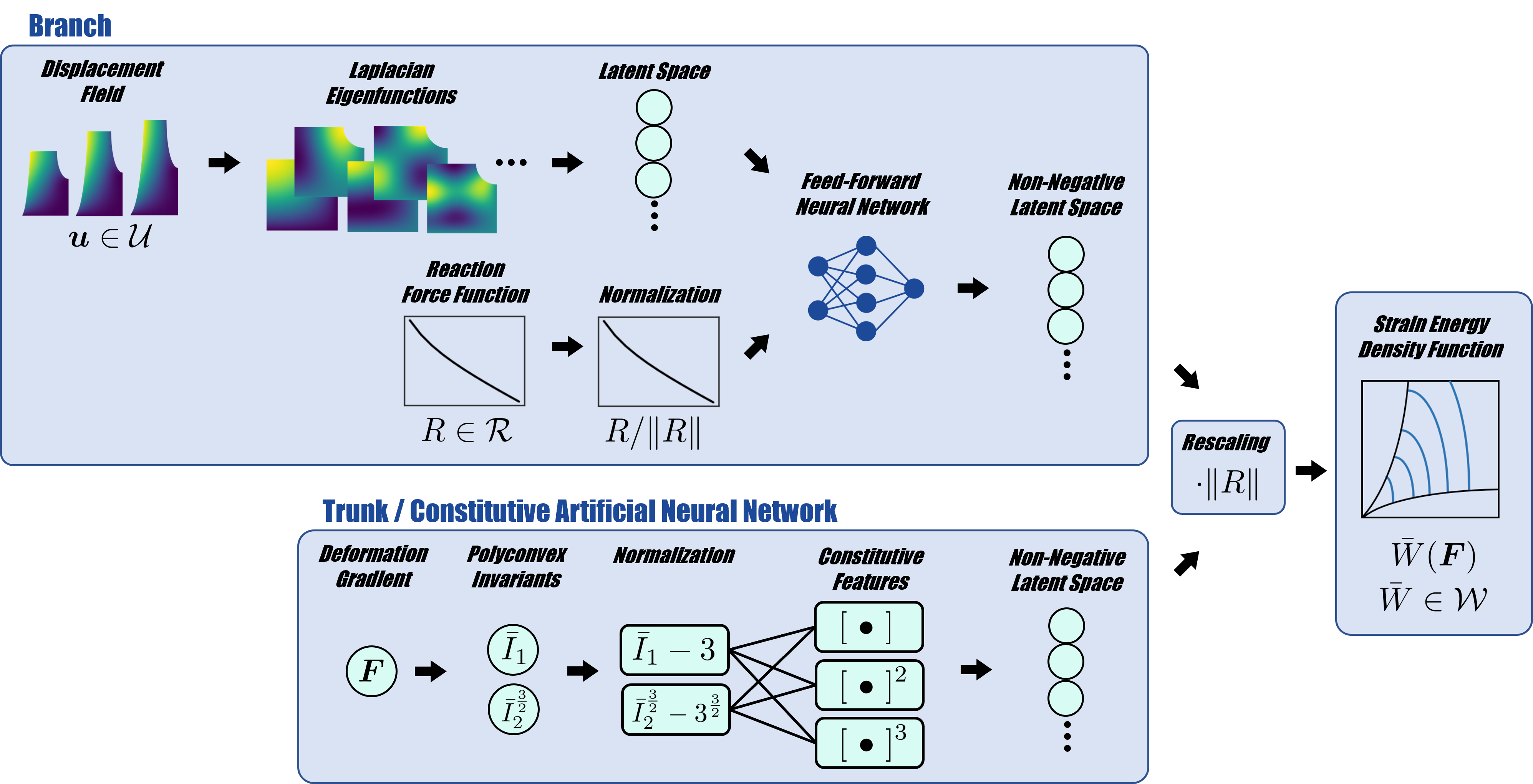}
    \caption{Constitutive Artificial Neural Operator (CANO) architecture.}
    \label{fig:cano}
\end{figure}

The PANN-inspired trunk architecture leaves the functional form of the latent features $\bfpsi \in \Rset_{\geq 0}^{N_b}$ flexible and determines it during training. As an alternative, we investigate a CANN-inspired trunk architecture in which $\bfpsi$ is composed of a priori chosen, well-established constitutive features. Specifically, we consider the first six features of a generalized Mooney–Rivlin material model, which can be derived from a Taylor expansion of the isochoric strain-energy density function about the undeformed configuration (\cref{app:Taylor_expansion})
\begin{equation}
\label{eq:features_cano}
% \begin{alignedat}{3}
% \psi_1(\bfF) &= [\bar I_1 - 3], \qquad &
% \psi_2(\bfF) &= [\bar I_1 - 3]^2, \qquad &
% \psi_3(\bfF) &= [\bar I_1 - 3]^3, \\
% \psi_4(\bfF) &= [\bar I_2^{3/2} - 3^{3/2}], \qquad &
% \psi_5(\bfF) &= [\bar I_2^{3/2} - 3^{3/2}]^2, \qquad &
% \psi_6(\bfF) &= [\bar I_2^{3/2} - 3^{3/2}]^3.
% \end{alignedat}
% \psi_1(\bfF) = \bar I_1^*, ~ 
% \psi_2(\bfF) = [\bar I_1^*]^2, ~ 
% \psi_3(\bfF) = [\bar I_1^*]^3, ~ 
% \psi_4(\bfF) = \bar I_2^*, ~ 
% \psi_5(\bfF) = [\bar I_2^*]^2, ~ 
% \psi_6(\bfF) = [\bar I_2^*]^3.
\bfpsi(\bfF) = [
\bar I_1^*,
~ \bar I_2^*,
~ [\bar I_1^*]^2,
~ [\bar I_2^*]^2,
~ [\bar I_1^*]^3,
~ [\bar I_2^*]^3
]\transposed.
\end{equation}
Since the polynomial features are convex and monotone for non-negative inputs, the resulting isochoric strain-energy density in \cref{eq:inner_bt} is polyconvex. We note that $\psi_i(\bfI)=0$. The CANO architecture is illustrated in \cref{fig:cano}. The trunk net of the CANO is less flexible in fitting the constitutive material response than that of the PANO. However, this reduced flexibility decreases the number of trainable weights and makes the material model description more interpretable.

\subsection{Experimental design}
\label{sec:experimental design}

In this work, we assume an experimental setup with a fixed geometry, prescribed Dirichlet boundary conditions, and a fixed number of time steps. The experiment is designed to be reproducible for different materials across laboratories to enable a neural operator that is trained on this setup to be used repeatedly for constitutive model discovery. While we consider a relatively simple specimen geometry here, future work may explore specimen geometries optimized for material characterization \citep{ghouli_topology_2025,bhattacharya_optimal_2026}.

As we will show later, a neural operator trained on the reference geometry can be used for inference on geometries with different side lengths and thicknesses. Furthermore, we will demonstrate that, although the discretization is assumed to be fixed during training data generation, the neural operators can be applied to different discretizations of the geometry and can accommodate situations in which only partial displacement field information is available over the specimen surface. Spatial discretization independence is important because we cannot always guarantee that the full displacement fields are captured at all time steps throughout an experiment. In contrast, temporal discretization independence is generally less critical, since displacements and reaction forces can typically be measured at high sampling frequencies \citep{abbasi_discovery_2026,flaschel_unsupervised_2026} such that we can construct data at arbitrary time points through interpolation.

\begin{figure}[!h]
    \centering
    \includegraphics[width=0.6\textwidth]{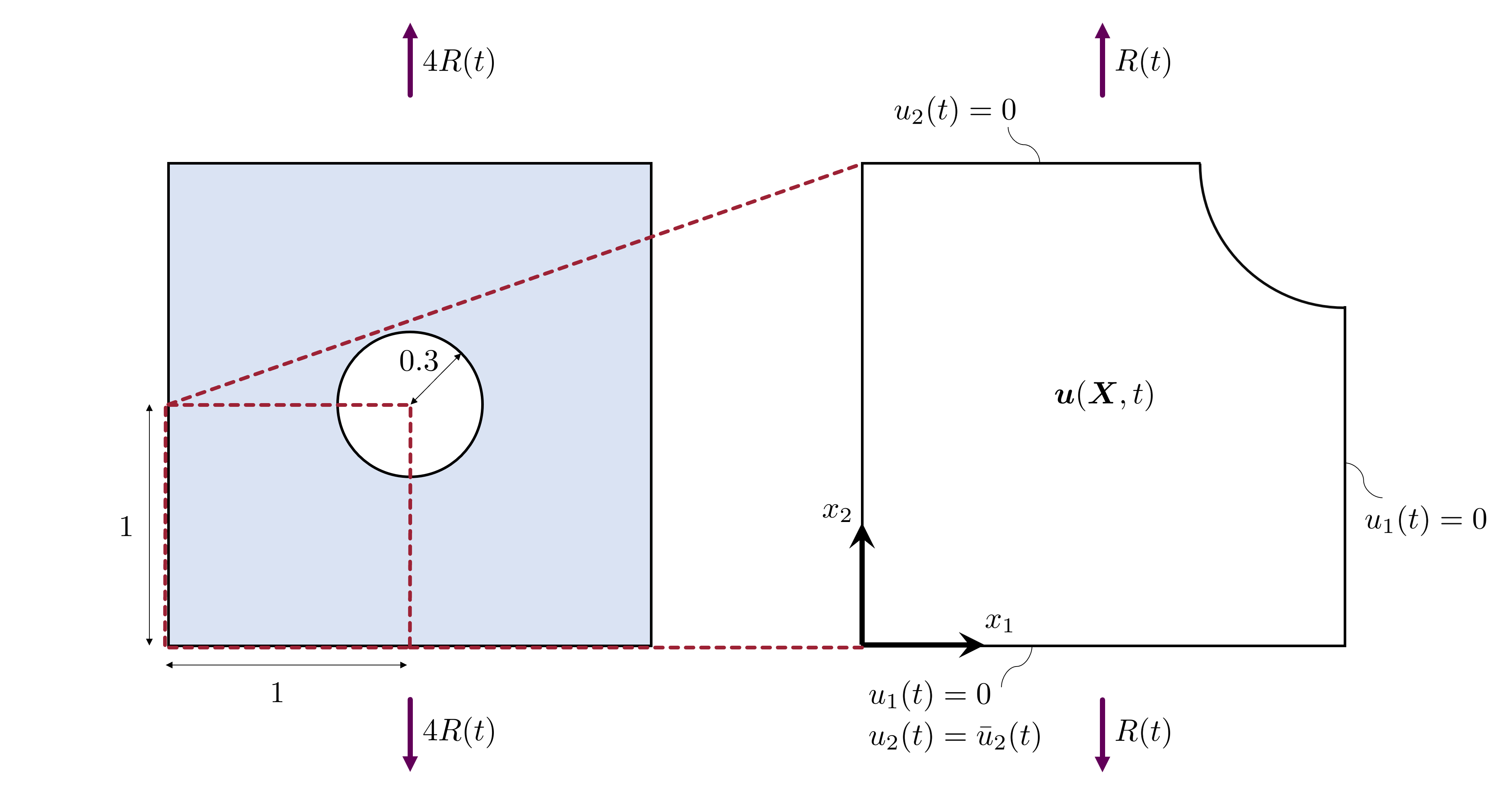}
    \caption{Specimen geometry (left) and symmetry-reduced computational domain with applied boundary conditions (right).}
    \label{fig:geometry}
\end{figure}

Here, we consider a thin plate with a central hole, as illustrated in \cref{fig:geometry}. In the undeformed configuration, the plate occupies the domain $\Omega^{\text{3D}}$. The length and height of the specimen are $2$, and the thickness is $0.01$. The plate is clamped at the top and bottom boundaries and is deformed under displacement control over $N_t$ time steps in constant increments until a maximum displacement of $\bar u_2 = -2.0$ is reached. Due to the symmetry of the problem, only one eighth of the plate is considered in the simulations, resulting in a computational domain with length and height equal to 1 and a thickness of 0.005. On the symmetry-reduced domain, the displacement in the $x_1$-direction is prescribed at the bottom and right boundaries, denoted by $\Gamma^{\text{3D}}_{1}$. The displacement in the $x_2$-direction is prescribed at the top and bottom boundaries, denoted by $\Gamma^{\text{3D}}_{2}$. The displacement in the $x_3$-direction is fixed at the back surface. We assume that the two-dimensional displacement field on the surface of the specimen, denoted by $\Omega^{\text{2D}}$, can be measured using image-based measurement techniques such as Digital Image Correlation (DIC), and the scalar-valued vertical reaction force can be measured using a load cell. In actual experiments, the symmetry condition may be slightly violated. In such cases, the displacement field over $\Omega^{\text{2D}}$ can be obtained by averaging the displacement fields of the four quadrants of the surface.

We note that, for the considered boundary value problem under displacement control and near plane stress conditions, certain rescalings of the geometry and boundary conditions leave the solution invariant up to a scaling factor. Specifically, scaling the plate thickness by a factor $s_{3}$ scales the reaction force by the same factor while leaving the displacements on the measurement surface unchanged, provided that the thickness remains sufficiently small compared to the length and height, so that the plane stress assumption remains valid. Furthermore, if the length, height, and hole radius are scaled by a factor $s_{1}$, and the prescribed displacement is scaled by the same factor while keeping the thickness fixed, both the displacement field and the reaction force scale by $s_{1}$. These scaling relations will be exploited later to enable a neural operator trained on a reference geometry to perform inference on data generated from geometries with different side lengths and thicknesses.

\subsection{Training}
\label{sec:Training}

To generate training data, we discretize the three-dimensional geometry and perform finite element simulations to solve the displacement-controlled boundary value problem for different strain-energy density functions. The finite element simulations are performed using FEniCSx. From the simulation results, we extract the two-dimensional displacement fields on the surface of the specimen $\Omega^{\text{2D}}$ at $N_{\bfX}$ points. For details, we refer to \cref{app:forward_problem}.

To obtain a meaningful dataset that covers a large portion of the constitutive modeling space, we consider the Taylor expansion of the isochoric strain-energy density function about the undeformed configuration (see \cref{app:Taylor_expansion} for details). Since classical material models are generally restricted to low-order polynomial terms, we restrict ourselves to terms up to third order
\begin{equation}
% \bar W(\bar I_1 - 3,\bar I_2^{3/2} - 3^{3/2}) =
% C_{10} [\bar I_1 - 3]
% + C_{01} [\bar I_2^{3/2} - 3^{3/2}]
% + C_{20} [\bar I_1 - 3]^2
% + C_{02} [\bar I_2^{3/2} - 3^{3/2}]^2
% + C_{30} [\bar I_1 - 3]^3 
% + C_{03} [\bar I_2^{3/2} - 3^{3/2}]^3.
\bar W(\bar I_1^*,\bar I_2^*) =
C_{10} \bar I_1^*
+ C_{01} \bar I_2^*
+ C_{20} [\bar I_1^*]^2
+ C_{02} [\bar I_2^*]^2
+ C_{30} [\bar I_1^*]^3 
+ C_{03} [\bar I_2^*]^3.
\end{equation}
We use Latin hypercube sampling to sample parameters in the six-dimensional space $[0,1]^6$ to create parameter samples $\bar C_{ij} \in [0,1]$. For inputs greater than one, the higher-order polynomial features exhibit a greater influence on the total model. The maximum invariant $\bar I_1^*$ reached in the simulations is about 3, and the maximum invariant $\bar I_2^*$ is about 5. Therefore, we scale the parameters obtained through Latin hypercube sampling by $C_{10}=\bar C_{10}/3$, $C_{01}=\bar C_{10}/5$, $C_{20}=\bar C_{20}/3^2$, $C_{02}=\bar C_{02}/5^2$, $C_{30}=\bar C_{30}/3^3$, $C_{03}=\bar C_{03}/5^3$. We note that, owing to the reaction force normalization (see \cref{fig:pano,fig:cano}), the proposed neural operators are invariant under scalar scaling of the strain-energy density. Consequently, only the relative magnitudes of the constants $C_{ij}$ are of interest, and it suffices to restrict the Latin hypercube sampling to the unit interval $[0,1]^6$.

The finite element simulations provide labeled data pairs of strain-energy density functions and the associated displacement and reaction force data. To train the neural operators, we sample the invariants uniformly from the domain $\mathcal{I}_{\text{sample}} = \{ (\bar I_1 - 3, \bar I_2^{3/2} - 3^{3/2}) \in \mathcal{I}_{\text{adm}} \, | \, \bar I_2^{3/2} - 3^{3/2} \leq I_1 \}$. This sampling domain is chosen heuristically to approximately cover the range of invariants encountered in the training dataset. For an illustration of the invariants sampling, the reader is referred to \cref{fig:invariants_sampling} in \cref{app:admissible_invariants}. In our numerical experiments, alternative sampling domains did not significantly affect the results.

% \cap [0,10]^2 

We obtain labeled data of the form $(\bfu_{\text{data}},\bfR_{\text{data}},\bar\bfI^*_{\text{data}},\bar \bfW_{\text{data}})_i$ with $i=1,\dots,N_{\text{data}}$, $\bfu_{\text{data}} \in \Rset^{2 \times N_{\bfX} \times N_t}$, $\bfR_{\text{data}} \in \Rset^{N_t}$, $\bar\bfI^*_{\text{data}} \in \Rset^{2 \times N_{\bar I}}$, and $\bar \bfW_{\text{data}} \in \Rset^{N_{\bar I}}$. In this work, we run $N_{\text{data}}=3000$ simulations. $80\%$ of the data is used for training ($N_{\text{data}}^{\text{train}}=0.8N_{\text{data}}$), $10\%$ for validation ($N_{\text{data}}^{\text{val}}=0.1N_{\text{data}}$), and $10\%$ for testing ($N_{\text{data}}^{\text{test}}=0.1N_{\text{data}}$).
The loss function is defined as
\begin{equation}
    % \text{Loss}(\bftheta) = \frac{1}{0.8 N_{\text{data}} N_{\bar I}} \sum_i^{0.8 N_{\text{data}}} \left\| \bar \bfW_{\text{NO}}\left((\bfu_{\text{data}},\bfR_{\text{data}},\bar\bfI^*_{\text{data}})_i;\bftheta\right) - (\bar \bfW_{\text{data}})_i \right\|^2,
    \text{Loss}(\bftheta) = \frac{1}{N_{\text{data}}^{\text{train}}} \sum_i^{N_{\text{data}}^{\text{train}}} \text{MSE}_i(\bftheta)
    \quad \text{with} \quad \text{MSE}_i(\bftheta) = \frac{1}{N_{\bar I}} \left\| \bar \bfW_{\text{NO}}\left((\bfu_{\text{data}},\bfR_{\text{data}},\bar\bfI^*_{\text{data}})_i;\bftheta\right) - (\bar \bfW_{\text{data}})_i \right\|^2,
\end{equation}
where $\bar \bfW_{\text{NO}}$ is the neural operator prediction, and $\bftheta$ represents all the trainable weights. The mean squared error $\text{MSE}_i$ will be used in the following to assess the prediction accuracy of the neural operator for one individual training or testing sample $i$. The hyperparameters of the training procedure are provided in \cref{app:hyperparameters}.

\section{Results}
\label{sec:Results}

In the following, we present the results obtained with the trained PANO and CANO on the previously described training dataset, comprising 3,000 simulations generated by Latin hypercube sampling of the parameter domain of the cubic Taylor expansion model. We then discuss the discretization independence and noise robustness of the neural operators. Furthermore, we demonstrate that the neural operators can infer material models from full-field data even when the geometry side lengths and thickness differ from those assumed during the generation of the training data.

\begin{figure}[!h]
    \centering
    \begin{subfigure}[b]{0.9\textwidth}
        \centering
        \includegraphics[width=\textwidth]{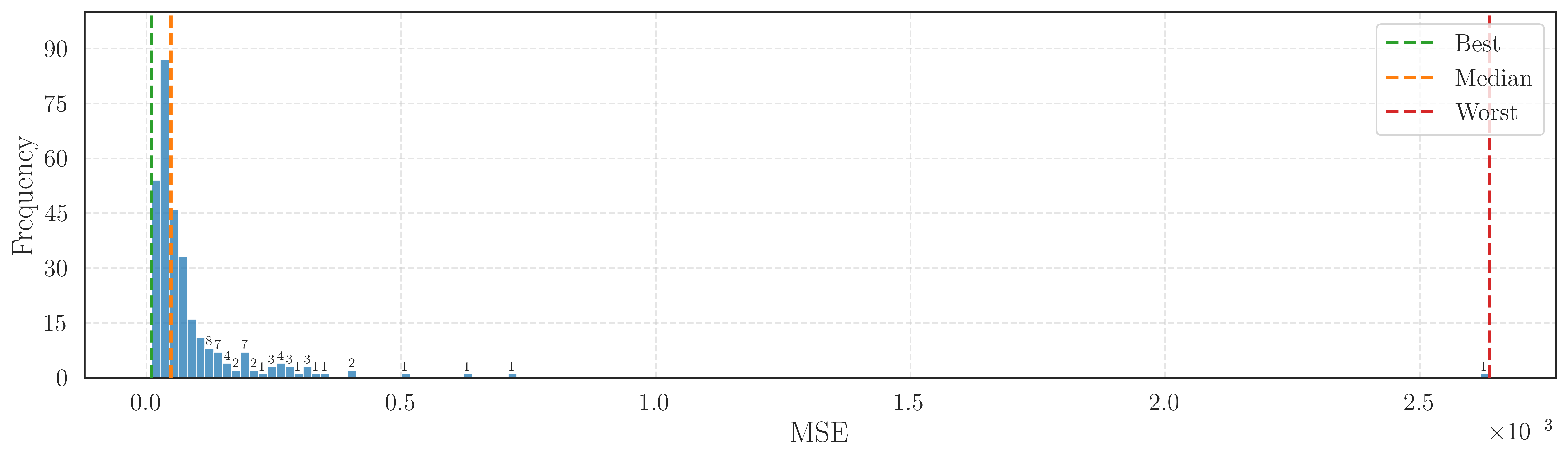}
        \caption{Distribution of MSE across the unseen testing data.}
        \label{fig:results_pano_mse}
    \end{subfigure}
    \centering
    \begin{subfigure}[b]{0.3\textwidth}
        \centering
        \includegraphics[width=\textwidth]{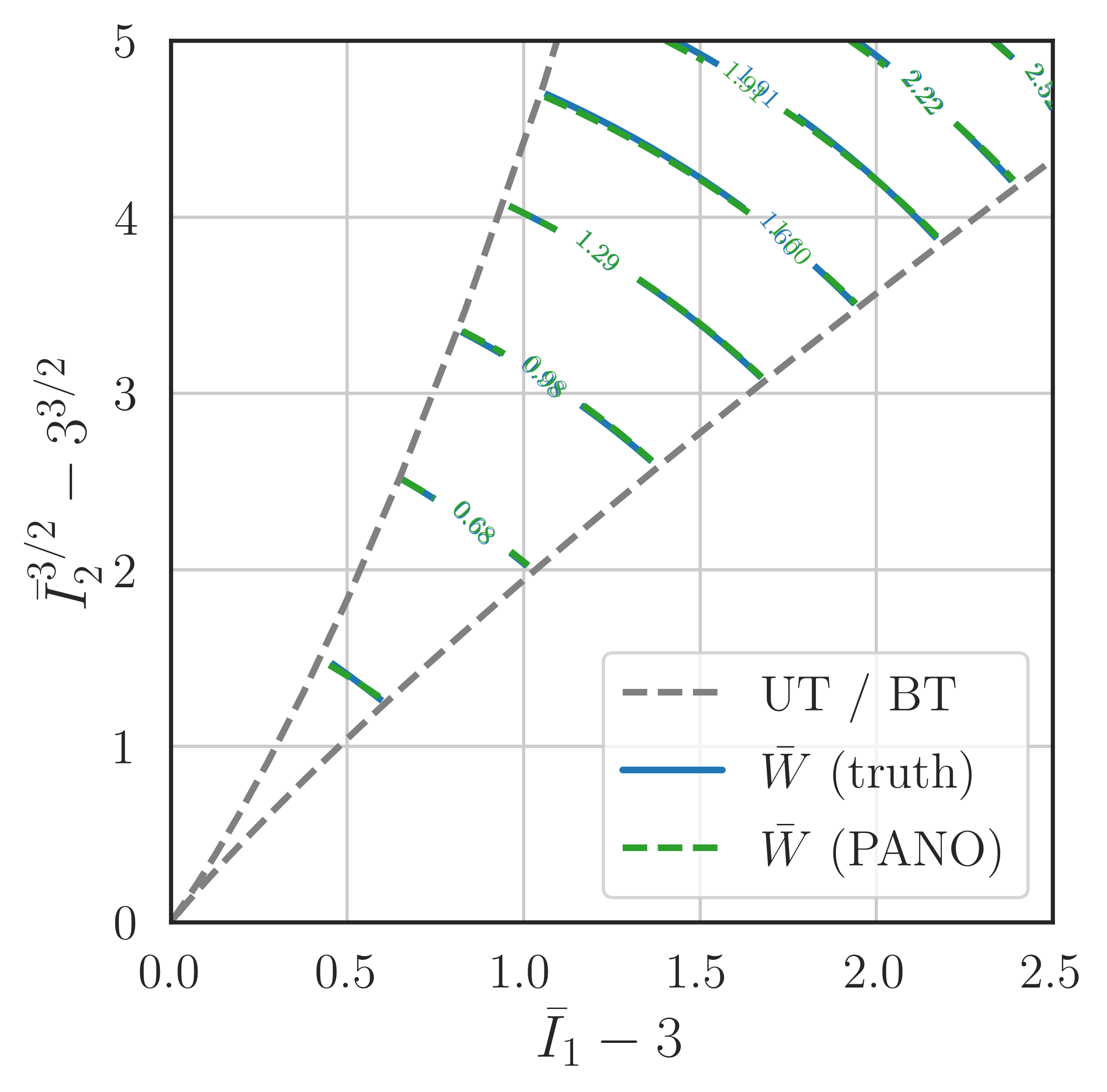}
        \caption{Best MSE.}
        \label{fig:results_pano_best}
    \end{subfigure}
    \begin{subfigure}[b]{0.3\textwidth}
        \centering
        \includegraphics[width=\textwidth]{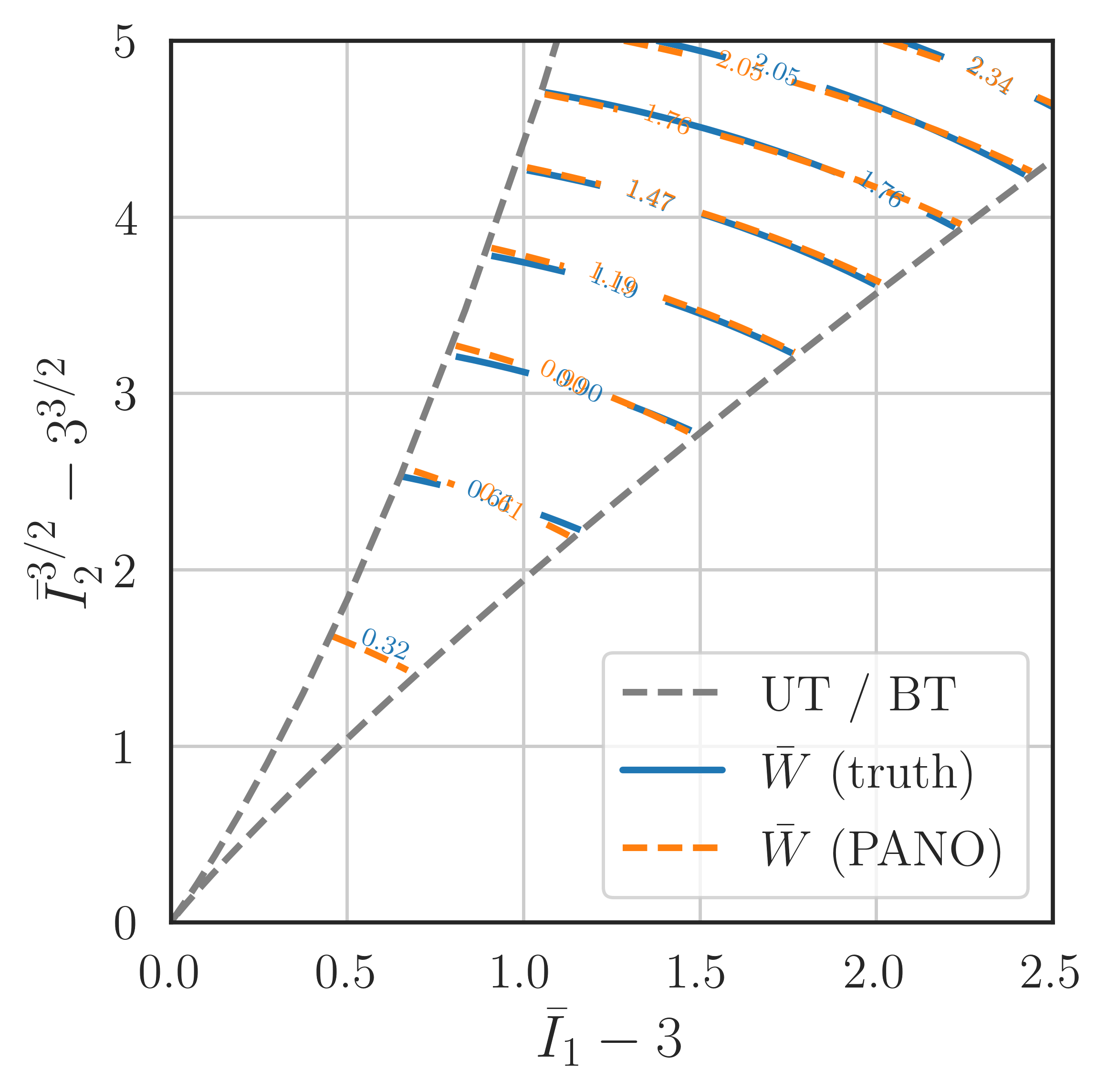}
        \caption{Median MSE.}
        \label{fig:results_pano_median}
    \end{subfigure}
    \begin{subfigure}[b]{0.3\textwidth}
        \centering
        \includegraphics[width=\textwidth]{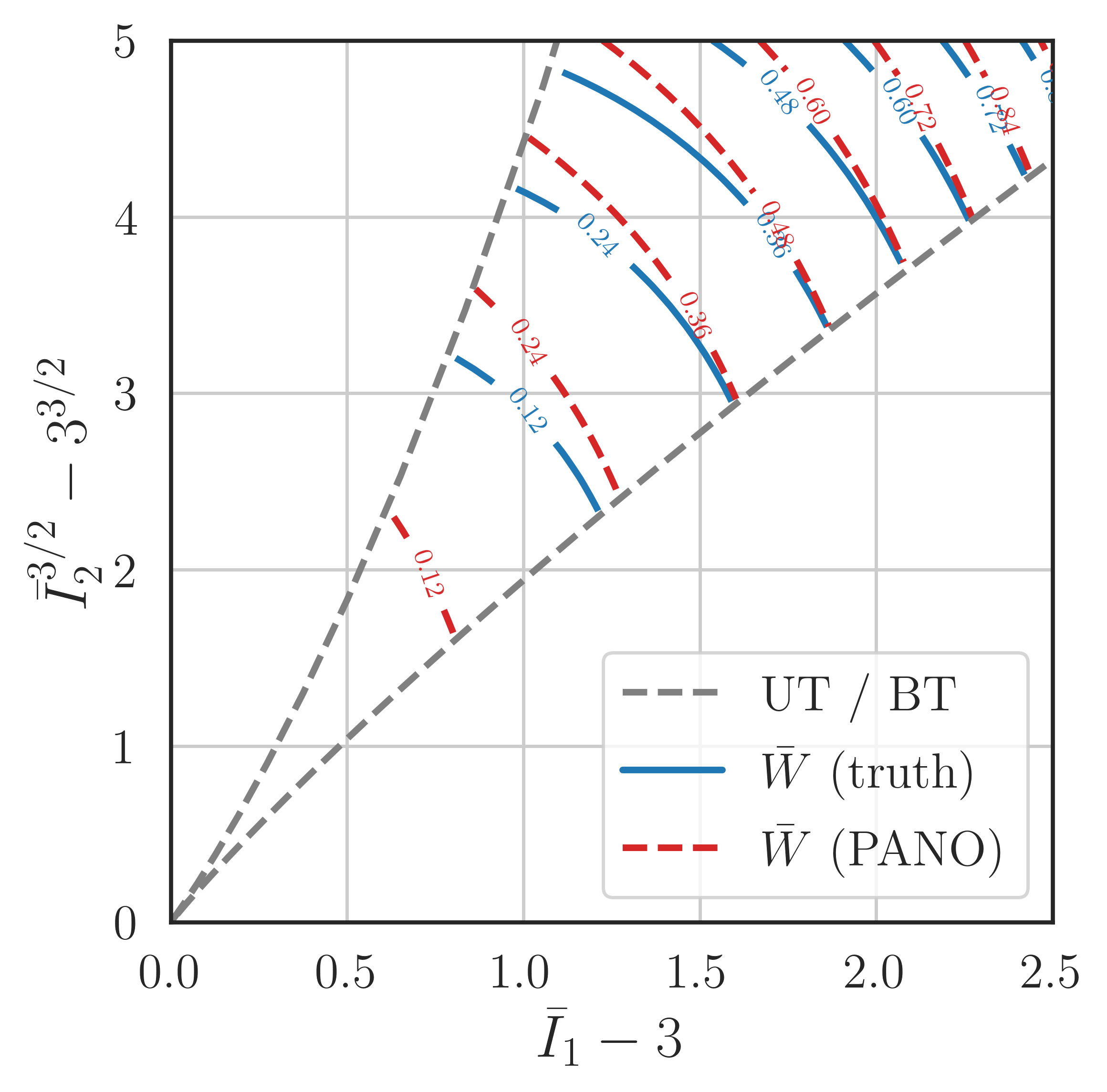}
        \caption{Worst MSE (outlier).}
        \label{fig:results_pano_worst}
    \end{subfigure}
    \caption{PANO results for \textbf{unseen testing data}.}
    \label{fig:results_pano}
\end{figure}

% \subsection{Physics-Augmented Neural Operator}
\subsection{PANO}

We use the Adam optimizer to train the PANO on $80\%$ of the simulation data. $10\%$ of the simulation data is used for validation during the training process. The corresponding validation loss is used to select the best model over all training epochs. Another $10\%$ of the simulation data is held aside and is entirely unseen during training. These testing data are used in the following to assess the accuracy and reliability of the trained networks. For a detailed description of the hyperparameters of the neural operator architecture and the training process, we refer to \cref{app:hyperparameters}.

\cref{fig:results_pano_mse} shows a histogram of the MSE computed for all simulations in the unseen testing dataset. The histogram indicates that the majority of samples exhibit low MSE values, while only a few outliers attain larger errors. We highlight the minimum, median, and maximum MSE values observed in the test dataset in the histogram. \cref{fig:results_pano_best,fig:results_pano_median,fig:results_pano_worst} show the corresponding true and discovered isochoric strain-energy densities for the minimum, median, and maximum MSE cases, respectively. Excellent agreement is observed for the minimum and median MSE cases. Only the prediction corresponding to the maximum MSE exhibits noticeable deviations from the true isochoric strain-energy density. However, as shown by the histogram in \cref{fig:results_pano_mse}, this case represents an exceptional outlier, which is not representative of the overall predictive performance of PANO. The second worst MSE is substantially smaller than the worst MSE, and shows a significantly improved agreement between the true and discovered models.

% \subsection{Constitutive Artificial Neural Operator}
\subsection{CANO}

\begin{figure}[!h]
    \centering
    \begin{subfigure}[b]{0.9\textwidth}
        \centering
        \includegraphics[width=\textwidth]{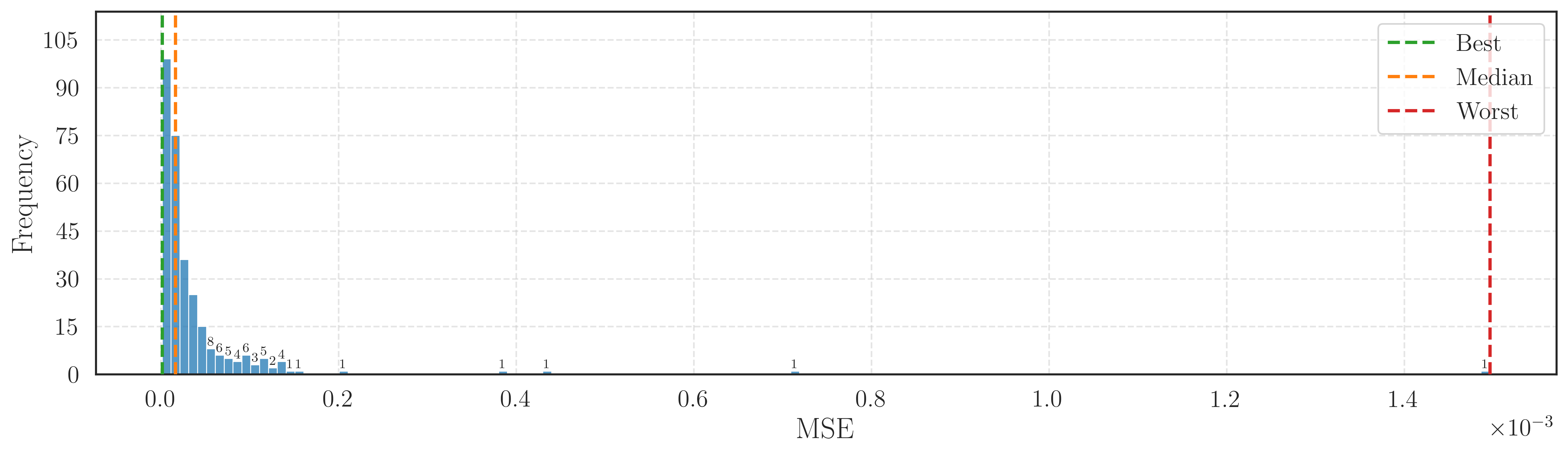}
        \caption{Distribution of MSE across the unseen testing data.}
        \label{fig:results_cano_mse}
    \end{subfigure}
    \centering
    \begin{subfigure}[b]{0.3\textwidth}
        \centering
        \includegraphics[width=\textwidth]{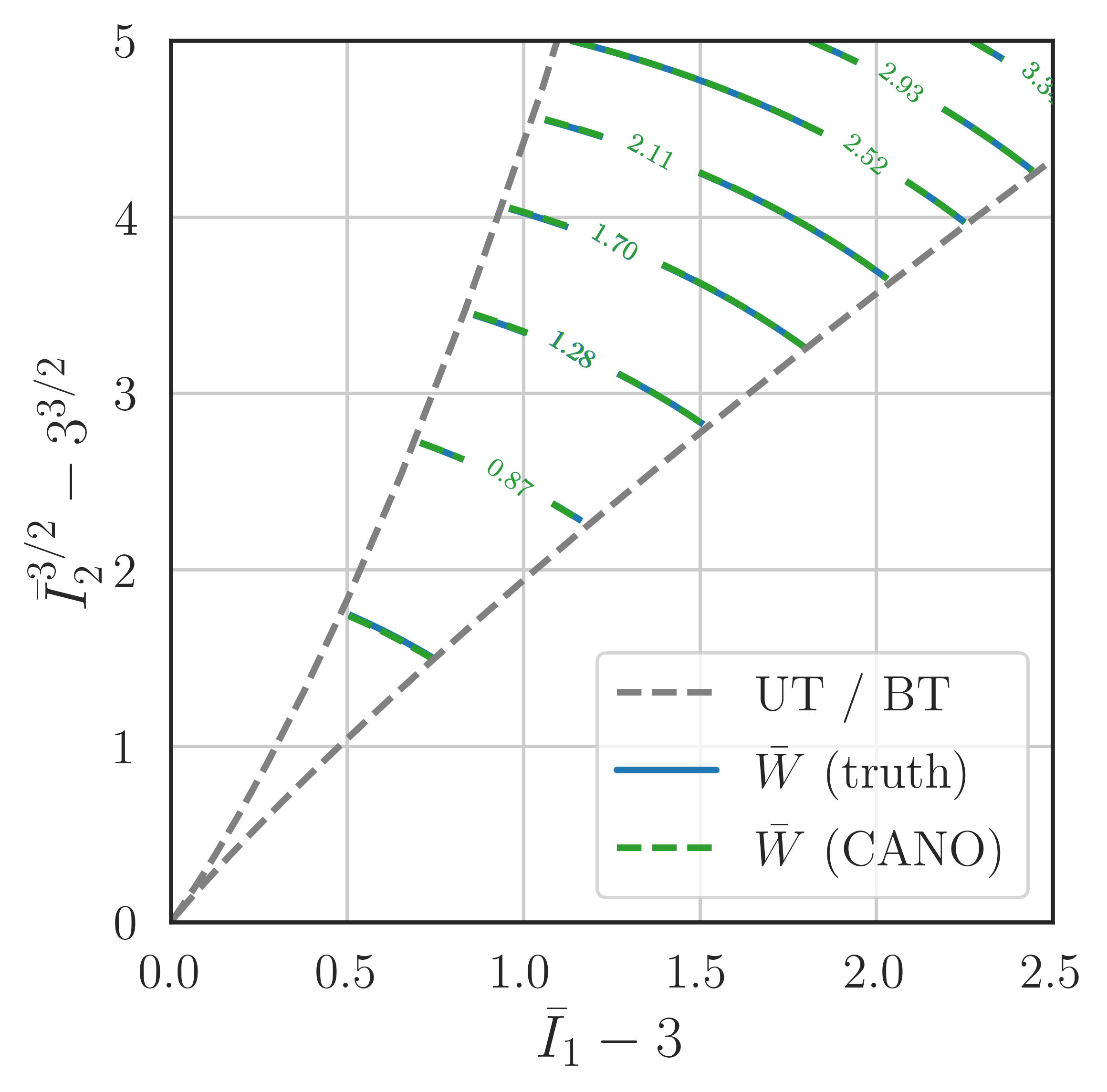}
        \caption{Best MSE.}
        \label{fig:results_cano_best}
    \end{subfigure}
    \begin{subfigure}[b]{0.3\textwidth}
        \centering
        \includegraphics[width=\textwidth]{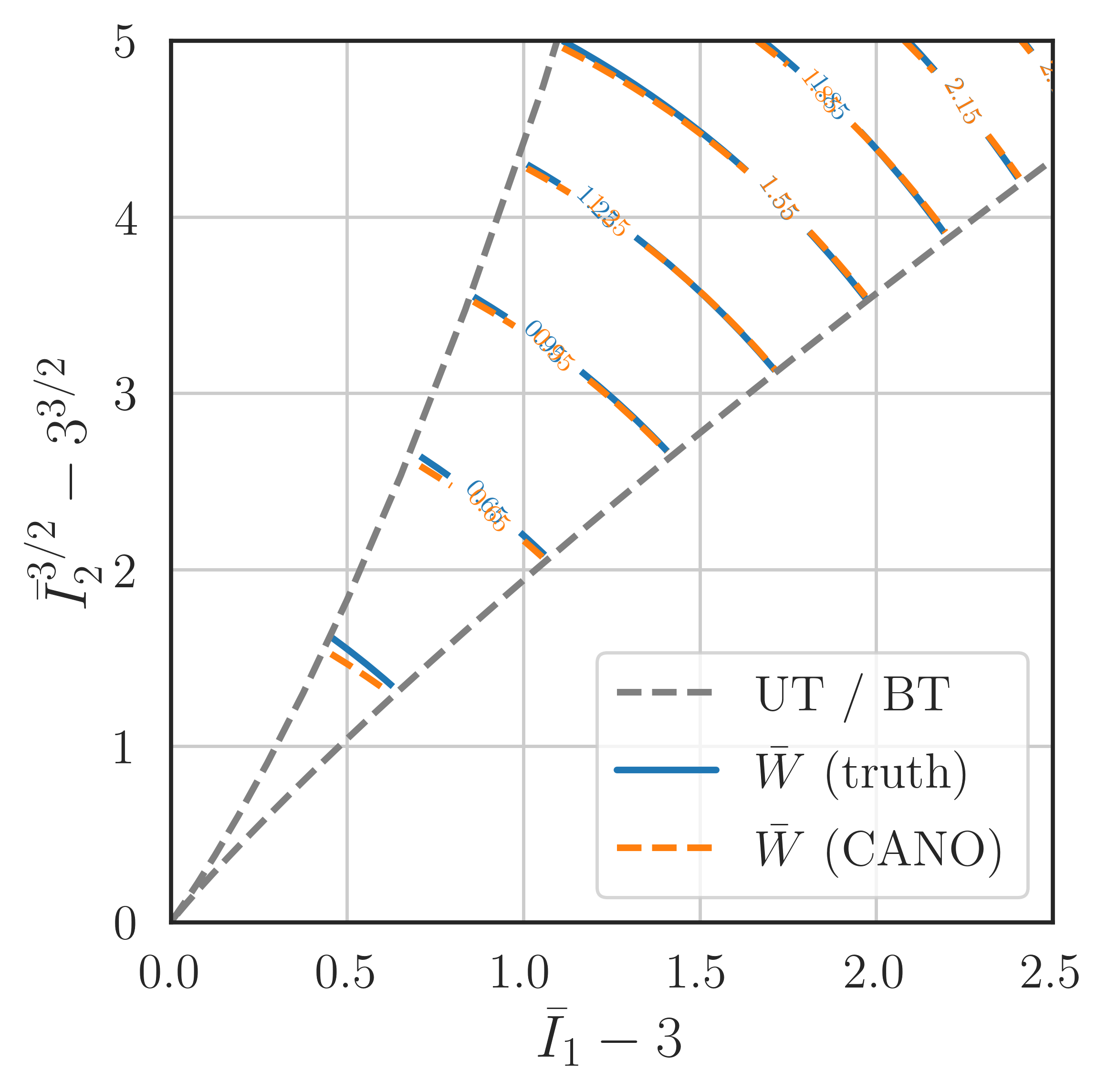}
        \caption{Median MSE.}
        \label{fig:results_cano_median}
    \end{subfigure}
    \begin{subfigure}[b]{0.3\textwidth}
        \centering
        \includegraphics[width=\textwidth]{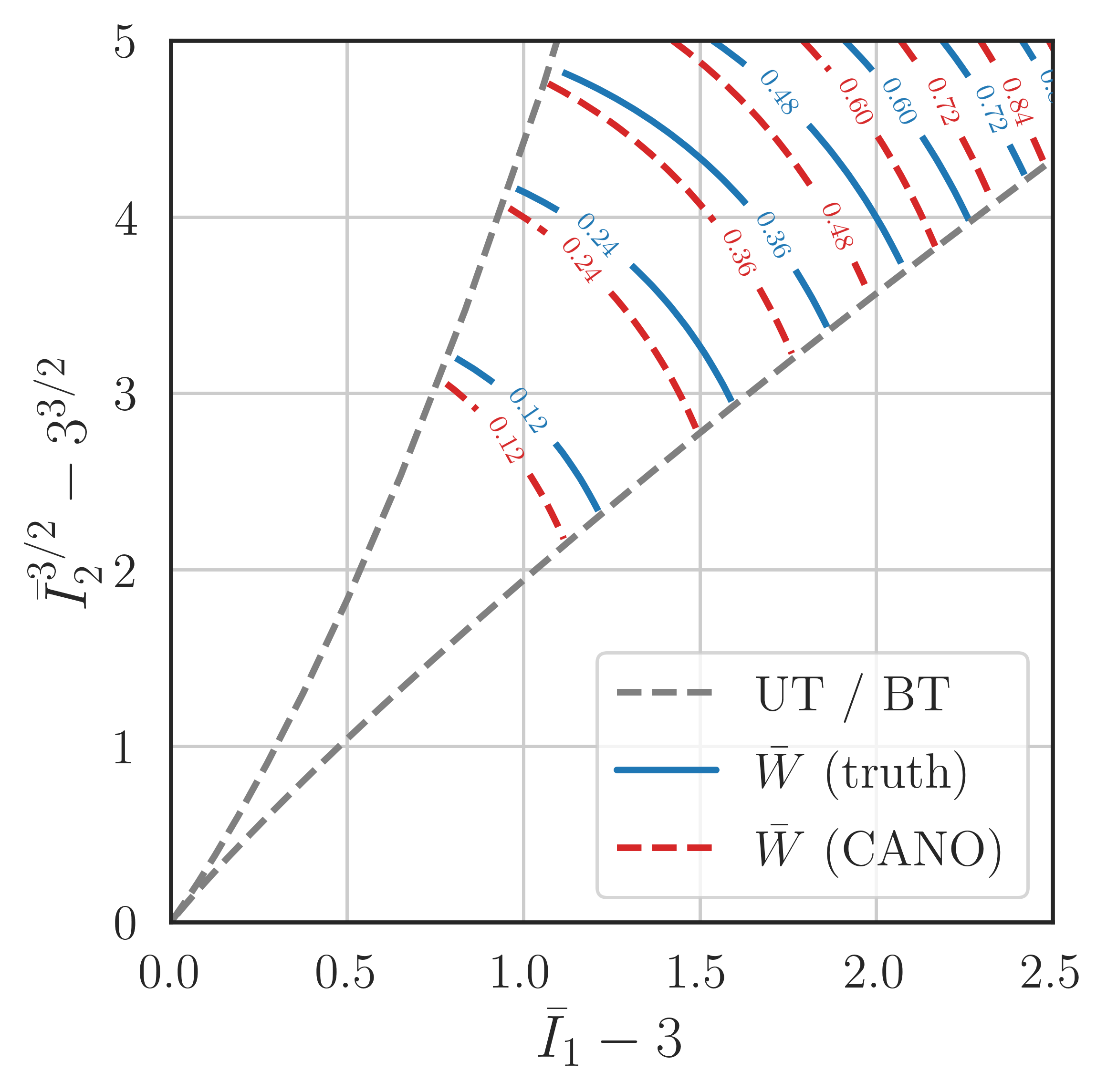}
        \caption{Worst MSE (outlier).}
        \label{fig:results_cano_worst}
    \end{subfigure}
    \caption{CANO results for \textbf{unseen testing data}.}
    \label{fig:results_cano}
\end{figure}

Next, we use the Adam optimizer to train the CANO on $80\%$ of the simulation data while using $10\%$ of the data for validation and $10\%$ of the data for testing. We refer to \cref{app:hyperparameters} for a detailed description of the hyperparameters. \cref{fig:results_cano_mse} shows a histogram of the MSE computed from the unseen testing dataset, which shows that the majority of samples exhibit low MSE values. \cref{fig:results_cano_best,fig:results_cano_median,fig:results_cano_worst} show the true and discovered models corresponding to the minimum, median, and maximum MSE. The agreement for the minimum and median MSE cases is excellent, while the prediction corresponding to the maximum MSE exhibits small deviations from the true isochoric strain-energy density.

A comparison of the PANO and CANO results reveals that the MSE distribution for CANO is shifted towards lower values, and the minimum and median MSE values are closer to one another than those obtained with PANO. This indicates that CANO predicts the underlying material models more consistently and accurately than PANO. Importantly, the maximum MSE achieved by CANO is lower than that of PANO. This improvement is reflected in the much closer agreement between the true and discovered isochoric strain-energy densities for the worst-performing case shown in \cref{fig:results_cano_worst}. These findings can be attributed to the constitutive structure that is explicitly embedded in the CANO architecture. By incorporating constitutive features directly into the network design, CANO is better equipped to learn accurate constitutive representations from the training data and to extrapolate to unseen data.

\subsection{Noisy data}

To assess the predictive performance of the neural operators under noisy measurements, we select a representative simulation from the test dataset and perturb the corresponding data with random noise. Specifically, we consider the simulation corresponding to the median MSE obtained by CANO. To emulate experimental measurements, uncorrelated Gaussian noise with a standard deviation $\sigma$ is added to the displacement fields of all load steps. Following the discussion in \cite{flaschel_unsupervised_2021}, realistic noise levels are represented by $\sigma=10^{-4}$ and $\sigma=10^{-3}$. In addition, we investigate the extreme case of $\sigma=10^{-2}$. Considering that the maximum absolute displacement in the first load step is $0.2$, this corresponds to an exceptionally high noise level and therefore represents a challenging test case for the neural operators.

\begin{figure}[!h]
    \centering
    \begin{subfigure}[b]{0.24\textwidth}
        \centering
        \includegraphics[width=\textwidth]{fig/results/Wbar_cano_2026-06-25-12-31-24/TaylorCubic_0816_median.png}
        \caption{No noise.}
    \end{subfigure}
    \begin{subfigure}[b]{0.24\textwidth}
        \centering
        \includegraphics[width=\textwidth]{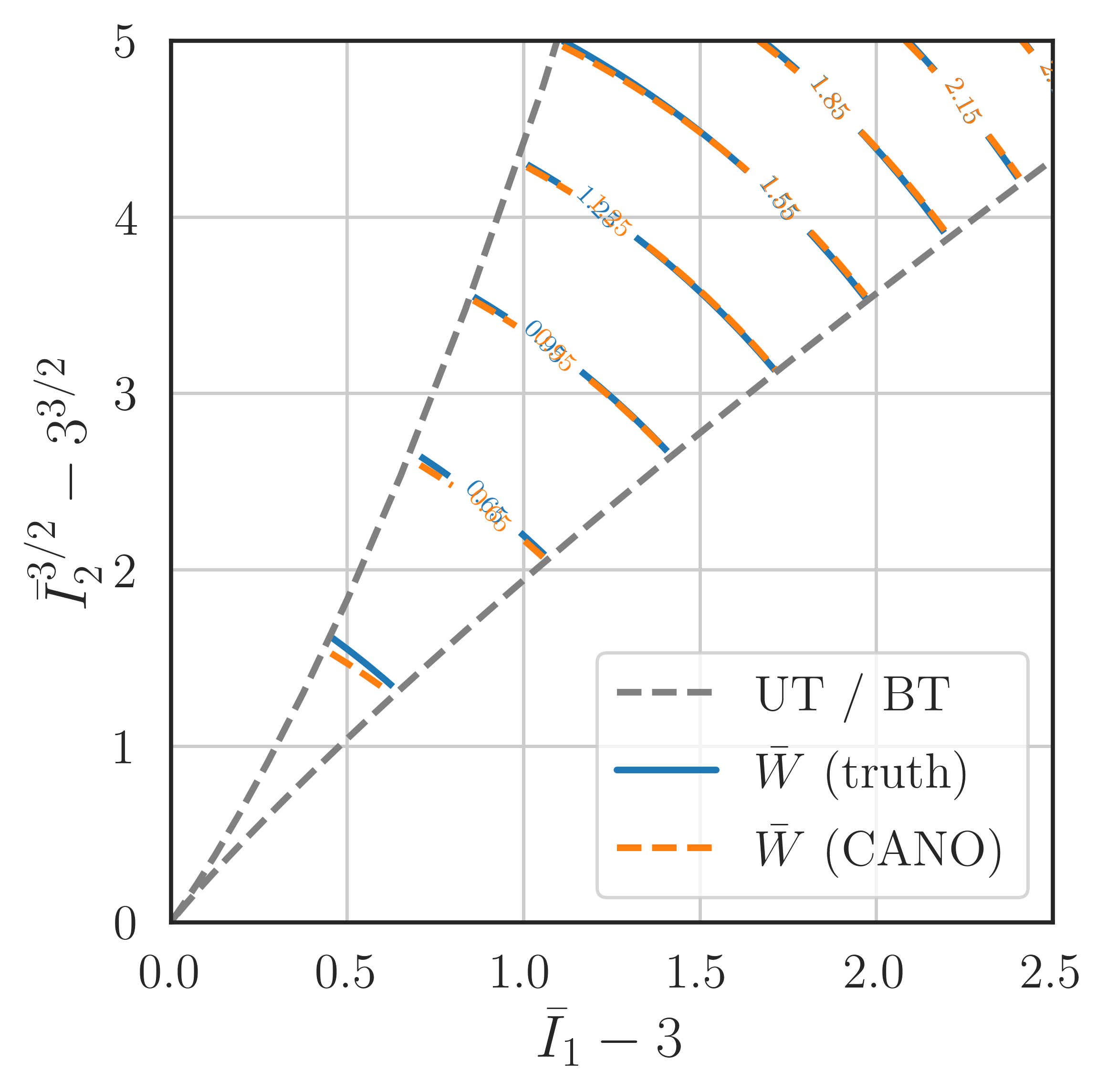}
        \caption{$\sigma=10^{-4}$.}
    \end{subfigure}
    \begin{subfigure}[b]{0.24\textwidth}
        \centering
        \includegraphics[width=\textwidth]{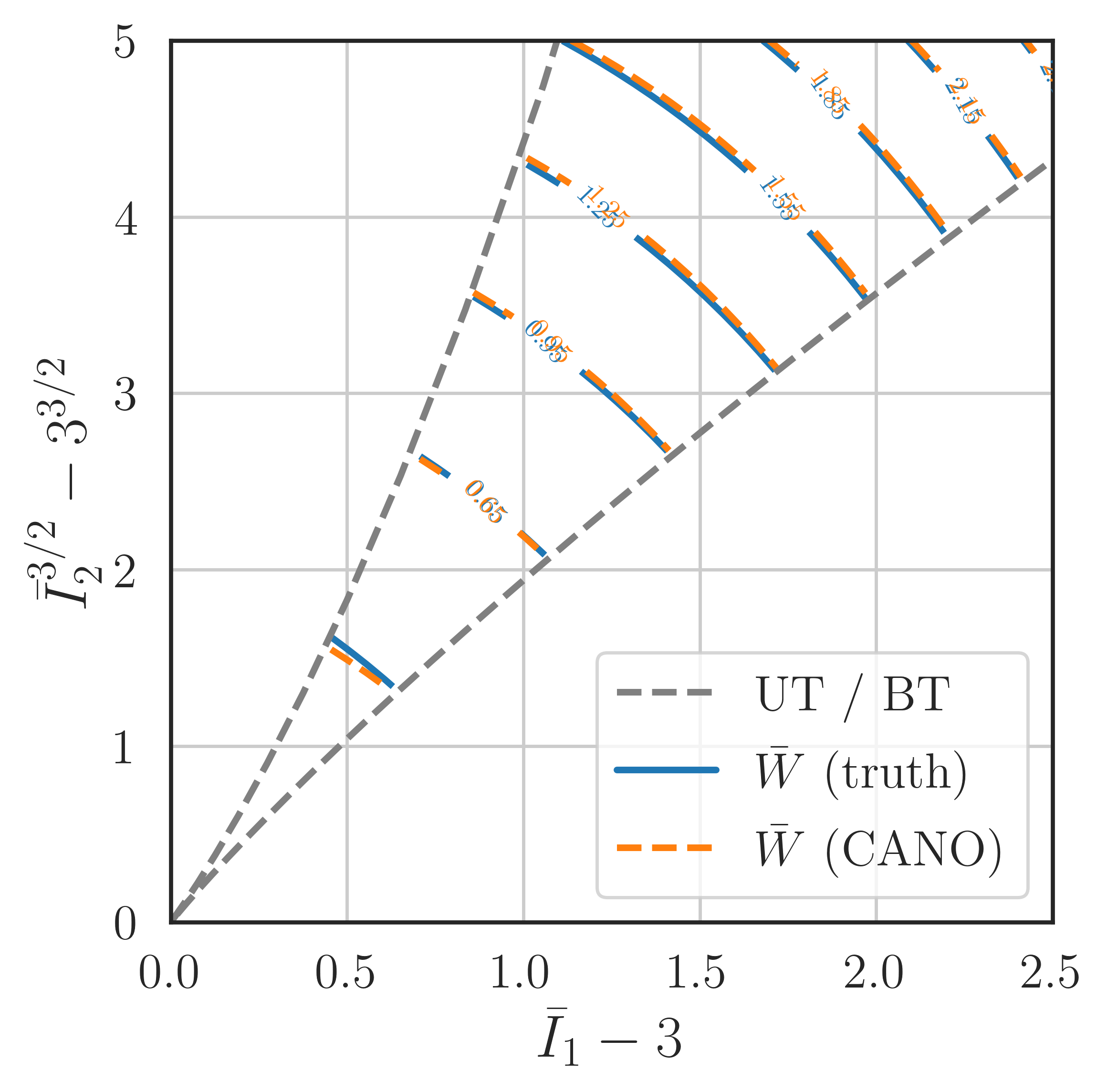}
        \caption{$\sigma=10^{-3}$.}
    \end{subfigure}
    \begin{subfigure}[b]{0.24\textwidth}
        \centering
        \includegraphics[width=\textwidth]{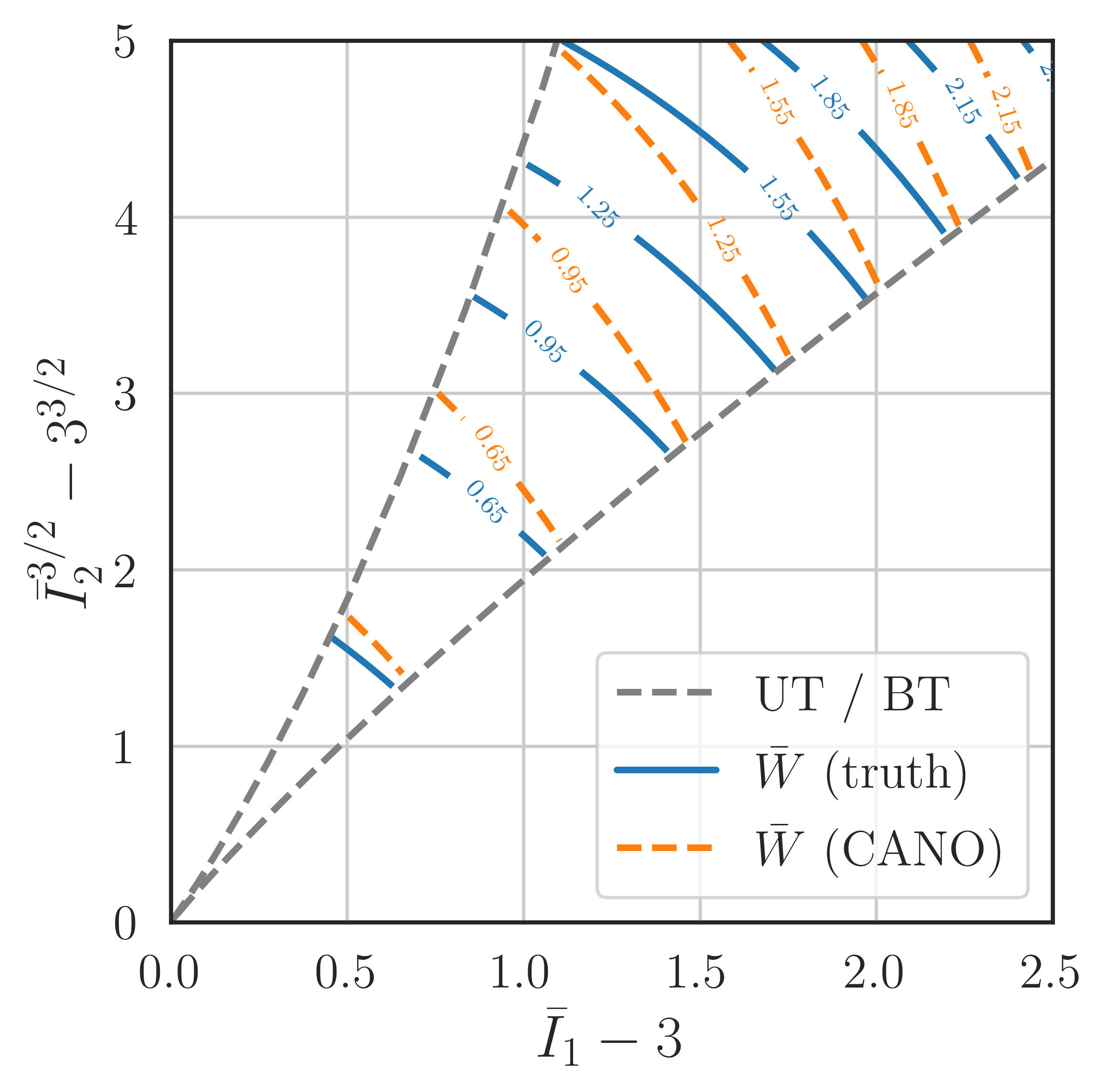}
        \caption{$\sigma=10^{-2}$ (extreme case).}
    \end{subfigure}
    \caption{CANO results for the median MSE sample of the \textbf{unseen testing data} with \textbf{noise} in the displacement data.}
    \label{fig:results_cano_noise}
\end{figure}

\Cref{fig:results_cano_noise} compares the true and discovered material models for all considered noise levels with the corresponding noise-free prediction. For the realistic noise levels of $\sigma=10^{-4}$ and $\sigma=10^{-3}$, the discovered material models exhibit only minor deviations from the ground truth, indicating that the neural operator is robust to measurement noise. Even for the extreme and rather unrealistic noise level of $\sigma=10^{-2}$, the predicted material model remains reasonably close to the ground truth, demonstrating a remarkable degree of robustness under severely perturbed input data. The observed robustness to measurement noise can likely be attributed to the first layer of the branch net, where the displacement field is represented using Laplacian eigenfunctions. Owing to their non-local support, the Laplacian eigenfunctions inherently smooth the displacement field, thereby attenuating high-frequency noise while preserving the underlying deformation patterns.

\subsection{Discretization independence and missing data}

In the following, we investigate the robustness of the neural operators to incomplete data, varying spatial discretizations, and geometries with different dimensions. Such robustness is essential for practical applications, as it enables a trained neural operator to be applied to experimental data acquired on different measurement grids and from specimens with geometries that differ from those used during training.

\begin{figure}[!h]
    \centering
    \begin{subfigure}[b]{0.24\textwidth}
        \centering
        \includegraphics[width=\textwidth]{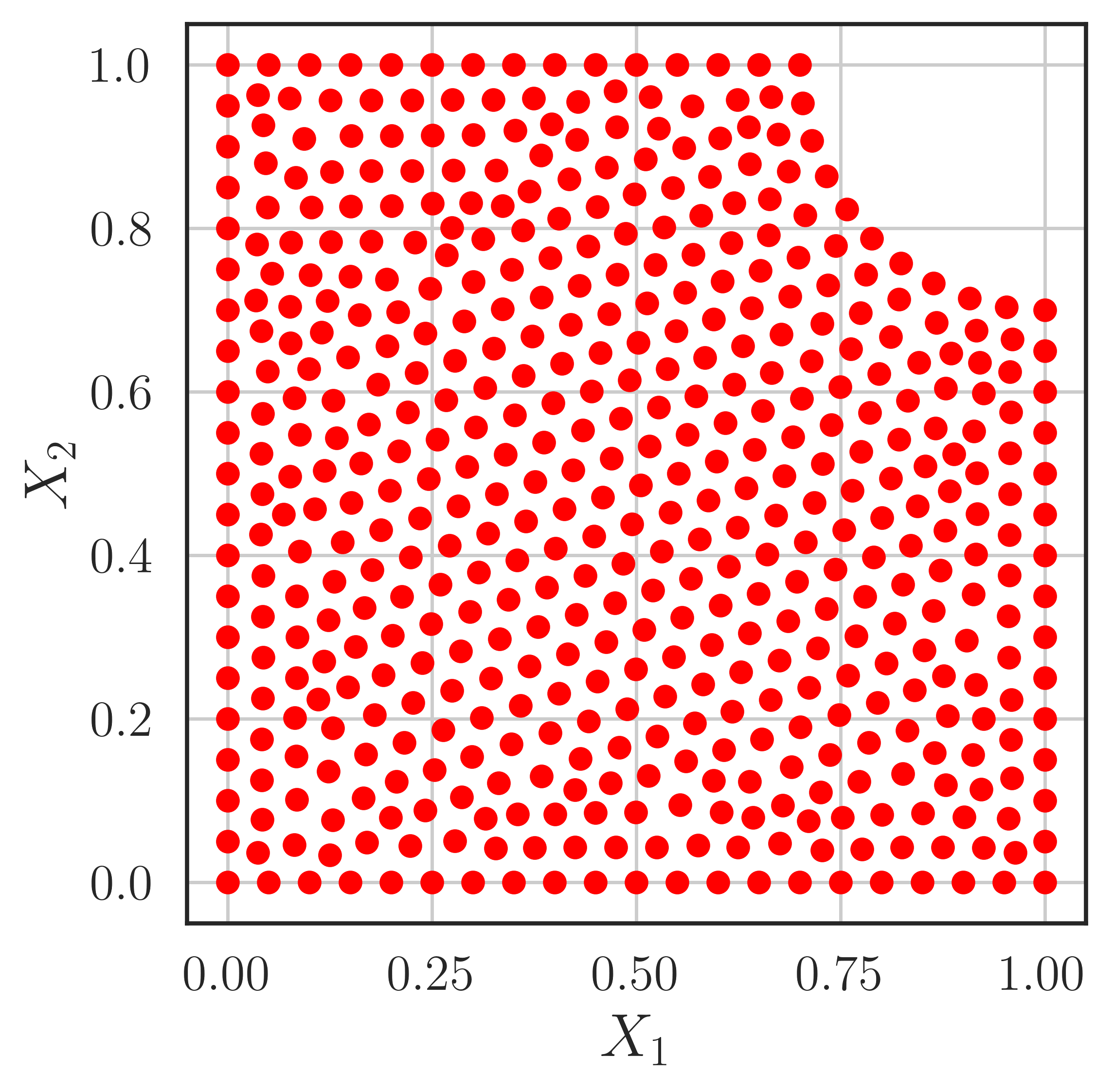}
        \caption{Full data ($N_{\bfX} = 499$).}
        \label{fig:results_cano_scarce_data1}
    \end{subfigure}
    \begin{subfigure}[b]{0.24\textwidth}
        \centering
        \includegraphics[width=\textwidth]{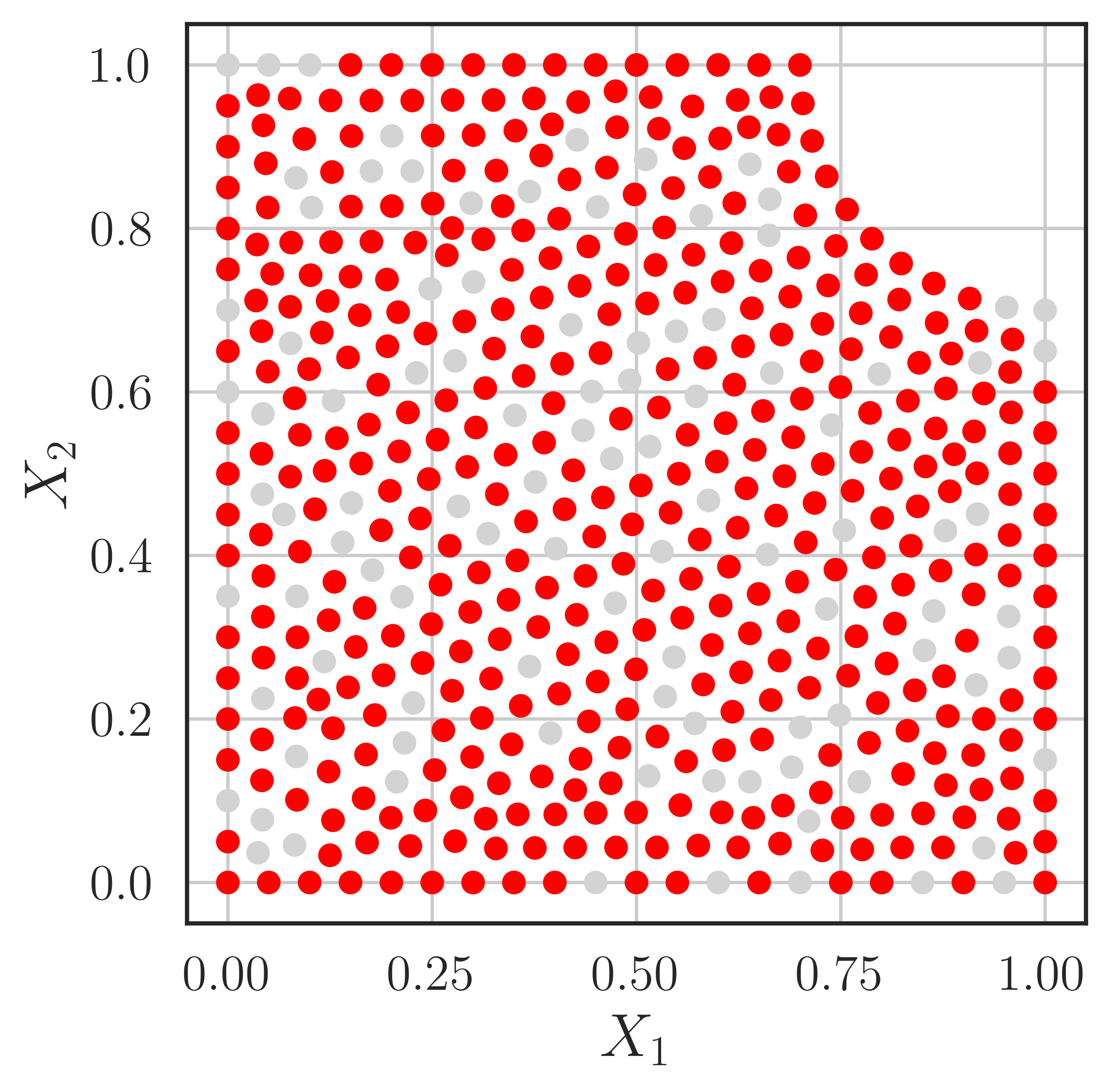}
        \caption{Scarce data ($N_{\bfX} = 400$).}
        \label{fig:results_cano_scarce_data2}
    \end{subfigure}
    \begin{subfigure}[b]{0.24\textwidth}
        \centering
        \includegraphics[width=\textwidth]{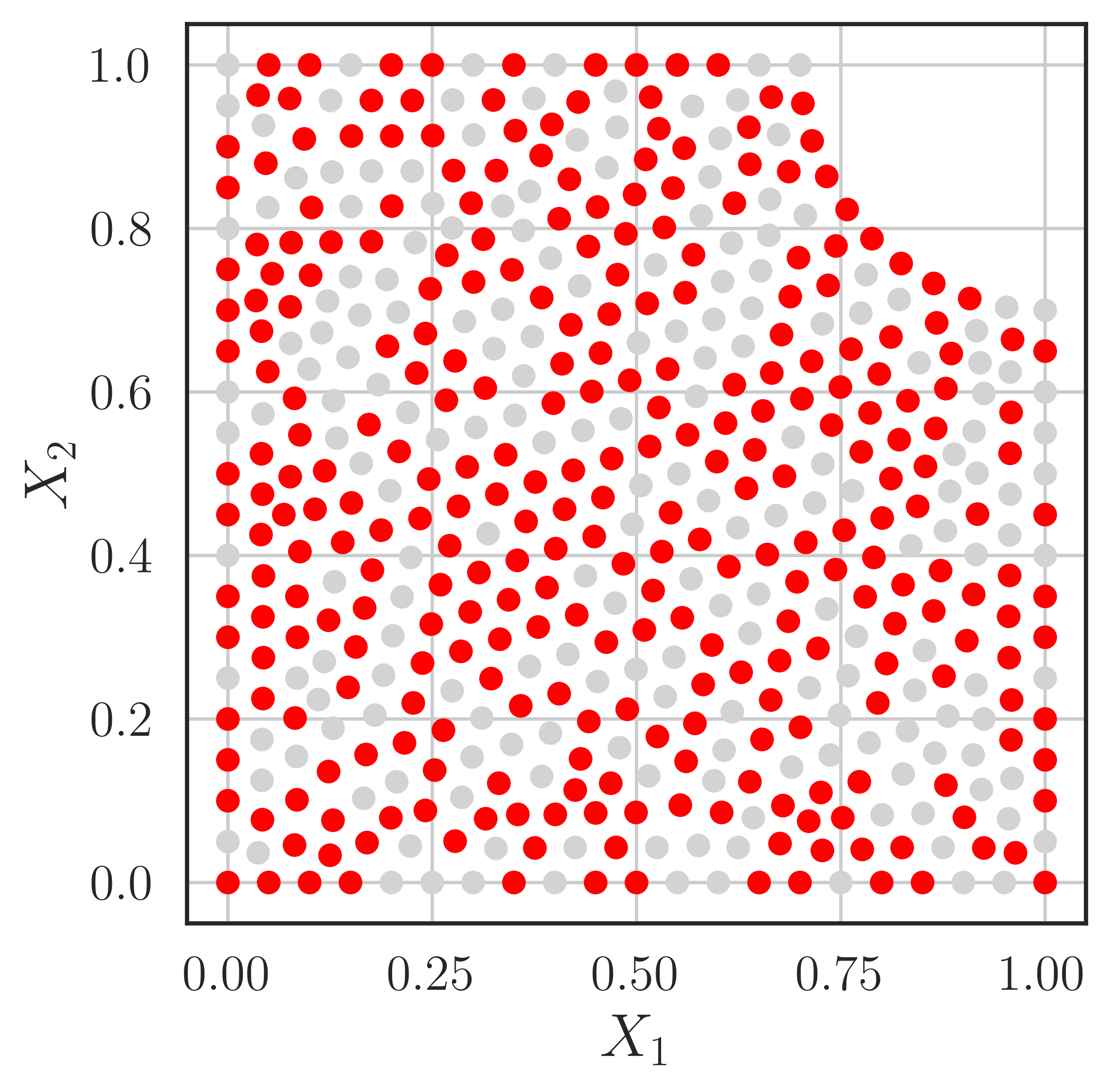}
        \caption{Scarce data ($N_{\bfX} = 300$).}
        \label{fig:results_cano_scarce_data3}
    \end{subfigure}
    \begin{subfigure}[b]{0.24\textwidth}
        \centering
        \includegraphics[width=\textwidth]{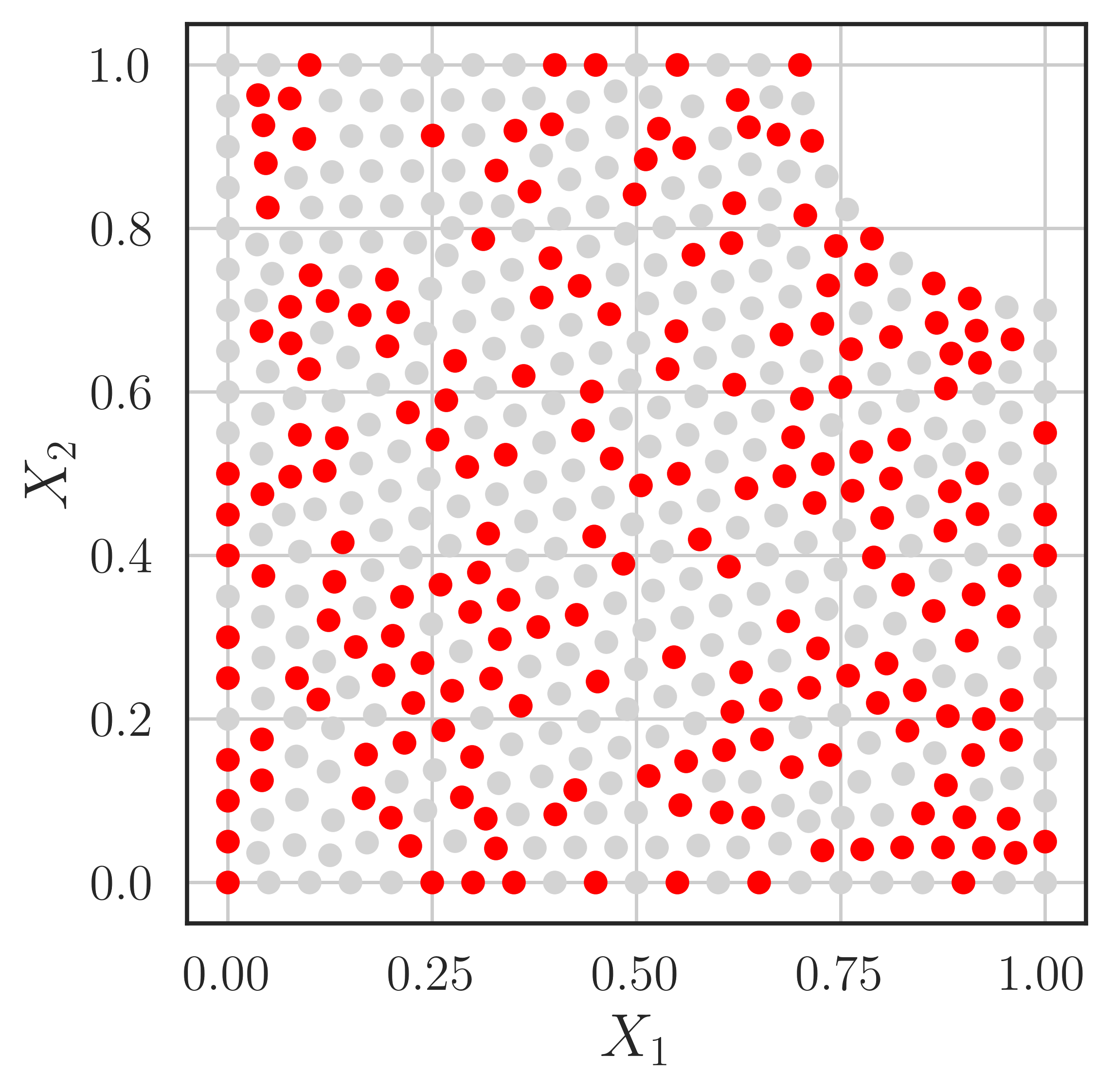}
        \caption{Scarce data ($N_{\bfX} = 200$).}
        \label{fig:results_cano_scarce_data4}
    \end{subfigure}
    \centering
    \begin{subfigure}[b]{0.24\textwidth}
        \centering
        \includegraphics[width=\textwidth]{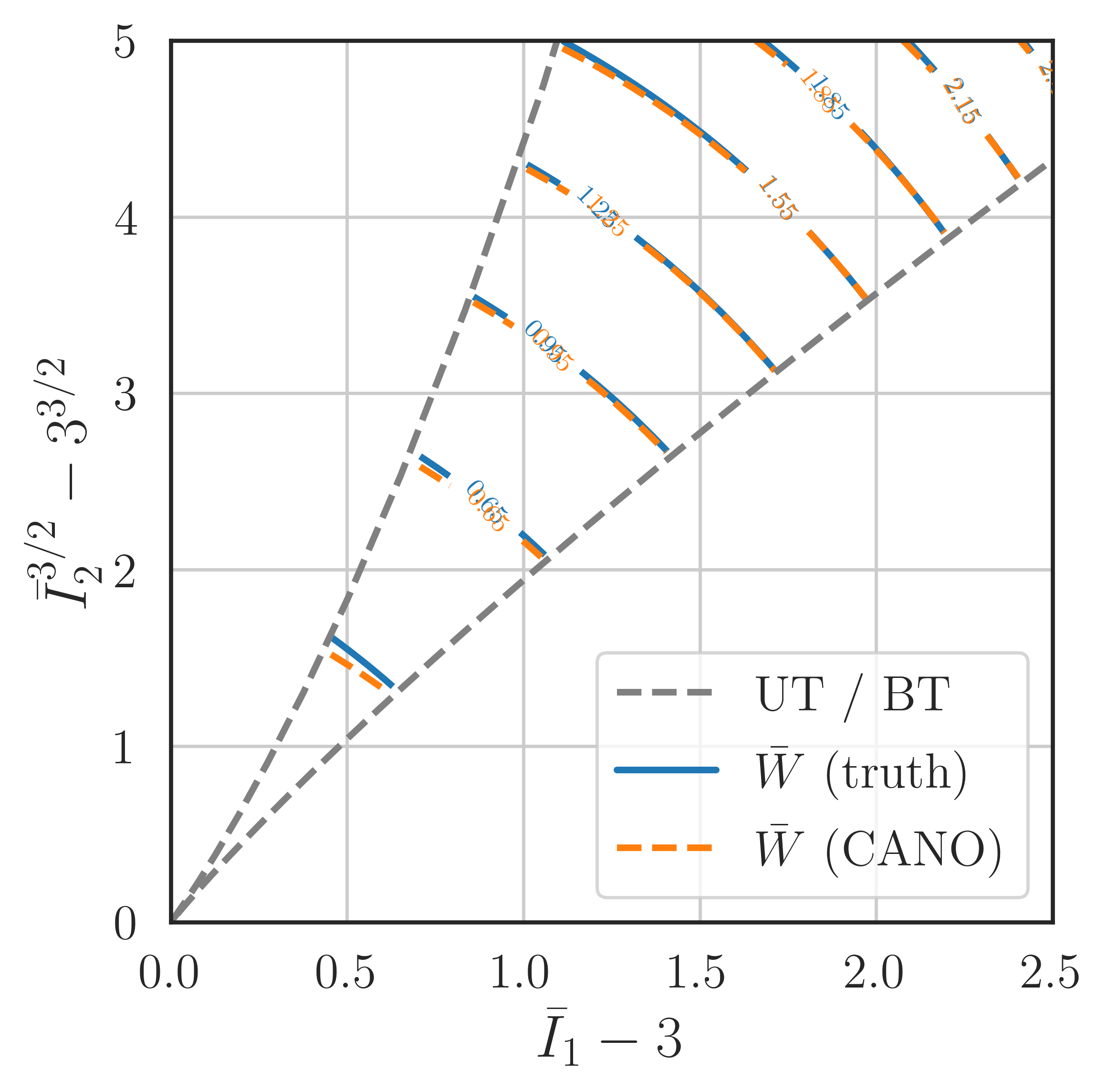}
        \caption{Model comparison ($N_{\bfX} = 499$).}
        \label{fig:results_cano_scarce1}
    \end{subfigure}
    \begin{subfigure}[b]{0.24\textwidth}
        \centering
        \includegraphics[width=\textwidth]{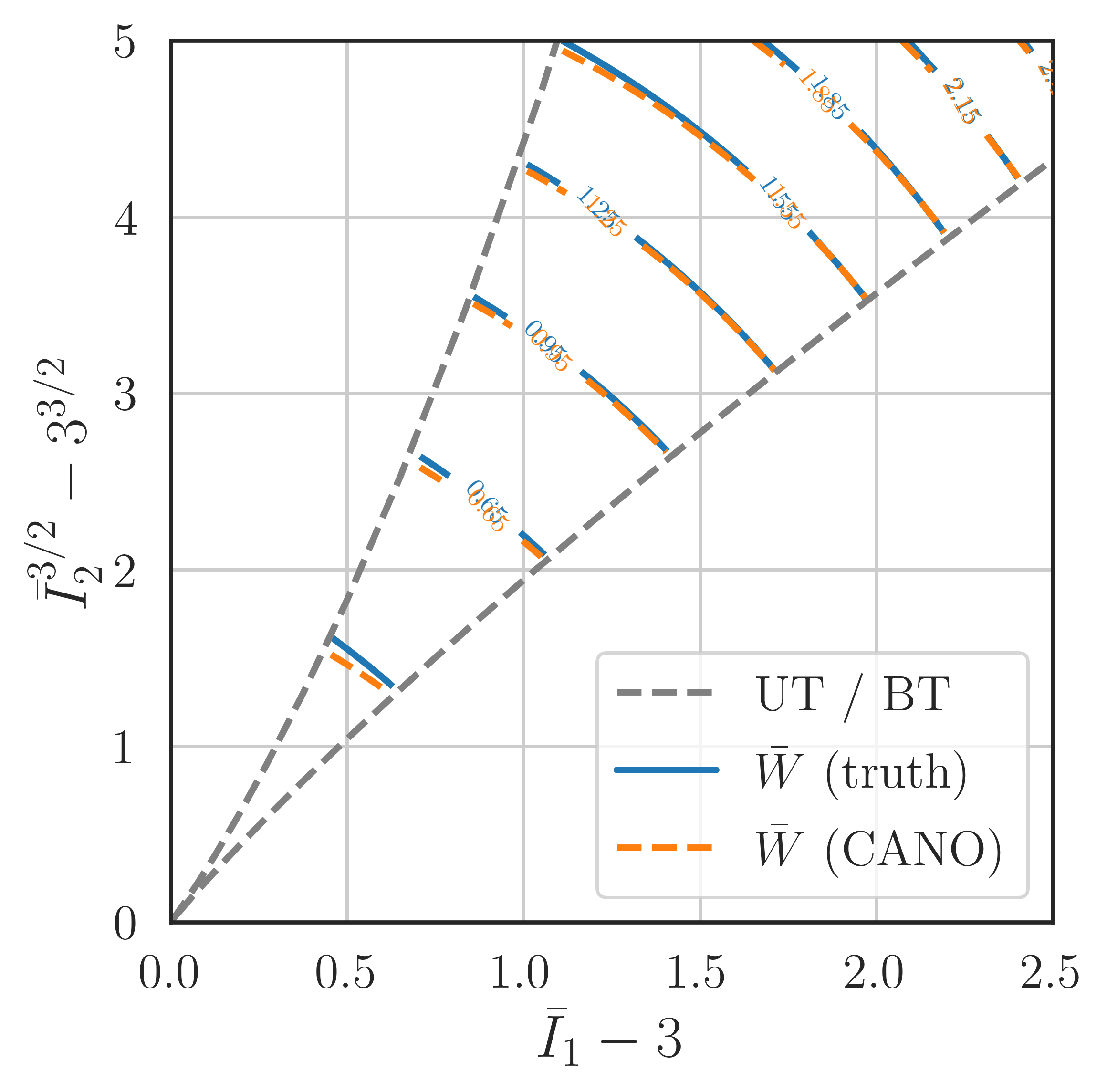}
        \caption{Model comparison ($N_{\bfX} = 400$).}
        \label{fig:results_cano_scarce2}
    \end{subfigure}
    \begin{subfigure}[b]{0.24\textwidth}
        \centering
        \includegraphics[width=\textwidth]{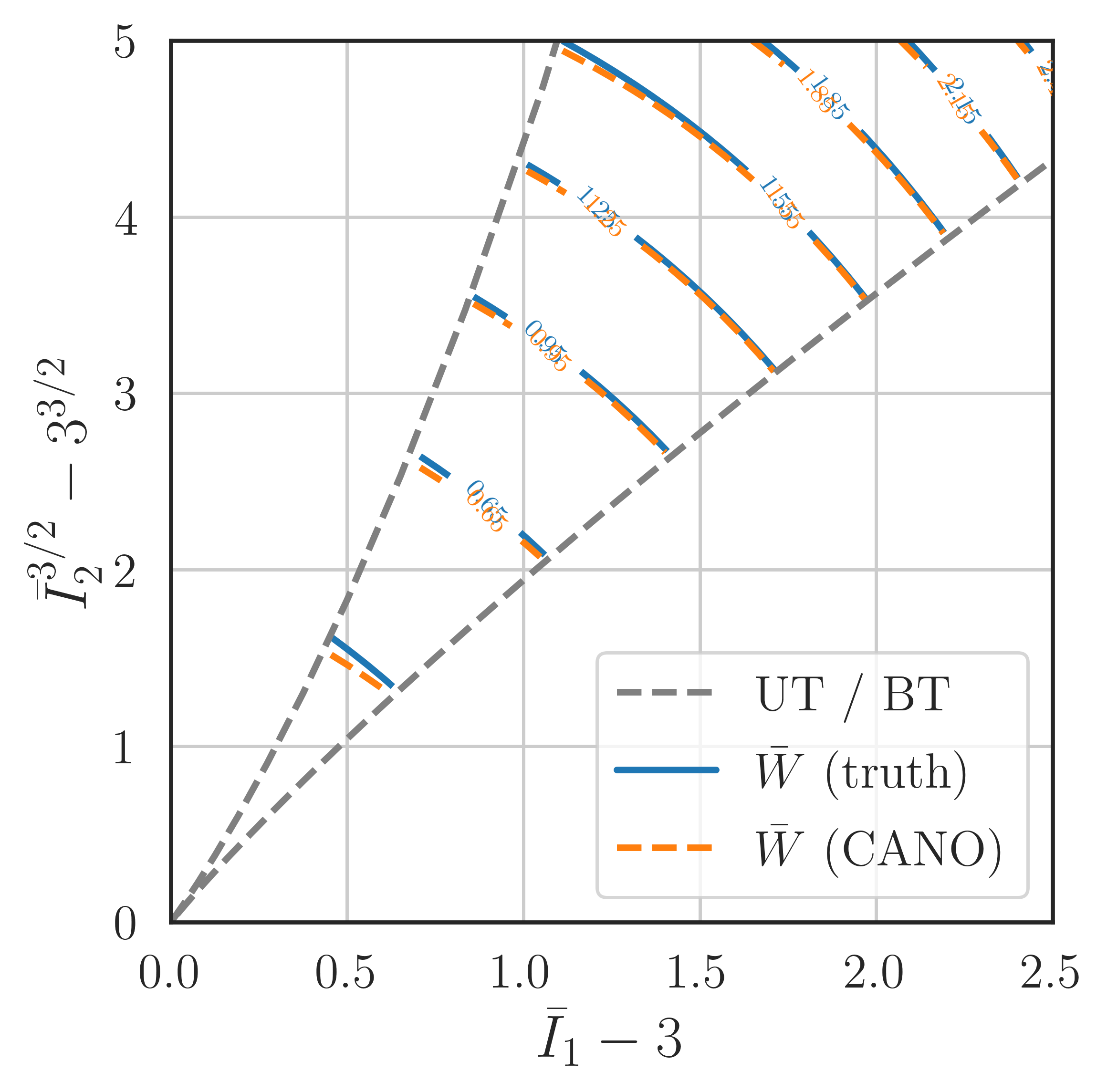}
        \caption{Model comparison ($N_{\bfX} = 300$).}
        \label{fig:results_cano_scarce3}
    \end{subfigure}
    \begin{subfigure}[b]{0.24\textwidth}
        \centering
        \includegraphics[width=\textwidth]{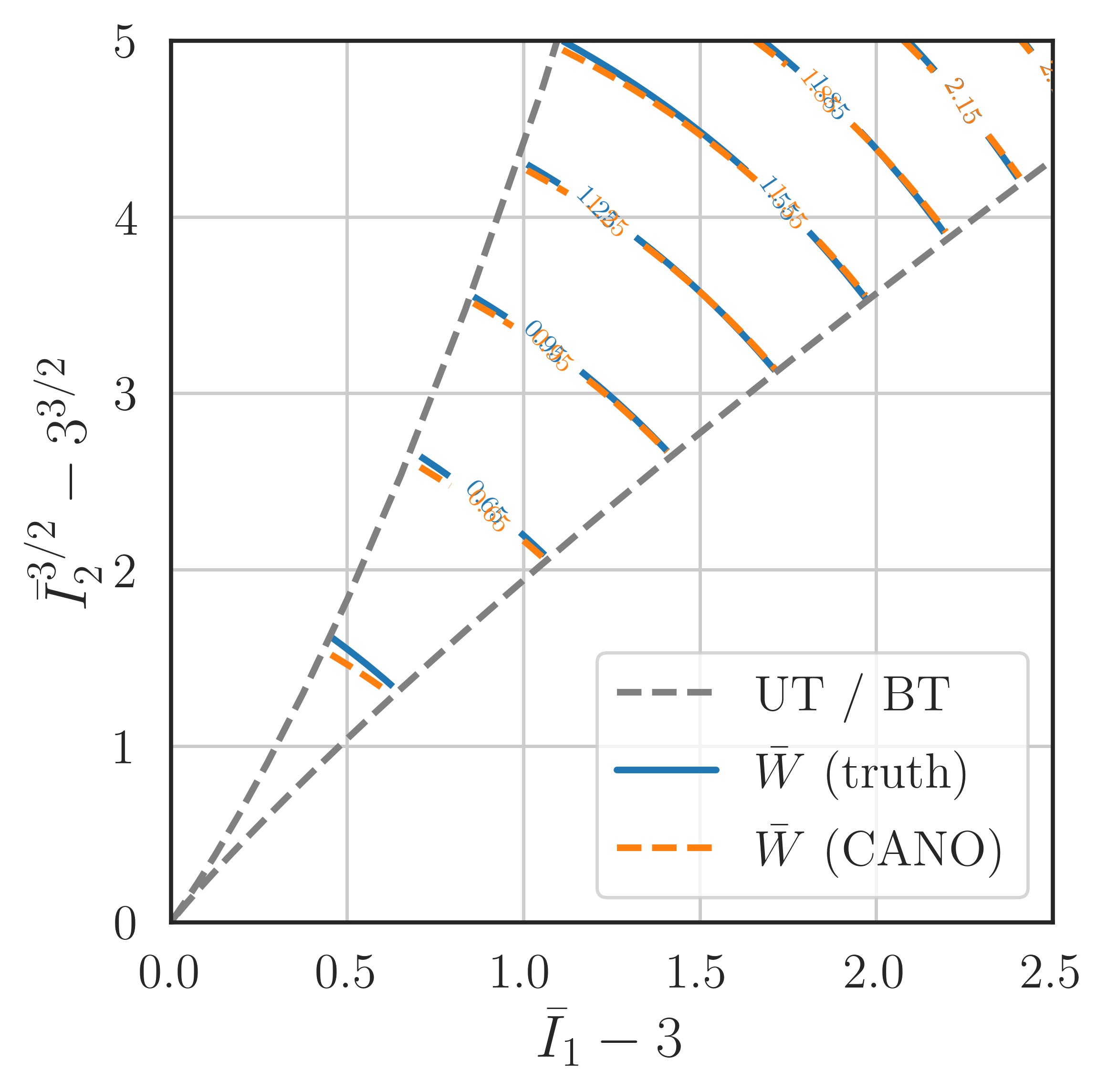}
        \caption{Model comparison ($N_{\bfX} = 200$).}
        \label{fig:results_cano_scarce4}
    \end{subfigure}
    \caption{CANO results for the median MSE sample of the \textbf{unseen testing data} with \textbf{missing displacement data}. Only the red points in (a-d) are considered for inferring the models in (e-h).}
    \label{fig:results_cano_scarce}
\end{figure}

First, we investigate the effect of incomplete measurement data on the predictions of the neural operators. This investigation is motivated by practical applications, where full-field measurements, such as those obtained from DIC, may contain missing or invalid data due to occlusions, poor image texture, or correlation failures. As a representative example, we consider the simulation corresponding to the median MSE obtained by CANO. From the full set of spatial measurement points, we randomly select a subset that is provided as input to the neural operator, while the remaining points are discarded. \cref{fig:results_cano_scarce_data1} shows the spatial distribution of the measurement points for the complete dataset, whereas \cref{fig:results_cano_scarce_data2,fig:results_cano_scarce_data3,fig:results_cano_scarce_data4} show the randomly selected subsets for progressively decreasing numbers of measurement points. The corresponding displacement fields are projected onto the Laplacian eigenfunction basis and subsequently serve as input to the branch net of the neural operators. The resulting true and discovered material models are shown in \cref{fig:results_cano_scarce1,fig:results_cano_scarce2,fig:results_cano_scarce3,fig:results_cano_scarce4}. The results demonstrate that the predictions of the neural operators remain remarkably robust, exhibiting only minor changes even when a substantial fraction of the measurement data is omitted.

\begin{figure}[!h]
    \centering
    \begin{subfigure}[b]{0.3\textwidth}
        \centering
        \includegraphics[width=\textwidth]{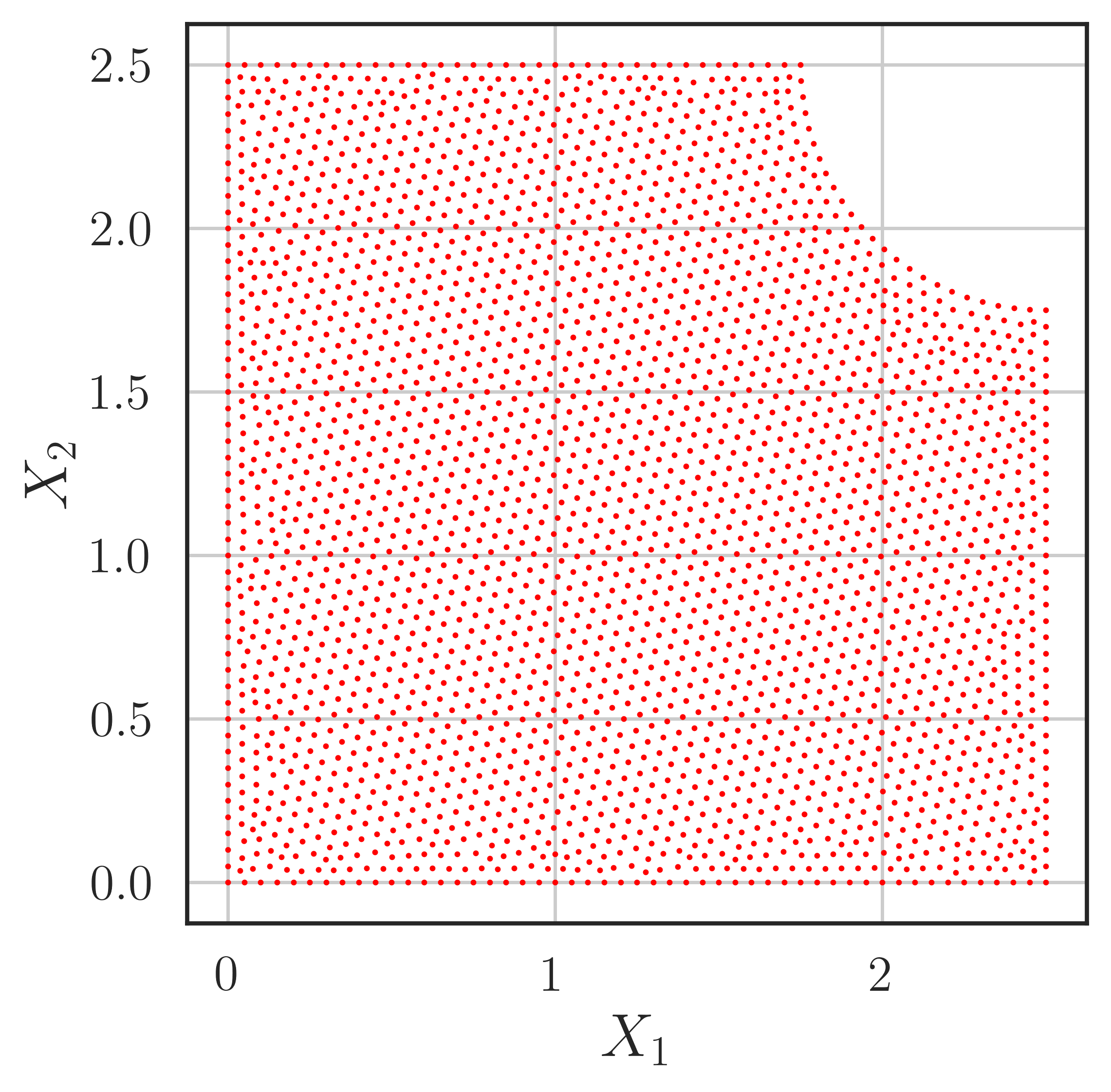}
        \caption{Spatially refined data ($N_{\bfX} = 2849$).} %on a scaled specimen geometry
        \label{fig:results_cano_mesh_large_data}
    \end{subfigure}
    \hspace{1cm}
    \begin{subfigure}[b]{0.3\textwidth}
        \centering
        \includegraphics[width=\textwidth]{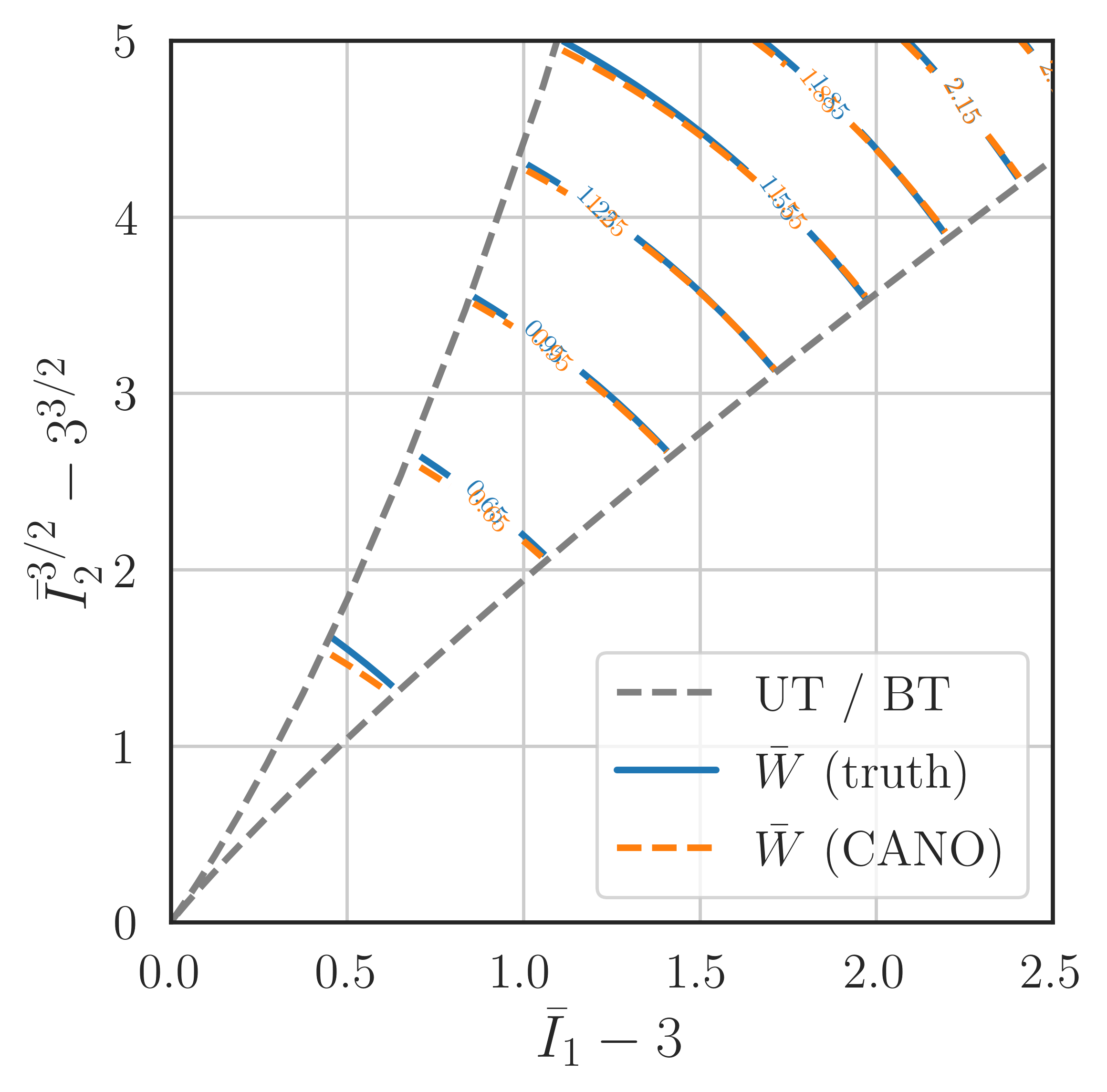}
        \caption{Model comparison.}
        \label{fig:results_cano_mesh_energy}
    \end{subfigure}
    \caption{CANO results for \textbf{unseen data} stemming from a \textbf{different discretization} and \textbf{different geometry dimensions}.}
    \label{fig:results_cano_mesh_large}
\end{figure}

In practical applications, it may not always be feasible to fabricate specimens with the same geometric dimensions as those used to generate the training data. Nevertheless, the trained neural operators can also be applied to specimens of different sizes. The only requirements are that the geometry remains a thin square plate with a central hole, that the ratio of the hole radius to the specimen side length is preserved, and that the same amount of total stretch is applied. The absolute specimen dimensions, however, remain arbitrary, and both the side length and the thickness may differ from those considered during training.
To demonstrate invariance with respect to specimen dimensions, we consider a specimen whose side length and thickness are scaled by factors of $s_{1}=2.5$ and $s_{3}=1.5$, respectively, relative to the geometry used for training. As a representative example, we select the material model corresponding to the median MSE obtained by CANO. The deformation of the scaled specimen is then simulated using a spatial discretization that deliberately differs from that employed during the generation of the training data. \cref{fig:results_cano_mesh_large_data} shows the spatial locations at which displacements are evaluated on the surface of the new specimen geometry. The resulting simulation data are then used as input to the neural operator. To this end, the displacement data are scaled by $s_{1}$, and the reaction force is scaled by $s_{1}$ and $s_{3}$ prior to being passed to the branch net. \cref{fig:results_cano_mesh_energy} shows the resulting discovered material model in comparison with the true material model. The good agreement indicates that the trained neural operators are not restricted to specific geometry dimensions or spatial discretizations.

\section{Ill-posedness of the infinite-dimensional inverse problem}
\label{sec:ill-posedness}

\subsection{General remarks}

It is well known that inverse problems, including those considered in this work, can be ill-posed.  Different material modeling functions in $\mathcal{W}$ may be equally consistent with a given set of observed data in $\mathcal{U}$ and $\mathcal{R}$ \citep{hartmann_identifiability_2018,moreno-mateos_learning_2026}. In the following, we discuss the ill-posed and infinite-dimensional nature of the problem and present numerical experiments suggesting that our proposed neural operator architecture regularizes the inverse problem and reduces its ill-posedness.

% bounded domains
A particularly challenging aspect of the learning problem considered in this work is that the functions in the image space $\mathcal{W}$ are defined over the unbounded domain $\mathcal{I}_{\text{adm}}$. This stands in contrast to operator learning for forward problems, where the quantities of interest are typically PDE solutions defined over a bounded domain. The deformations contained in any experimental dataset used to train our neural operators are necessarily bounded. Regardless of how the standardized experiment is designed, there will always exist deformations that lie outside the range encountered during training. This renders the unique identification of a function in $\mathcal{W}$ for a given experimental dataset impossible. We note, however, that this characteristic of the inverse problem is not specific to our operator learning framework, but is inherent to material characterization frameworks in general, including traditional optimization-based and machine learning-based approaches.

% infinite-dimensionality
Even if the domain of the functions in $\mathcal{W}$ is assumed to be bounded, uniquely identifying a function in $\mathcal{W}$ from a given experimental dataset remains challenging because of the infinite-dimensional nature of the problem. The experimental observations provide only finite-dimensional information about the underlying material model, whereas the sought function belongs to an infinite-dimensional space. Consequently, multiple functions in $\mathcal{W}$ may be consistent with the same set of observations, even when their domains are restricted to a bounded set $\mathcal{I}_{\text{bounded}} \subset \mathcal{I}_{\text{adm}}$.

In practice, additional assumptions or regularization are required to restrict the admissible solution space and obtain meaningful reconstructions. Hence, in our study, we restrict the design space of the constitutive functions. As described previously, the space $\mathcal{W}$ is constrained to only include physically admissible functions. This reduces the space of admissible constitutive behaviors and thereby mitigates the ill-posedness of the problem.
Furthermore, while neural operators are designed to address infinite-dimensional learning problems, their practical implementations are always restricted to finite-dimensional latent representations, such as the latent feature vector $\bfpsi$ in our PANO and CANO architectures. The principal difference between PANO and CANO lies in the construction of these latent features. In CANO, the features are prescribed a priori, whereas in PANO they are learned during training.
The universal approximation properties of neural operators have been studied theoretically \citep{chen_universal_1995}. However, these results are established in the asymptotic regime of infinitely many parameters and therefore provide only theoretical guarantees regarding the expressive power of the architectures. They do not directly address the approximation capabilities or regularization properties of practical implementations with finite-dimensional latent representations and finite network capacity.

The fact that the practical implementations of the PANO and CANO operate on finite-dimensional feature vectors $\bfpsi$ provides an additional source of regularization. In the following, we present a numerical investigation to demonstrate that the proposed neural operator architectures regularize the learning problem sufficiently to render it well-posed.

\subsection{Material model identifiability through the weak equilibrium equations}

As an attempt to assess the well-posedness of the learning problem within a constrained class of material models, we investigate the sensitivity of the weak equilibrium equations and the measured reaction forces to perturbations in the material model. We restrict our analysis to the class of material models that can be represented by our CANO architecture and comment on the PANO architecture subsequently.

We consider CANO's constitutive modeling features $\psi_i(\bfF)$ (see \cref{eq:features_cano}) obtained from the cubic Taylor expansion of the isochoric strain-energy density (see \cref{app:Taylor_expansion}). Within this constrained class of material models, the constitutive behavior is fully characterized by the material parameters $C_{ij}$, which we collect in the parameter vector $\bfc \in \Rset^6$. Consequently, the strain-energy density and the first Piola-Kirchhoff stress can be expressed as
\begin{equation}
    W(\bfF) = c_i \psi_i(\bfF) - p [J-1], \quad \bfP(\bfF) = c_i \frac{\partial \psi_i}{\partial \bfF} - p \bfF\mtransposed,
\end{equation}

Our goal is to prove the identifiability of the unknown parameters $\bfc$ through the weak equilibrium equations for the given displacement field and reaction force data. We consider the reduced plane stress problem and assume that the displacement fields $\bfu_t(\bfX): \Omega^{\text{2D}}\rightarrow\Rset^2$ and reaction forces $R_t$ are known from measurements at time steps $t$. Under the incompressibility condition, the deformation gradient computes to % $\bfF_t(\bfX) = \bfI + \nabla \bfu_t(\bfX)$.
\begin{equation}
    \bfF_t(\bfX) =\begin{bmatrix}
        % 1 + u_{1t,1}(\bfX) & u_{1t,2}(\bfX) & 0 \\
        \bfI + \nabla_{\bfX} \bfu_t(\bfX) & 0 \\
        0 & \frac{1}{\det(\bfI + \nabla \bfu_t(\bfX))} \\
    \end{bmatrix}.
\end{equation}

Under the assumption of plane stress conditions, the first Piola-Kirchhoff stress components $P_{13}$, $P_{31}$, $P_{23}$, $P_{32}$, and $P_{33}$ vanish identically. From $P_{33}=0$, we compute the pressure term $p={c_i \, {\partial \psi_i}/{\partial F_{33}}} \cdot {1}/{F^{-1}_{33}}$, and the stress at time step $t$ is computed to
% \begin{equation}
%     p={ \frac{\partial \bar W}{\partial F_{33}}} \frac{1}{F^{-1}_{33}} 
% \end{equation}
% \begin{equation}
%     \bfP(\bfF_t) = \left.\frac{\partial \bar W}{\partial \bfF}\right|_{\bfF=\bfF_t} - \left.{\frac{\partial \bar W}{\partial F_{33}}}\right|_{\bfF=\bfF_t} \frac{\bfF_t\mtransposed}{F^{-1}_{t33}} 
% \end{equation}
\begin{equation}
    \bfP(\bfF_t) = c_i \left[ \left.\frac{\partial \psi_i}{\partial \bfF}\right|_{\bfF=\bfF_t} - \left.{\frac{\partial \psi_i}{\partial F_{33}}}\right|_{\bfF=\bfF_t} \frac{\bfF_t\mtransposed}{F^{-1}_{t33}}\right].
\end{equation}

Next, we show that the parameters $\bfc$ can be uniquely identified from the given data through the weak equilibrium equations. This procedure is closely aligned with weak formulation-based inverse parameter identification methods from full-field measurements, such as the Virtual Fields Method (VFM) \citep{grediac_principle_1989,pierron_virtual_2012}, the Equilibrium Gap Method (EGM) \citep{claire_finite_2004}, or EUCLID \citep{flaschel_unsupervised_2021,flaschel_discovering_2022}. We discretize the weak form of the linear momentum balance by introducing linear shape functions $N_k(\mathbf{X})$ for the discretization of the test functions. Denoting basis vectors by $\bfe_j$, this leads to the following expression for the internal nodal forces
\begin{equation}
    f^{\text{int}}_{tjk} = \int_\Omega^{\text{2D}} \, \bfP(\bfF_t) \colon \nabla (N_k \bfe_j) \, \mathrm{d}\bfX,
\end{equation}
where $t$ denotes the time step, $j$ the spatial dimension, and $k$ the node in the finite element mesh. By substituting the stress, we obtain
\begin{equation}
    f^{\text{int}}_{tjk} = c_i \int_\Omega^{\text{2D}} \, \left[ \left.\frac{\partial \psi_i}{\partial \bfF}\right|_{\bfF=\bfF_t} - \left.{\frac{\partial \psi_i}{\partial F_{33}}}\right|_{\bfF=\bfF_t} \frac{\bfF_t\mtransposed}{F^{-1}_{t33}}\right] \colon \nabla (N_k \bfe_j) \, \mathrm{d}\bfX,
\end{equation}
where all integrals are evaluated using standard numerical quadrature. All internal forces are collected in the force vector $\bff^{\text{int}}\in\Rset^{2N_tN_{\bfX}}$ which is written as
\begin{equation}
    \bff^{\text{int}} = \bfA^{\text{int}} \bfc,
\end{equation}
where the matrix $\bfA^{\text{int}} \in \Rset^{2N_tN_{\bfX} \times 6}$ collects contributions arising from the integrals above. For a detailed description of the assembly of this matrix, we refer to \cite{flaschel_unsupervised_2021}. At all free degrees of freedom, the internal nodal forces should vanish. We introduce the selection matrix $\bfS^{\text{free}}$ that selects the appropriate entries from the internal nodal force vector $\bff^{\text{free}}=\bfS^{\text{free}}\bff^{\text{int}}=\bfS^{\text{free}}\bfA^{\text{int}} \bfc = \bf0$. Further, the internal nodal forces at the boundary must sum up to the measured reaction forces. We introduce the matrix $\bfS^{\text{react}}$ such that $\bfS^{\text{react}}\bff^{\text{int}}=\bfS^{\text{react}}\bfA^{\text{int}} \bfc = \bfR^{\text{react}}$, where $\bfR^{\text{react}}$ collects the reaction forces of all load steps. We concatenate the equations and obtain
\begin{equation}
\label{eq:equilibrium_gap}
    \bfA \bfc = \bfR,
    \quad \text{with} \quad \bfA =
    \begin{bmatrix}
        \bfS^{\text{free}}\bfA^{\text{int}} \\
        \bfS^{\text{react}}\bfA^{\text{int}} \\
    \end{bmatrix}, ~
    \bfR = 
    \begin{bmatrix}
        \bf0 \\
        \bfR^{\text{react}} \\
    \end{bmatrix}.
\end{equation}
The system above can be used to identify the parameters $\mathbf{c}$ in a least-squares sense. The system of equations clarifies why the reaction forces are needed for the inverse problem. The equations corresponding to the free degrees of freedom are homogeneous. They may constrain the relative combination of material parameters, but they do not determine their absolute value. The measured reaction force provides this scale information.

In the following, we use the system of equations above to assess the well-posedness of the inverse problem by studying the properties of $\bfA$. We consider two parameter vectors $\bfc$ and $\bfc+\Delta \bfc$. They fulfill the equilibrium equations if $\bfA \bfc = \bfR$ and $\bfA [\bfc+\Delta \bfc] = \bfR$. Subtracting the two corresponding systems gives
\begin{equation}
    \bfA \Delta \bfc = \bf0.
\end{equation}
Thus, the first-level non-uniqueness within the restricted material class is
\begin{equation}
    \ker\bfA
    =
    \left\{
    \Delta \bfc\in\mathbb{R}^6:
    \bfA \Delta \bfc = \bf0
    \right\}.
\end{equation}
We compute the singular value decomposition
\begin{equation}
    \bfA = \bfU\bfSigma \bfV\transposed.
\end{equation}
The numerical rank is defined as
\begin{equation}
    \operatorname{rank}_{\tau}(\bfA)
    =
    \#\left\{
    i:
    \frac{\sigma_i}{\sigma_{\max}}>\tau
    \right\},
\end{equation}
where $\tau$ is a prescribed relative tolerance, $\sigma_i=\Sigma_{ii}$ (no Einstein summation) are the singular values of $\bfA$, and $\sigma_{\max}$ is the largest singular value. The numerical nullity is then
\begin{equation}
    \dim\ker_{\tau}\bfA
    =
    6-\operatorname{rank}_{\tau}(\bfA).
\end{equation}
If $\bfA$ possesses right singular vectors associated with zero or sufficiently small singular values, these vectors represent material parameter perturbations that produce no observable change in the measured displacement and reaction force data. Such directions correspond to non-identifiable parameter combinations, indicating that the inverse problem is ill-posed. Conversely, if all singular values exceed the prescribed relative tolerance, every material parameter perturbation produces a detectable change in the available observations. In this case, no non-identifiable directions exist, and the inverse problem is well-posed.

We assemble the matrix $\bfA$ and compute its singular values for representative datasets in the training data. For all cases considered, the smallest singular value is at least $15\%$ of the largest singular value, and is therefore well above the numerical tolerance
\begin{equation}
\frac{\sigma_{\min}}{\sigma_{\max}}
\gtrsim
0.15
\gg
\tau.
    % 0.15
    % \lesssim
    % \frac{\sigma_{\min}}{\sigma_{\max}}.
    % \lesssim
    % 0.2.
\end{equation}
Thus, $\bfA$ has full numerical rank, which implies that its numerical nullity is zero
\begin{equation}
    \operatorname{rank}_{\tau}(\bfA)=6,
    \qquad
    \dim\ker_{\tau} \bfA=0.
\end{equation}
Consequently, no non-identifiable material parameter directions are detected within the prescribed numerical tolerance, and our standardized experiment exhibits no detectable first-level non-uniqueness. We note that this conclusion is restricted to the chosen material class with six parameters and to the fixed-displacement first-level analysis. It does not prove the uniqueness of the full constitutive-function inverse problem. Non-uniqueness may reappear when the material class is enriched by introducing coupling terms, higher-order terms, or more general constitutive features, such as those in the trunk net of the PANO architecture. For the latter case, the numerical rank and nullity of the matrix $\bfA$ depend on the feature functions $\psi_i(\bfF)$ learned during PANO training.

\section{Conclusions}
\label{sec:Conclusions}

We presented a novel operator learning framework for directly inferring constitutive functions from experimentally measurable quantities, namely displacement fields and reaction force functions. The proposed network architectures, PANO and CANO, successfully learn the mapping from experimental data to constitutive functions using the training dataset and demonstrate accurate predictions on unseen test data. Importantly, the projection onto Laplacian eigenfunctions enables the trained neural operators to perform robust inference from noisy measurements, incomplete data, varying spatial discretizations, and specimens with different dimensions. At the same time, the discovered constitutive functions can be evaluated at arbitrary deformation states, which makes the predictions independent of the discretization used during training.

Several directions for future research emerge from the present work. The proposed framework was developed and validated for a thin plate with a central hole. Extending the approach to varying specimen geometries represents a natural next step and would further enhance its applicability to experimental settings.
The present framework employs Laplacian eigenfunctions to encode the displacement fields. Future work may systematically investigate alternative basis functions and assess their influence on the accuracy, robustness, and generalization capabilities of the neural operators.
Although our proposed approach is independent of the spatial discretization, it is not invariant with respect to the temporal discretization. This is a reasonable assumption, as experimental data can generally be acquired at high frequencies in time. Nevertheless, future work could achieve time-discretization independence by representing both the displacement fields and the reaction force functions via an appropriate temporal basis.
In future work, we further aim to broaden the applicability of neural operators across different experimental configurations by making them independent of specimen geometry and topology.
% Alternative sampling strategies for generating the training data in the invariant space also merit further investigation.
Furthermore, the present study focused on data obtained from experiments on complex specimen geometries. Extending the framework to simpler experimental configurations with homogeneous deformation states is expected to be straightforward and may facilitate applications in standard material characterization procedures.
Finally, an important direction for future research concerns the design of the constitutive model space used during training data generation. We expect that the generality and predictive capabilities of the proposed framework can be further improved by enriching the training database with a broader range of constitutive models. In particular, future work will consider more complex material classes, including anisotropic and dissipative constitutive behavior.

In summary, our work demonstrates that the inverse discovery of constitutive functions from experimental data can be significantly accelerated by leveraging the information contained in previously generated simulation data. In conventional approaches, such as Finite Element Model Updating, solving an inverse problem requires a large number of forward simulations, which are typically discarded once the optimization has converged. We envision a paradigm in which these simulation datasets are instead retained and used to train general neural operators that can subsequently solve inverse problems directly from experimental measurements to eliminate the need for repeated computationally expensive forward simulations.

\section*{Code and data availability}

The code and data will be publicly available upon publication at \url{https://github.com/mflaschel}.
% \begin{center}
%     \url{https://github.com/mflaschel/neural-operators-hyperelasticity} .
% \end{center}

\section*{Acknowledgments}

The authors gratefully acknowledge the use of an existing finite element method implementation in FEniCSx developed by Miguel Angel Moreno-Mateos and Simon Wiesheier.
The authors acknowledge support from the European Research Council (ERC) Grant 101141626 DISCOVER. Funded by the European Union. Views and opinions expressed are, however, those of the authors only and do not necessarily reflect those of the European Union or the European Research Council
Executive Agency. Neither the European Union nor the granting authority can be held responsible for them.
% The authors acknowledge the scientific support and HPC resources provided by the Erlangen National High Performance Computing Center NHR@FAU of the Friedrich-Alexander-Universität Erlangen-Nürnberg (FAU). The hardware is funded by the German Research Foundation (DFG).
% The authors utilized generative artificial intelligence to enhance the writing style in certain sections of the manuscript.
% After using these tools, the authors reviewed and edited the content as needed and take full responsibility for the content of the publication.

% \clearpage

\appendix
\crefalias{section}{appendix}

\section{Forward boundary value problem}
\label{app:forward_problem}

In the following, we detail the forward boundary value problem that is solved numerically to generate the simulation data used for training the neural operators. We consider an incompressible hyperelastic material that occupies the domain $\Omega^{\text{3D}}$ in the undeformed configuration and deforms over the time interval $[0,T]$. We neglect the moment of inertia and body forces throughout this work. Further, we assume a specimen under full displacement control, such that the Neumann boundary conditions are homogeneous. The boundary value problem then states
\begin{equation}
    \begin{aligned}
        &\bfF(\bfX,t) = \nabla_{\bfX} \bfu(\bfX,t) + \bfI \quad \text{in } \Omega^{\text{3D}} \times [0,T],\\
        &J(\bfX,t) = \det(\bfF(\bfX,t)) = 1, \quad \text{in } \Omega^{\text{3D}} \times [0,T],\\
        &\bfP(\bfX,t) = \frac{\partial \bar W}{\partial \bfF} - p(\bfX,t) \bfF\mtransposed(\bfX,t), \quad \text{in } \Omega^{\text{3D}} \times [0,T],\\
        &\nabla_{\bfX} \cdot \bfP(\bfX,t) = \boldsymbol{0}, \quad \text{in } \Omega^{\text{3D}} \times [0,T],\\
        &u_1(\bfX,t) = \bar u_1(\bfX,t), \quad \text{on } \Gamma^{\text{3D}}_1 \times [0,T],\\
        &u_2(\bfX,t) = \bar u_2(\bfX,t), \quad \text{on } \Gamma^{\text{3D}}_2 \times [0,T],\\
        &P_{1j}(\bfX,t) N_j(\bfX,t) = 0, \quad \text{on } \Gamma^{\text{3D}}_{N1} \times [0,T],\\
        &P_{2j}(\bfX,t) N_j(\bfX,t) = 0, \quad \text{on } \Gamma^{\text{3D}}_{N2} \times [0,T],\\
    \end{aligned}
\end{equation}
where $\Gamma^{\text{3D}}_i$ denote the Dirichlet boundaries at which the displacements $\bar u_i$ are prescribed, and $\Gamma^{\text{3D}}_{Ni}$ are the homogeneous Neumann boundaries with outward unit normal vector $\bfN$. Note that the full boundary is composed of the Dirichlet and Neumann boundaries $\Gamma^{\text{3D}} = \Gamma^{\text{3D}}_i \cup \Gamma^{\text{3D}}_{Ni}$. For generating the simulation data, we use an existing finite element method implementation. We refer to \cite{flaschel_unsupervised_2026} for details. The simulations are performed on a tetrahedral mesh. The displacements on the surface, $\Omega^{\text{2D}}$, can then be extracted on the corresponding triangular surface mesh, whose nodes coincide with a subset of the tetrahedral mesh nodes.

\section{Admissible range and sampling of the invariants}
\label{app:admissible_invariants}
The admissible region of the isochoric principal invariants has been studied by \cite{sawyers_possible_1977}, see also the recent works by \cite{baaser_reformulation_2013,dammas_when_2025,wiesheier_data-adaptive_2026-1}, who found that the admissible region is bounded by the uniaxial tension (UT) and equibiaxial tension (BT) curves in the invariants plane. Assuming incompressibility and uniaxial tension, the deformation gradient is $\bfF=\text{diag}(\lambda,\lambda^{-1/2},\lambda^{-1/2})$ and consequently $\bfC=\text{diag}(\lambda^2,\lambda^{-1},\lambda^{-1})$. The invariants compute to 
\begin{equation}
\label{eq:curve_UT}
    \bar I_1 = \lambda^2 + 2\lambda^{-1}, \quad \bar I_2 = 2 \lambda + \lambda^{-2}. \quad \text{(uniaxial tension/compression)}
\end{equation}
For incompressible equibiaxial tension, we have $\bfF=\text{diag}(\lambda,\lambda,\lambda^{-2})$ and $\bfC=\text{diag}(\lambda^2,\lambda^2,\lambda^{-4})$. The invariants compute to 
\begin{equation}
\label{eq:curve_EBT}
    \bar I_1 = \lambda^{-4} + 2\lambda^{2}, \quad \bar I_2 = 2 \lambda^{-2} + \lambda^4. \quad \text{(equibiaxial tension/compression)}
\end{equation}
It is easy to see that the possible deformations of uniaxial tension and equibiaxial tension, and hence also the corresponding invariants, are equivalent if we substitute $\lambda=\hat \lambda^{-1/2}$ into the equibiaxial deformation. Hence, \cref{eq:curve_UT,eq:curve_EBT} describe the same curves in the invariants plane, which are parametrized by $\lambda$. Specifically, the curve for uniaxial tension with $\lambda \geq 1$ equals the curve for equibiaxial tension with $\lambda \leq 1$, and vise versa. As discussed by the aforementioned works, these curves bound the admissible region of the invariants \citep{sawyers_possible_1977,baaser_reformulation_2013,dammas_when_2025,wiesheier_data-adaptive_2026-1}.

\begin{figure}[!h]
    \centering
    \includegraphics[width=0.3\textwidth]{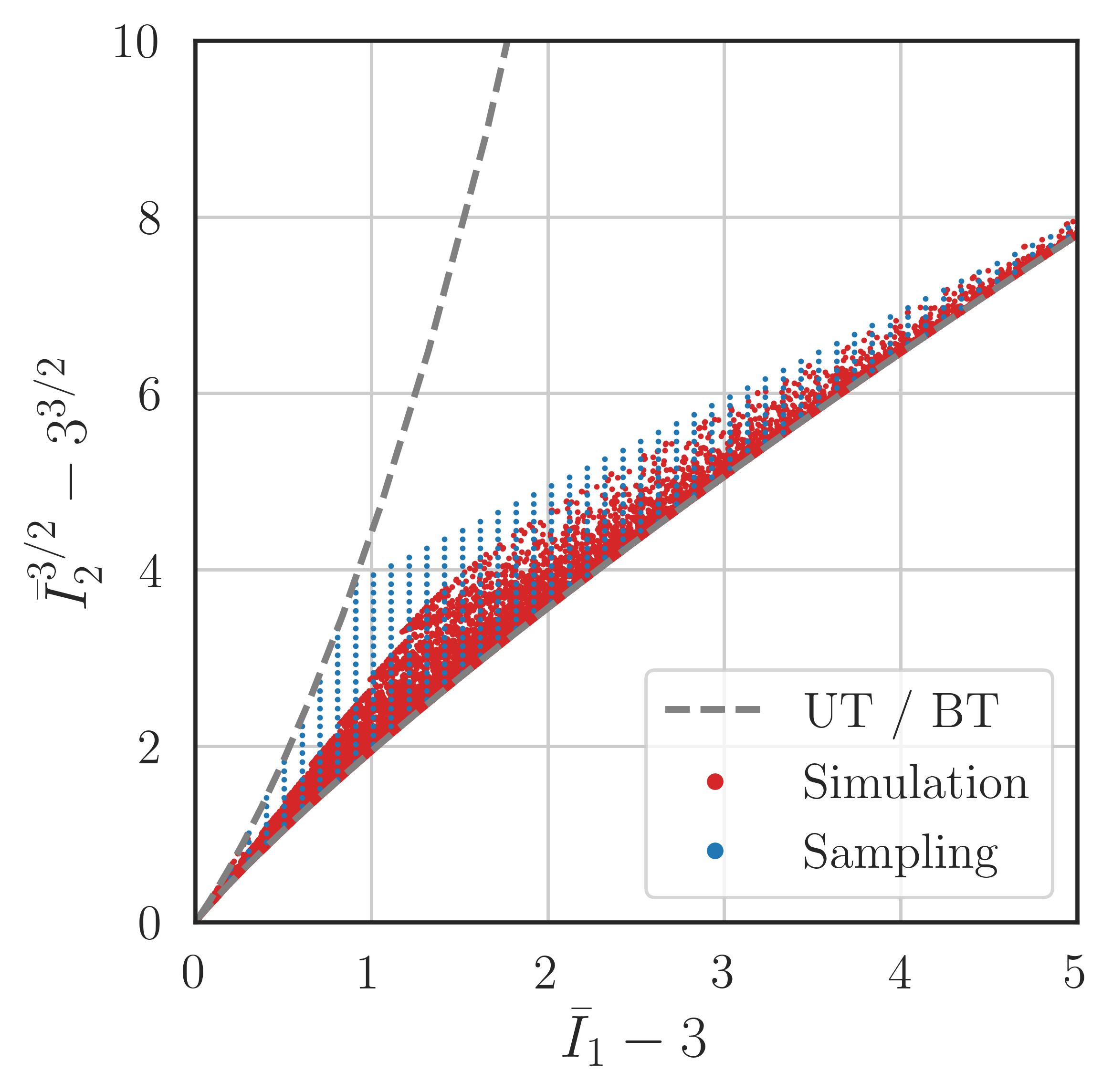}
    \caption{Invariants for a representative simulation and the corresponding sampling used for training data generation.}
    \label{fig:invariants_sampling}
\end{figure}

The mapping $(\bar I_1, \bar I_2) \mapsto (\bar I_1 - 3, \bar I_2^{3/2} - 3^{3/2})$ preserves the topology of the admissible domain, and hence the curve defined by 
\begin{equation}
    \bar I_1 = \lambda^2 + 2\lambda^{-1} - 3, \quad \bar I_2^{3/2} = [2 \lambda + \lambda^{-2}]^{3/2} - 3^{3/2},
\end{equation}
describes the bounds of the admissible region of the polyconvex invariants set $(\bar I_1 - 3, \bar I_2^{3/2} - 3^{3/2})$. We denote the admissible region by $\mathcal{I}_{\text{adm}}$ and refer to \cref{fig:invariants_sampling} for an illustration.

The generation of training data for the neural operators requires sampling the invariants. In this work, we sample invariants in the domain $\mathcal{I}_{\text{sample}} = \{ (\bar I_1^*, \bar I_2^*) \in \mathcal{I}_{\text{adm}} \, | \, \bar I_2^* \leq \bar I_1^* + 3 \}$. This sampling domain is chosen heuristically to capture the range of invariants observed in a representative simulation (see \cref{fig:invariants_sampling}). We note that in our numerical experiments, different sampling strategies did not significantly affect the training results. The primary purpose of sampling the invariants is to provide a quantifiable measure of the mismatch between the true and predicted isochoric strain-energy density functions. Since these functions are defined on unbounded domains, there is no universal way to numerically quantify the discrepancy between them.

% \cap [0,10]^2

\section{Taylor expansion of the strain-energy density}
\label{app:Taylor_expansion}

We recall $I_1^*=\bar I_1 - 3$ and $I_2^*=\bar I_2^{3/2} - 3^{3/2}$, and write the isochoric strain-energy density function as $\bar W(I_1^*,I_2^*)$. Any general, sufficiently smooth function $\bar W(I_1^*,I_2^*)$ can be approximated through a Taylor expansion about the undeformed state $(I_1^*,I_2^*)=(0,0)$
\begin{equation}
\begin{aligned}
\bar W(I_1^*,I_2^*) =\;& \bar W(0,0)
+ I_1^*\, \bar W_{,I_1^*}(0,0)
+ I_2^*\, \bar W_{,I_2^*}(0,0) \\
&+ \frac{1}{2}\left[
[I_1^*]^2 \bar W_{,I_1^* I_1^*}(0,0)
+ 2 I_1^* I_2^* \bar W_{,I_1^* I_2^*}(0,0)
+ [I_2^*]^2 \bar W_{,I_2^* I_2^*}(0,0)
\right] \\
&+ \frac{1}{6}\left[
[I_1^*]^3 \bar W_{,I_1^* I_1^* I_1^*}(0,0)
+ 3 [I_1^*]^2 I_2^* \bar W_{,I_1^* I_1^* I_2^*}(0,0)
+ 3 I_1^* [I_2^*]^2 \bar W_{,I_1^* I_2^* I_2^*}(0,0)
+ [I_2^*]^3 \bar W_{,I_2^* I_2^* I_2^*}(0,0)
\right] \\
&+ \mathcal{O}\!\left(\|(I_1^*,I_2^*)\|^4\right).
\end{aligned}
\end{equation}
Here, we used the comma convention for denoting partial derivatives.
We note that $\bar W(0,0) = 0$, and by introducing the respective constants $C_{ij}$, this can be written as
\begin{equation}
\begin{aligned}
\bar W(I_1^*,I_2^*) =\;&
C_{10} \, I_1^*
+ C_{01} \, I_2^*
+ C_{20} [I_1^*]^2
+ C_{11} I_1^* I_2^*
+ C_{02} [I_2^*]^2
 \\
&+ C_{30} [I_1^*]^3 
+ C_{21} [I_1^*]^2 I_2^*
+ C_{12} I_1^* [I_2^*]^2
+ C_{03} [I_2^*]^3
+ \mathcal{O}\!\left(\|(I_1^*,I_2^*)\|^4\right).
\end{aligned}
\end{equation}
This is a modified version of the generalized Mooney-Rivlin model, which depends directly on the invariants $I_1^*$ and $I_2^*$ instead of $I_1$ and $I_2$. As higher-order contributions are generally negligible for most practical applications, only terms up to third order are considered for training data generation in the present work. Further, since we restrict our attention to separable strain-energy densities, we neglect coupling terms and consider functions of the form
\begin{equation}
\bar W(I_1^*,I_2^*) =
C_{10} \, I_1^*
+ C_{01} \, I_2^*
+ C_{20} [I_1^*]^2
+ C_{02} [I_2^*]^2
+ C_{30} [I_1^*]^3 
+ C_{03} [I_2^*]^3,
\end{equation}
in which we set $C_{ij} \geq 0$.

\section{Laplacian displacement field encoding}
\label{app:Laplacian_encoding}

To construct a discretization-independent neural operator, we map the input displacement fields to a finite basis in the first layer of the branch net. We consider an interpolation with basis functions that have non-local support on the domain $\Omega^{\text{2D}}$ such that the interpolation function fulfills the Dirichlet boundary conditions of the displacement fields. We additively split the displacement fields
\begin{equation}
\label{eq:Laplacian_lifting}
    u_1(\bfX,t) = u_1^{\text{lift}}(\bfX,t) + u_1^{\text{hom}}(\bfX,t), \quad
    u_2(\bfX,t) = u_2^{\text{lift}}(\bfX,t) + u_2^{\text{hom}}(\bfX,t), \\
\end{equation}
into a priori chosen lifting fields $u_i^{\text{lift}}(\bfX,t)$, which fulfill the Dirichlet boundary conditions, and the remaining fields $u_i^{\text{hom}}(\bfX,t)$, which vanish on the Dirichlet boundaries. Specifically, for the geometry and loading conditions described in \cref{sec:experimental design}, we choose $u_1^{\text{lift}}(\bfX,t) = 0$ and $u_2^{\text{lift}}(\bfX,t) = \bar u_2(t) [1 - X_2]$, which fulfill the boundary conditions shown in \cref{fig:geometry}.

It remains to interpolate the functions $u_i^{\text{hom}}(\bfX,t)$
\begin{equation}
\label{eq:Laplacian_interpolation_hom}
    u_1^{\text{hom}}(\bfX,t) \approx \sum_{j=1}^{N_{\phi}} a_{1jt} \, \phi_{1j}(\bfX), \quad
    u_2^{\text{hom}}(\bfX,t) \approx \sum_{j=1}^{N_{\phi}} a_{2jt} \, \phi_{2j}(\bfX), \\
\end{equation}
where $a_{ijt}$ are interpolation coefficients and $\phi_{ij}(\bfX)$ are basis functions that vanish on the respective Dirichlet boundaries. We note that due to the lifting of the displacement fields, the basis functions are assumed to be independent of time.

\begin{figure}[!h]
    \centering
    \begin{subfigure}[b]{0.19\textwidth}
        \centering
        \includegraphics[width=\textwidth]{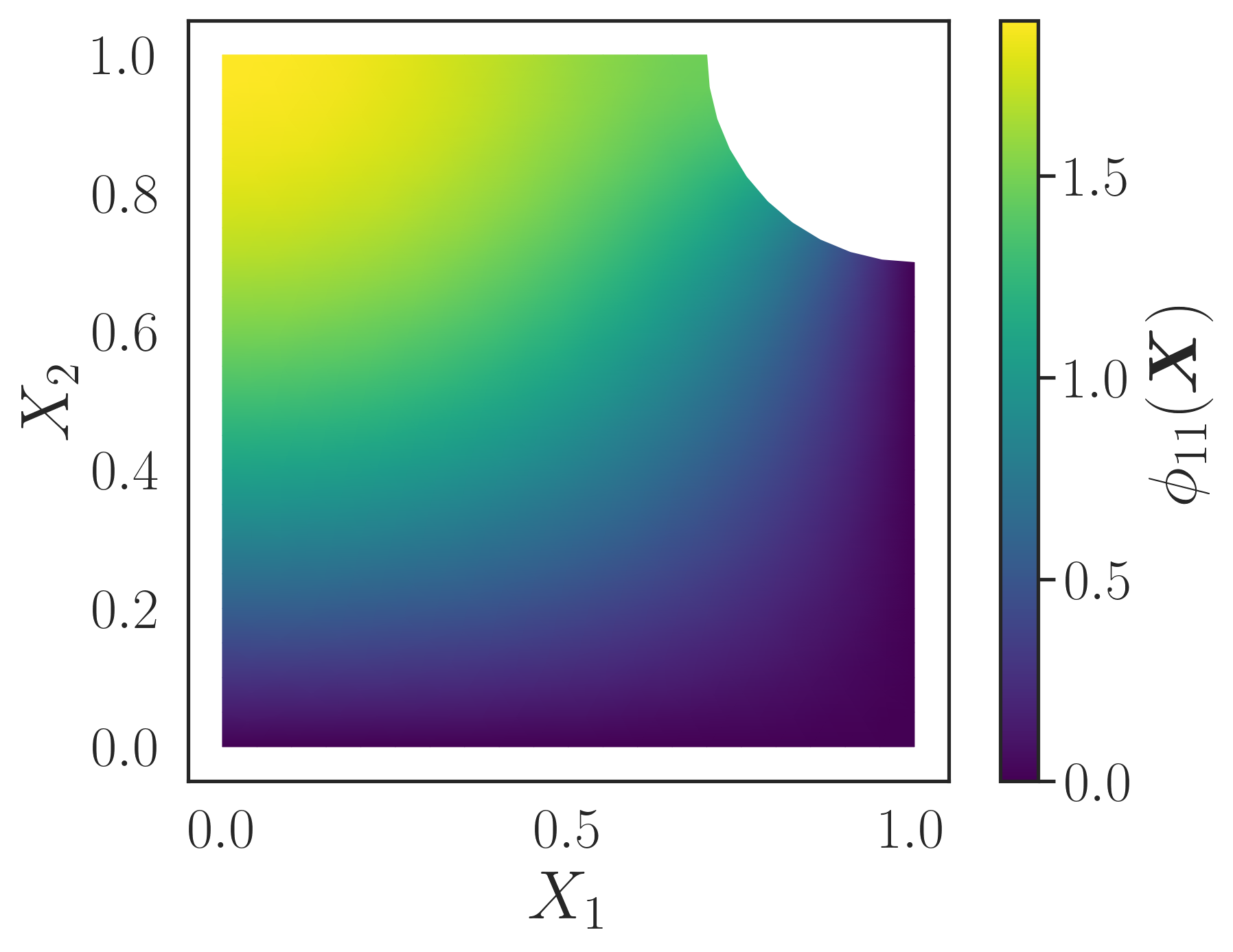}
        \caption{$\phi_{11}(\bfX)$}
    \end{subfigure}
    \begin{subfigure}[b]{0.19\textwidth}
        \centering
        \includegraphics[width=\textwidth]{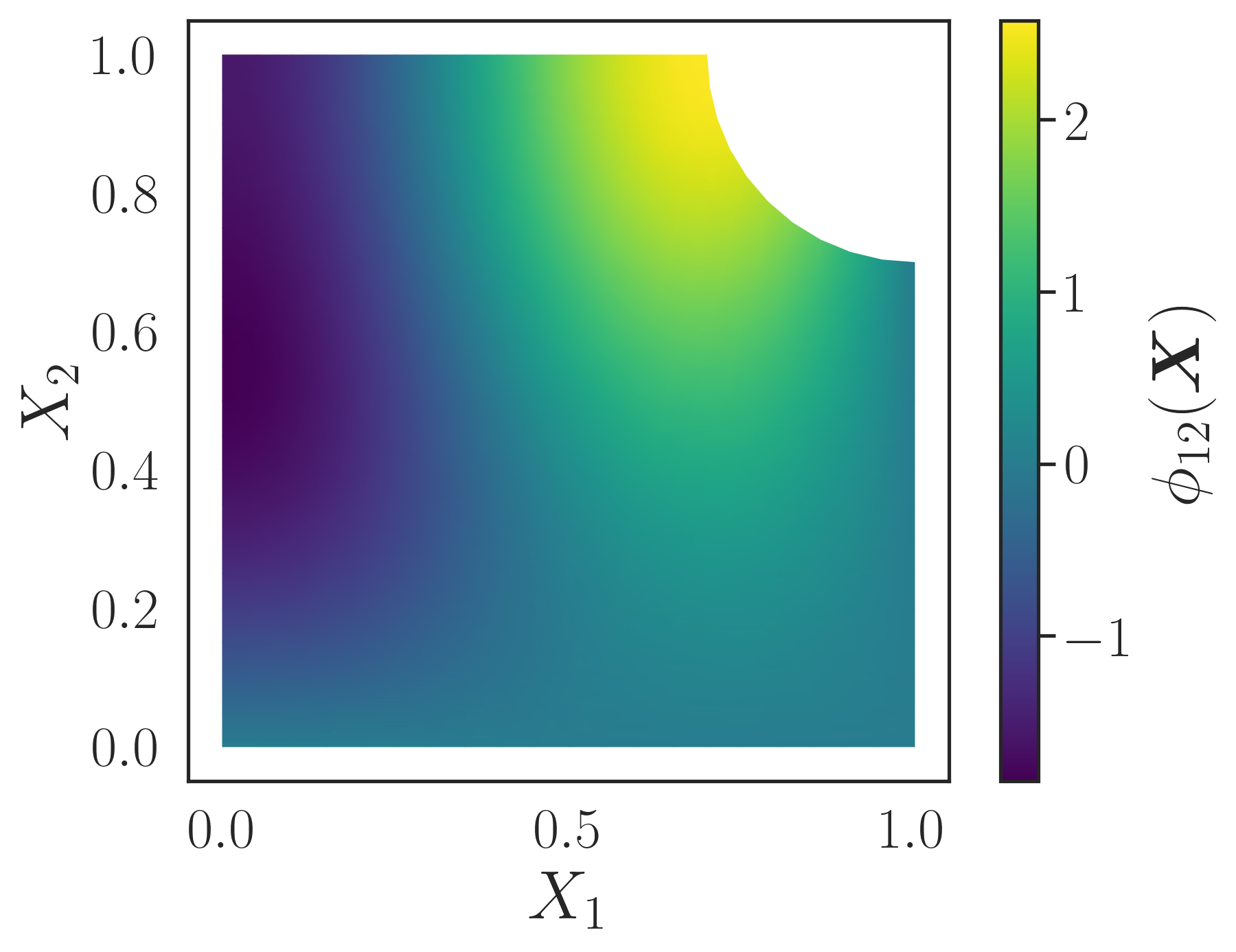}
        \caption{$\phi_{12}(\bfX)$}
    \end{subfigure}
    \begin{subfigure}[b]{0.19\textwidth}
        \centering
        \includegraphics[width=\textwidth]{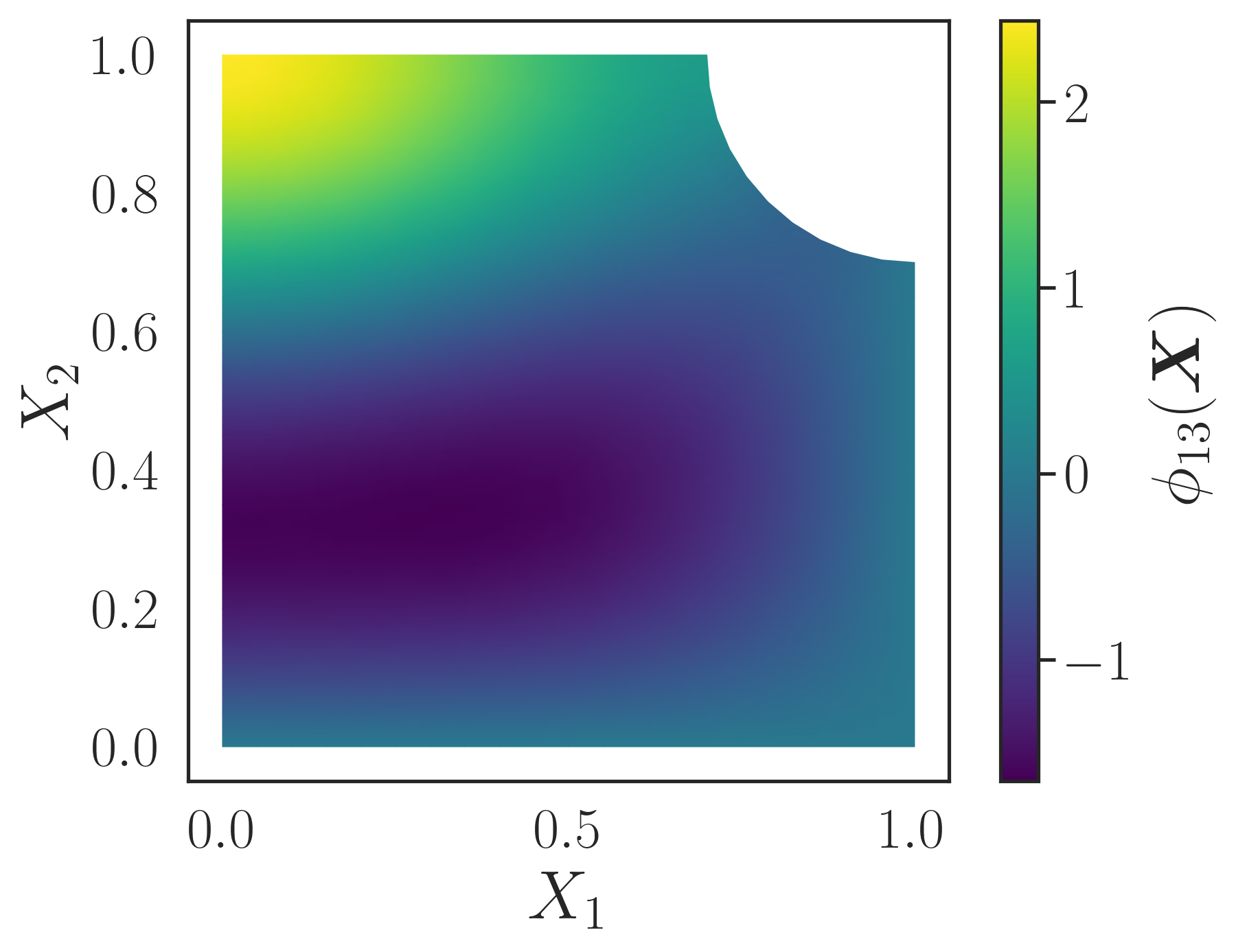}
        \caption{$\phi_{13}(\bfX)$}
    \end{subfigure}
    \begin{subfigure}[b]{0.19\textwidth}
        \centering
        \includegraphics[width=\textwidth]{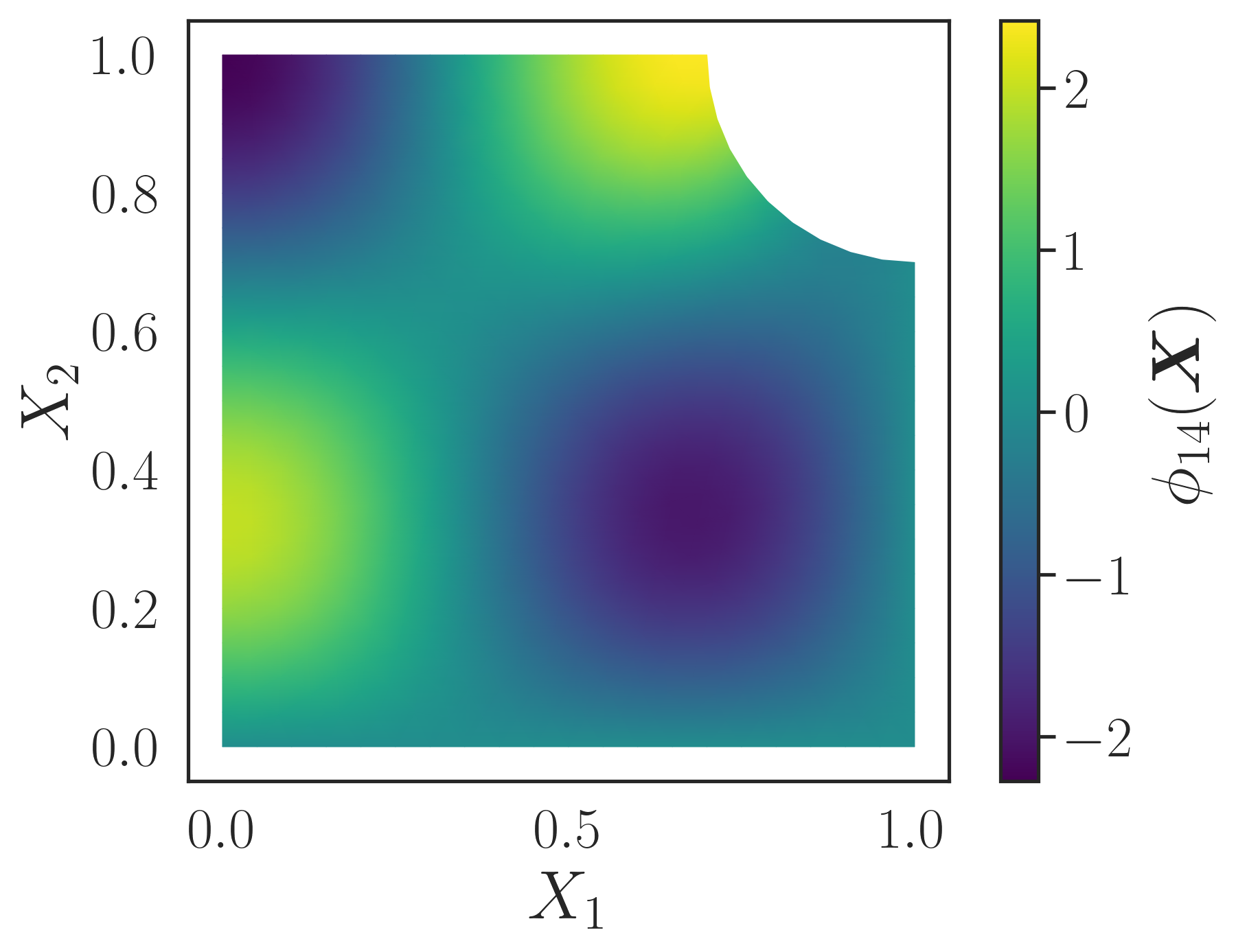}
        \caption{$\phi_{14}(\bfX)$}
    \end{subfigure}
    \begin{subfigure}[b]{0.19\textwidth}
        \centering
        \includegraphics[width=\textwidth]{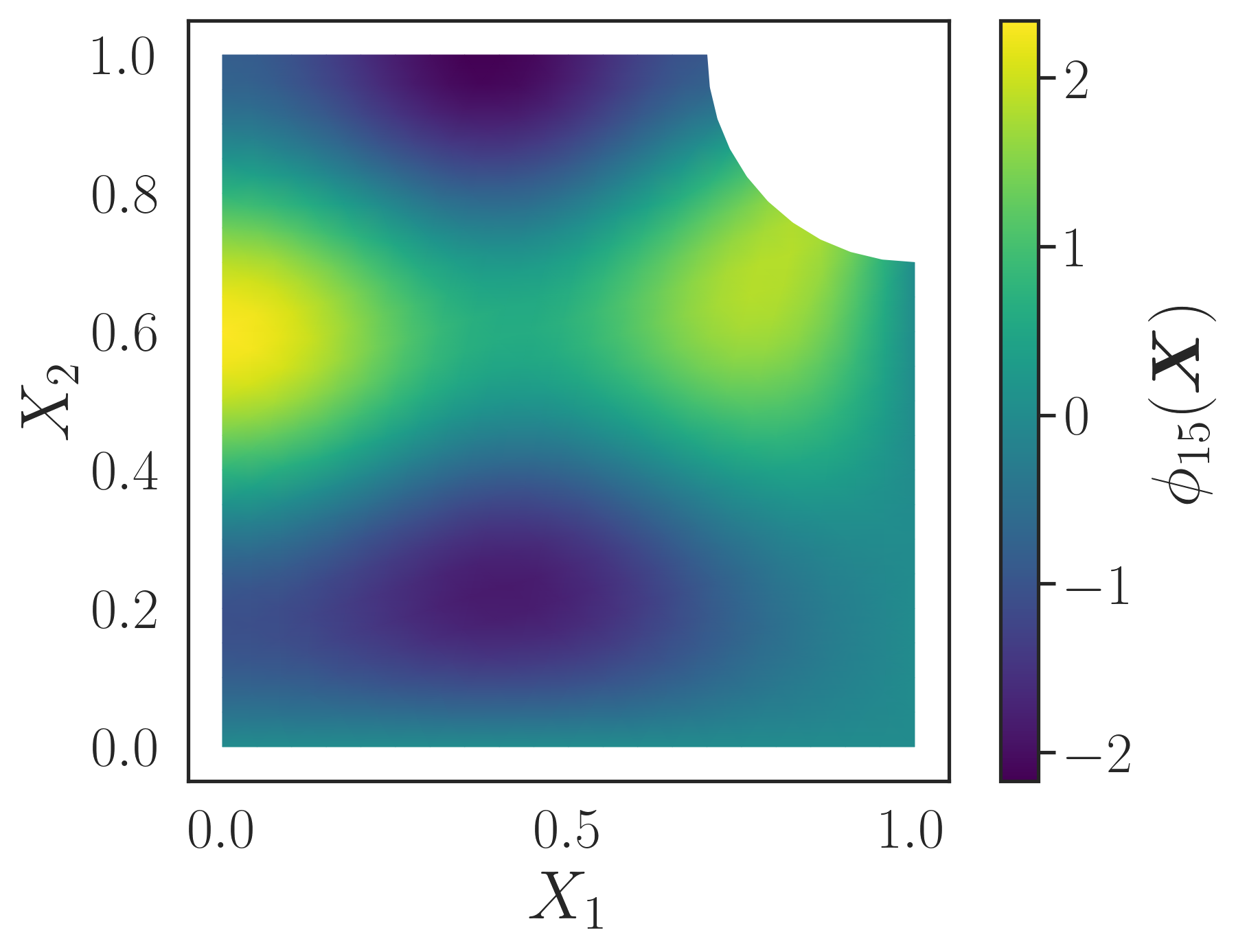}
        \caption{$\phi_{15}(\bfX)$}
    \end{subfigure}

    \centering
    \begin{subfigure}[b]{0.19\textwidth}
        \centering
        \includegraphics[width=\textwidth]{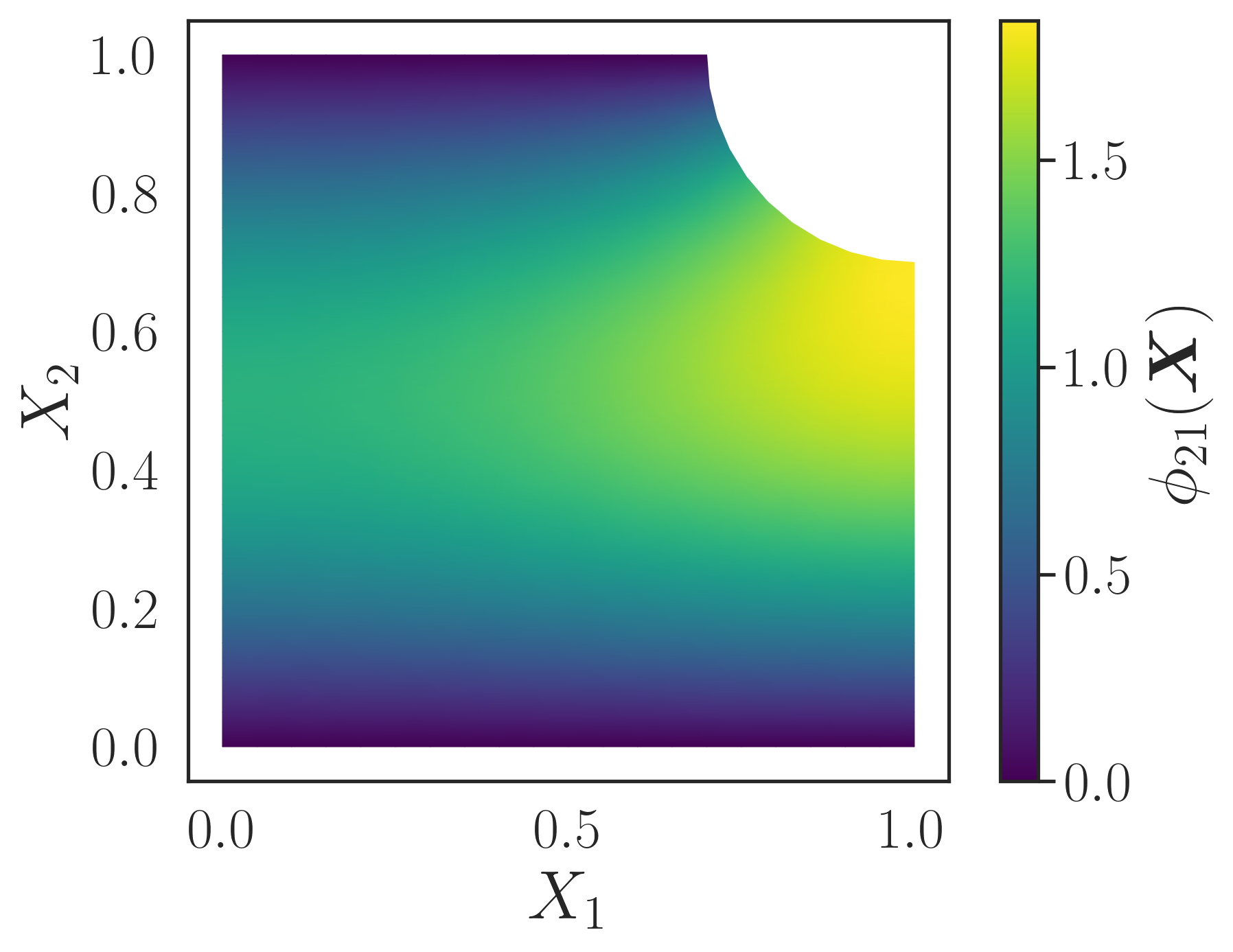}
        \caption{$\phi_{21}(\bfX)$}
    \end{subfigure}
    \begin{subfigure}[b]{0.19\textwidth}
        \centering
        \includegraphics[width=\textwidth]{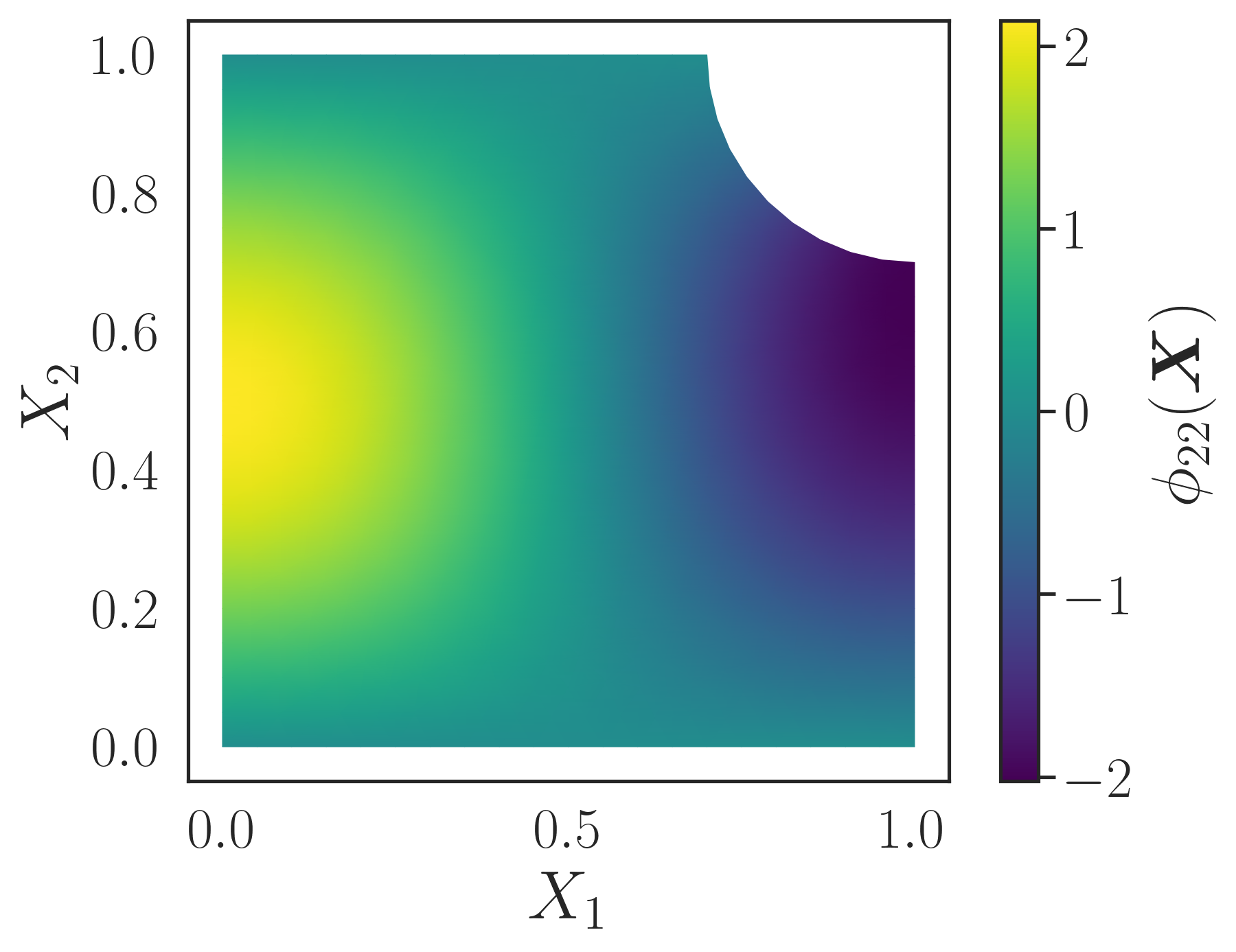}
        \caption{$\phi_{22}(\bfX)$}
    \end{subfigure}
    \begin{subfigure}[b]{0.19\textwidth}
        \centering
        \includegraphics[width=\textwidth]{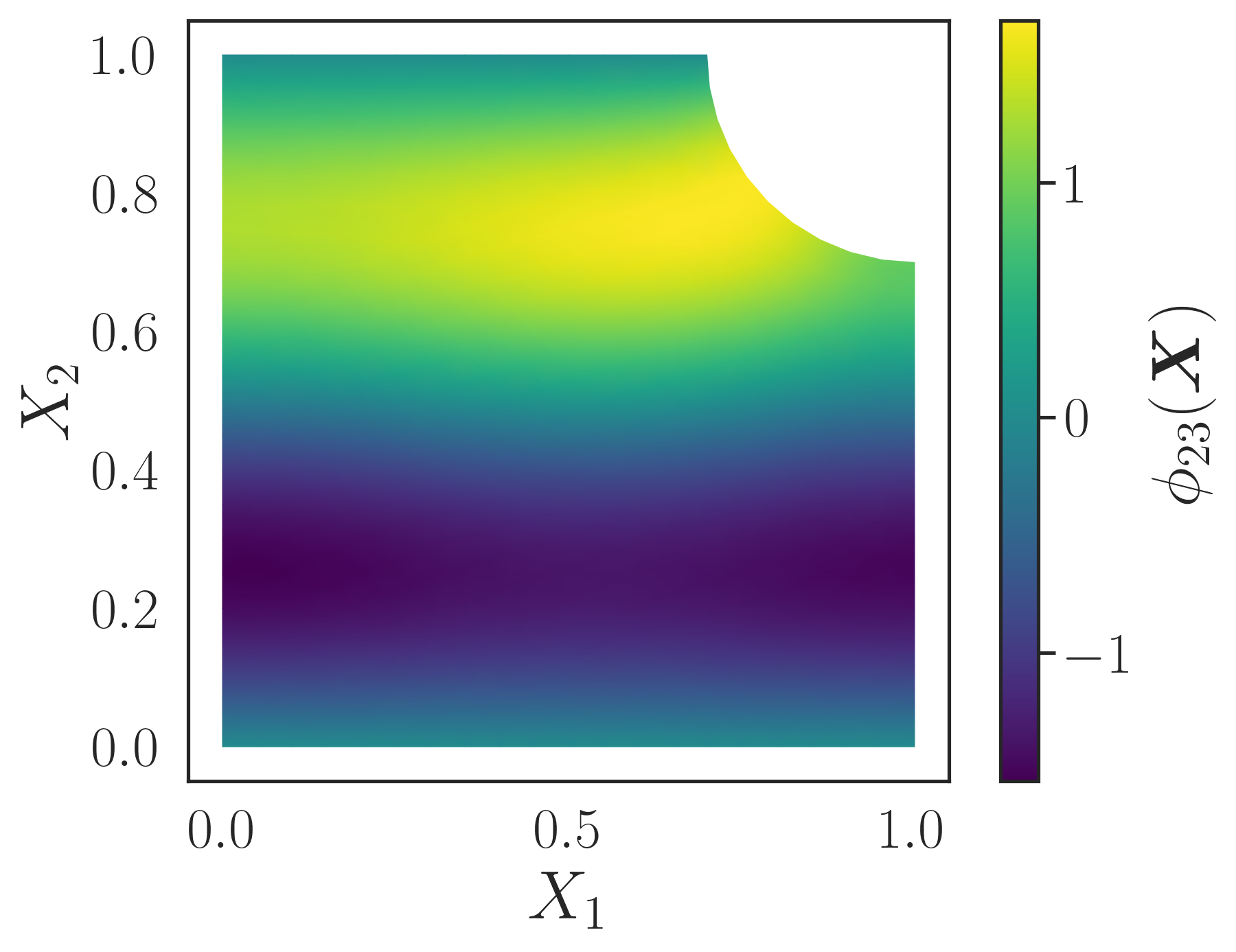}
        \caption{$\phi_{23}(\bfX)$}
    \end{subfigure}
    \begin{subfigure}[b]{0.19\textwidth}
        \centering
        \includegraphics[width=\textwidth]{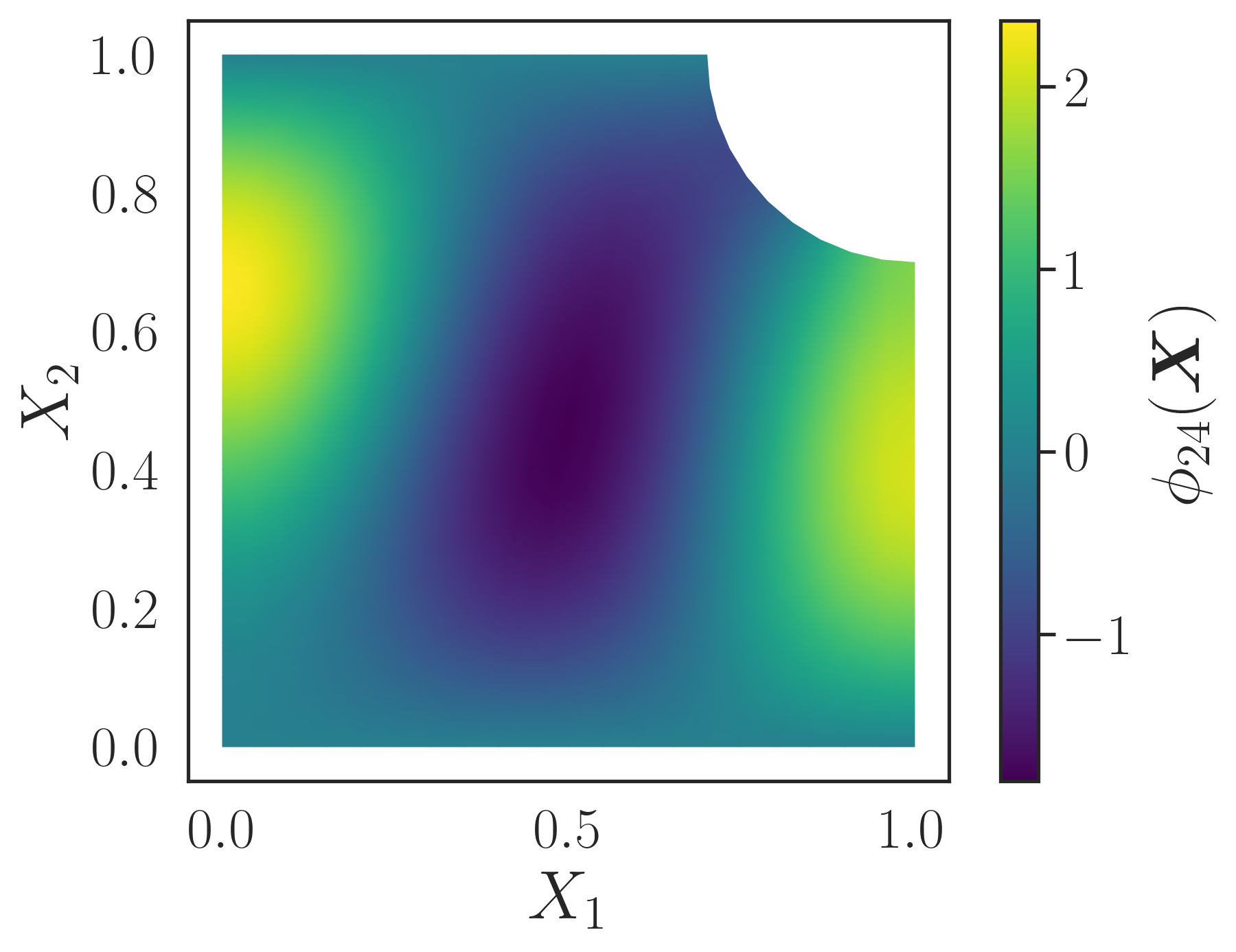}
        \caption{$\phi_{24}(\bfX)$}
    \end{subfigure}
    \begin{subfigure}[b]{0.19\textwidth}
        \centering
        \includegraphics[width=\textwidth]{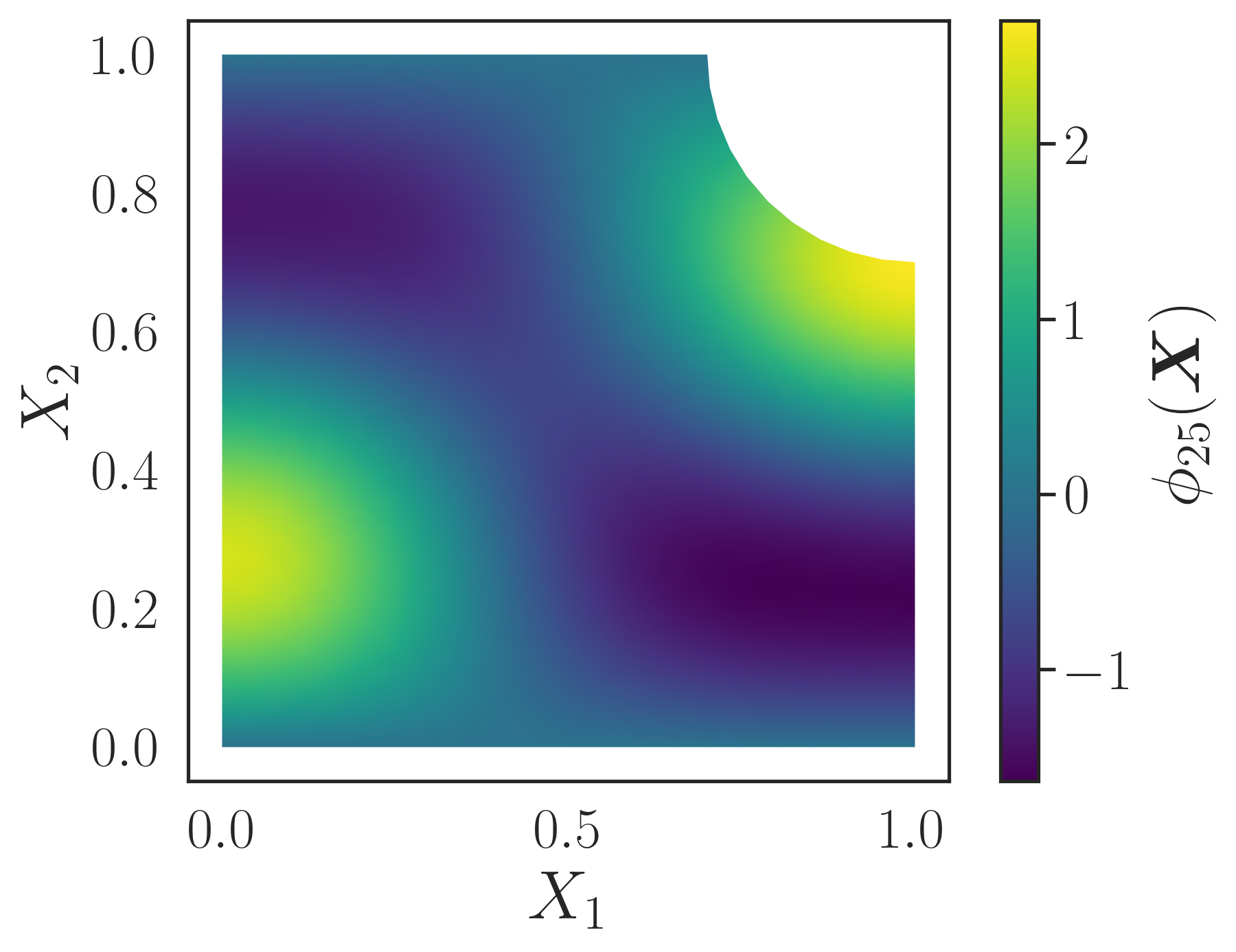}
        \caption{$\phi_{25}(\bfX)$}
    \end{subfigure}
    \caption{Laplacian eigenfunctions $\phi_{ij}(\bfX)$.}
    \label{fig:eigenfunctions}
\end{figure}

While different choices of basis functions are possible and are expected to yield comparable neural operator accuracy, we focus here on a specific set of basis functions that proved particularly effective in our numerical studies. The basis functions are chosen as the eigenfunctions of the Laplace operator
\begin{equation}
    - \nabla \phi_{ij}(\bfX) = \lambda_{ij} \phi_{ij}(\bfX), \quad \text{(no Einstein summation)}
\end{equation}
subject to homogeneous Dirichlet conditions on the displacement-constrained boundaries and homogeneous Neumann conditions on the remaining parts of the boundary.
To generate the basis functions, we discretize the eigenvalue problem above on a triangular finite element mesh.
This yields the generalized eigenvalue problem
\begin{equation}
    \bfK \bfphi^h_{ij} = \lambda_{ij} \bfM \bfphi^h_{ij}, \quad \text{(no Einstein summation)}
\end{equation}
where $\bfphi^h_{ij} \in \Rset^{N_X}$ denotes the vector of nodal values of the basis function on the chosen reference mesh, and $\bfK$ and $\bfM$ denote the associated finite element stiffness and mass matrices.
Since $u_1$ and $u_2$ generally do not share the same Dirichlet boundaries, we impose the homogeneous Dirichlet boundary conditions separately for each displacement component, yielding two generalized eigenvalue problems
\begin{equation}
    \bfK_1^{\text{free}} \bfphi^h_{1j} = \lambda_{1j} \bfM_1^{\text{free}} \bfphi^h_{1j}, \quad
    \bfK_2^{\text{free}} \bfphi^h_{1j} = \lambda_{2j} \bfM_2^{\text{free}} \bfphi^h_{2j}. \quad \text{(no Einstein summation)}
\end{equation}
Solving the eigenvalue problems finally yields the basis functions $\bfphi^h_{1j}$ and $ \bfphi^h_{2j}$, from which we retain the first $N_\phi$ eigenfunctions, corresponding to the lowest-frequency spatial modes. The first five eigenfunctions are illustrated in \cref{fig:eigenfunctions}. These functions are used to interpolate the displacement fields, as shown in \cref{eq:Laplacian_interpolation_hom}. Since $N_\phi < N_X$, in general, this reduces the dimensionality of the displacement data. For our training dataset, we have $N_X=499$ and $N_\phi=100$. We further expect that the interpolation improves the neural operators' robustness to spatial noise, as small local oscillations in the displacement fields have only a minor effect on the interpolation coefficients.

\section{Hyperparameters}
\label{app:hyperparameters}

The hyperparameters of the PANO and CANO architectures, together with the corresponding training settings, are summarized in \cref{tab:hyperparameters_pano_cano}. Most hyperparameters are chosen heuristically. However, for training the CANO, the number of neurons in the branch net, the initial learning rate, the weight decay, and the batch size, are optimized using a random search with the Python library \texttt{optuna}. Similarly, for training the PANO, the number of neurons in the branch and trunk nets, the number of latent features, the initial learning rate, the scheduler parameter \texttt{T\_0}, the weight decay, and the batch size, are optimized with \texttt{optuna}. To select the optimal parameters, we only use the validation data, not the testing data. Finally, for the PANO, we observed that restarting the training process three times for the selected hyperparameters further reduced the validation loss.

\begin{table}[h!]
\caption{
Hyperparameters of the PANO and CANO architectures and the training process.
}
\label{tab:hyperparameters_pano_cano}
\centering
\resizebox{\textwidth}{!}{
\begin{tabular}{|l|c|c|}
\hline
\multicolumn{3}{|l|}{\textbf{Network architecture}} \\
\hline
\textbf{Hyperparameter} & \textbf{PANO} & \textbf{CANO} \\
\hline

\# spatial points during training $N_{\bfX}$ & 499 & 499 \\
\# Laplacian eigenfunctions $N_\phi$ & 100 & 100 \\
\# time steps $N_{t}$ & 10 & 10 \\
\# invariant samples $N_{\bar I}$ & 481 & 481 \\
\# latent features $N_b$ & 12 & 6 \\
\# layers branch net & 1 & 1 \\
\# neurons branch net & 1024 & 4096 \\
Activation functions branch net & \texttt{ReLU} & \texttt{ReLU} \\
\# layers trunk net & 1 & 1 \\
\# neurons trunk net & 16 & N/A \\
Activation function trunk net & \texttt{Softplus} & N/A \\
\hline

\multicolumn{3}{|l|}{\textbf{Training configuration}} \\
\hline
\textbf{Hyperparameter} & \textbf{PANO} & \textbf{CANO} \\
\hline

Optimizer & \texttt{Adam} & \texttt{Adam} \\
\# epochs & 10,000 $\times$ 4 & 10,000 \\
Initial learning rate & \num{1.07e-05} & \num{3.50e-04} \\
Learning rate scheduler & \texttt{CosineAnnealingWarmRestarts} & \texttt{CosineAnnealingWarmRestarts} \\
Scheduler parameter \texttt{T\_0} & 4000 & 1000 \\
Scheduler parameter \texttt{T\_mult} & 1 & 1 \\
Scheduler parameter \texttt{eta\_min} & 0.0 & 0.0 \\
Weight decay & \num{3.36e-06} & \num{2.78e-06} \\
\# simulations $N_{\text{data}}$ & 3000 & 3000 \\
\# training simulations $N_{\text{data}}^{\text{train}}$ & 2400 & 2400 \\
\# validation simulations $N_{\text{data}}^{\text{val}}$ & 300 & 300 \\
\# testing simulations $N_{\text{data}}^{\text{test}}$ & 300 & 300 \\
Batch size & 32 & 32 \\
\hline

\end{tabular}
}
\end{table}

% \section{Convex monotone neural networks}
% \label{app:CMNN}
% \cite{amos_input_2017}

% \section{Finite element method implementation}
% \label{app:fem}
% \cite{flaschel_unsupervised_2026}

%\nocite{*}
\bibliographystyle{elsarticle-harv}
\bibliography{bib_Moritz}

%%%%%%%%%%%%%%%%%%%%%%%%%%%%%%%%%%%%%%%%%%%%%%%%%%%%%%%%%%%%%%%%%%%%%%%%%%%%%%%%%%%%%%%%%%%%%%%%%%%%%%%%%%%%%%%%%%%%%%%

\end{document}